\documentclass[lettersize,journal]{IEEEtran}
\usepackage{amsmath,amsfonts}
\usepackage{algorithmic}
\usepackage{algorithm}
\usepackage{amsmath}
\usepackage{array}
\usepackage[caption=false,font=normalsize,labelfont=rm,textfont=rm]{subfig}
\usepackage{textcomp}
\usepackage{stfloats}
\usepackage{url}
\usepackage{booktabs}
\usepackage{svg}
\usepackage{threeparttable}
\usepackage{verbatim}
\usepackage{xcolor}
\usepackage{multirow}
\usepackage{graphicx}
\usepackage{cite}
\begin{document}

\title{STF-GCN: A Multi-Domain Graph Convolution Network Method for Automatic Modulation Recognition via Adaptive Correlation}

\author{Mingyuan Shao,~\IEEEmembership{Graduate Student Member,~IEEE}, Zhengqiu Fu, Dingzhao Li, Fuqing Zhang, Yilin Cai, Shaohua Hong,~\IEEEmembership{Member,~IEEE}, Lin Cao, Yuan Peng
        and Jie Qi
\thanks{This work was supported in part by the National Natural Science Foundation of China under Grant 62371405 and Grant 62271426, and in part by Science and Technology on Underwater Test and Control Laboratory under Grant 2023-JCJQ-LB-030. \emph{(Corresponding author: Jie Qi).}}
\thanks{Mingyuan Shao, Fuqing Zhang and Jie Qi is with the School of Electronic Science and Engineering, Xiamen University, Xiamen, Fujian 361005, China (e-mail: shaomy666@stu.xmu.edu.cn; fuqingzhang1110@163.com; qijie@xmu.edu.cn).}
\thanks{Zhengqiu Fu, Dingzhao Li, Yilin Cai and Shaohua Hong are with the School of Informatics, Xiamen University, Xiamen, Fujian 361005, China, and also with the Key Laboratory of Southeast Coast Marine Information Intelligent Perception and Application, Ministry of Natural Resources, Xiamen 361005, China (email: fuzhengqiu@stu.xmu.edu.cn; lidingzhao@stu.xmu.edu.cn; ylcai@stu.xmu.edu.cn;
hongsh@xmu.edu.cn).}
\thanks{Lin Cao and Yuan Peng are with Science and Technology on Underwater Test and Control Laboratory, Dalian, Liaoning 116013, China (email: 150004269@qq.com; py970418@126.com).}
}

\maketitle

\begin{abstract}
Automatic Modulation Recognition (AMR) is an essential part of Intelligent Transportation System (ITS) dynamic spectrum allocation. However, current deep learning-based AMR (DL-AMR) methods are challenged to extract discriminative and robust features at low signal-to-noise ratios (SNRs), where the representation of modulation symbols is highly interfered by noise. 
Furthermore, current research on GNN methods for AMR tasks generally suffers from issues related to graph structure construction and computational complexity.
In this paper, we propose a Spatial-Temporal-Frequency Graph Convolution Network (STF-GCN) framework, with the temporal domain as the anchor point, to fuse spatial and frequency domain features embedded in the graph structure nodes. 
On this basis, an adaptive correlation-based adjacency matrix construction method is proposed, which significantly enhances the graph structure's capacity to aggregate local information into individual nodes. 
In addition, a PoolGAT layer is proposed to coarsen and compress the global key features of the graph, significantly reducing the computational complexity. 
The results of the experiments confirm that STF-GCN is able to achieve recognition performance far beyond the state-of-the-art DL-AMR algorithms, with overall accuracies of 64.35\%, 66.04\% and 70.95\% on the RML2016.10a, RML2016.10b and RML22 datasets, respectively. 
Furthermore, the average recognition accuracies under low SNR conditions from -14dB to 0dB outperform the state-of-the-art (SOTA) models by 1.20\%, 1.95\% and 1.83\%, respectively.
\end{abstract}
\section{Introduction}
\begin{figure*} 
  \centering            
\includegraphics[width=0.8\textwidth]{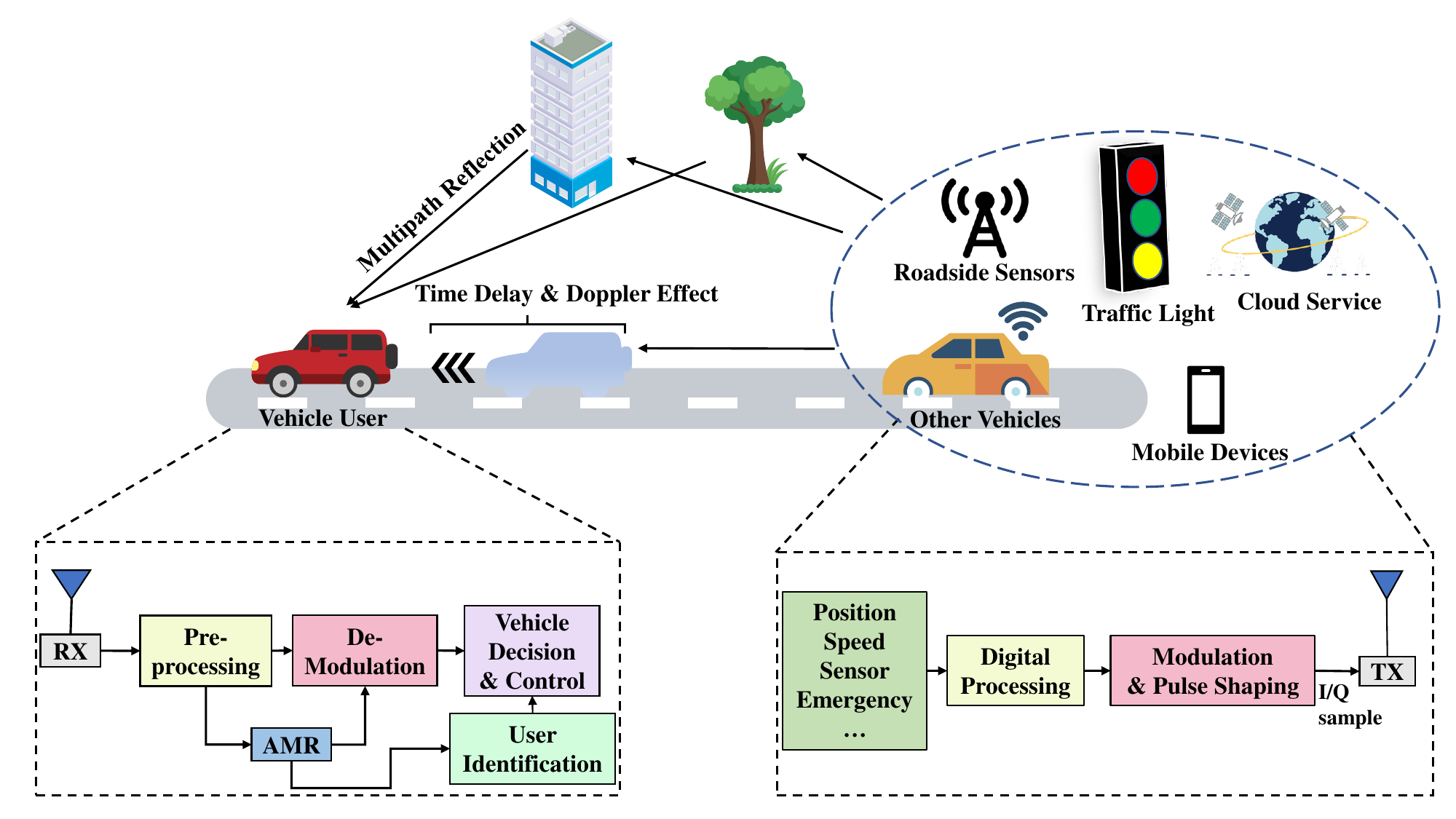}
  \caption{
A Vehicle-Road Collaboration scenario. The major tasks of AMR are to identify users and assist demodulation to achieve dynamic spectrum monitoring and efficient allocation of spectrum resources. In addition, it helps to detect and prevent malicious communication interference.} 
  \label{fig:task}
\end{figure*}

Vehicle-to-everything (V2X) technology represents a significant milestone in the evolution of Intelligent Transportation Systems (ITS) \cite{v2x1,6GITS,guiguan}. 
It is designed to enable seamless interaction and communication between vehicles and their environment, including connections between vehicles, road infrastructure, mobile devices, and network resources, shown in Fig. \ref{fig:task}. 
Recently, the 3rd Generation Partnership Project (3GPP) released new standards for New Radio Vehicle-to-Everything (NR-V2X) technology \cite{NR1,NR2}.
Many of these Vehicle-Road Collaboration (VRC) scenarios substantially raise the requirements for wireless communication, which poses new challenges for V2X in terms of spectrum resource management and safety assurance \cite{v2xamr,APP2}.
However, the unique characteristics of urban vehicular communication environments pose significant challenges to NR-V2X systems. Unlike static or less dynamic scenarios, urban environments are characterized by rapidly fluctuating channel conditions, severe multipath fading, and pronounced Doppler effects caused by high mobility. These factors, coupled with the dense coexistence of heterogeneous devices, result in highly dynamic and unpredictable spectral usage.
Furthermore, the stringent real-time and reliability requirements in VRC scenarios impose even greater demands on solutions. Consequently, there is an urgent need for more advanced methods capable of adapting to highly dynamic and challenging environments.

Automatic Modulation Recognition (AMR) \cite{AMR2, xiaoyang} serves as a core component of Cognitive Radio (CR) and plays a pivotal role in dynamic spectrum allocation and link adaptation.
In ITS, AMR quickly identifies the modulation types of both primary and secondary users \cite{user}, enabling dynamic spectrum monitoring and efficient allocation of spectrum resources to optimize utilization.
Additionally, it assists in detecting and preventing malicious communication interference \cite{LightAMC}.

Deep learning-based AMR (DL-AMR) has become a research hotspot in recent years due to its flexible architecture and excellent classification performance. The most common way to extract modulation features is to use convolutional neural network (CNN) \cite{STTMC, gui2, yang} to extract spatial features or recurrent neural network (RNN) \cite{LSTM2,DAE} to extract temporal features.
Some researchers have reported that hybrid convolutional recurrent neural network (CRNN) frameworks \cite{MCLDNN,PETCGDNN}, which combine CNN and RNN to achieve better recognition performance than single network structure. 
Although the feature extraction capability is enhanced, CRNN still experiences global feature loss due to the local generalization issue of CNN and the unidirectional cyclic iteration of RNN.
To extract more robust global features, some studies have introduced transformer-based frameworks \cite{zhaocaidan,FCA} or self-attention mechanisms \cite{FEA-T,GAFMAE}, which prevent local induction bias and unidirectional errors along with improving the global sensitivity of the model. 
dwever, the self-attention mechanism suffers from quadratic complexity, which results in a relatively slow inference time with same space complexity.

In conclusion, the above DL-AMR methods are capable of extracting features with a certain degree of differentiation. 
However, for low SNRs or modulation types within the same family, these methods still have limitations in feature extraction. This is because the modulated signal consists of locally and globally correlated modulation symbols, and the symbols are sensitive to noise interference when being mapped into a waveform. Current research primarily employs image and time analysis methods to extract features of modulated signals. 
However, these methods do not completely explore the relationships among modulation symbols. 
As a result, the extraction of modulation symbol information is susceptible to noise interference at low SNRs, which affects recognition performance.

Graph Neural Network (GNN) \cite{GNN} is a set of models for processing graph-structured data. 
Its variants are now widely used in time series applications such as traffic prediction \cite{Traffic1,graphwave}, indoor localization \cite{dingwei} and anomaly detection \cite{han}. 
Since graph structures excel at aggregating local node information and mining potential features, they are well-suited for mapping the connection between local and global features among modulation symbols. 
GNN has been applied to the field of AMR in several studies. 

However, there are still some problems in applying graph neural networks to AMR tasks. On the one hand, existing studies expose the problem of insufficiently rich information of graph feature nodes, for which the graph structure tends to focus only on its single-domain features (e.g., time domain, frequency domain, or spatial domain dominated by constellation diagrams). This leads to the graph structure that are not applicable to all modulation types, with limited recognition performance. On the other hand, the existing research has high complexity and structural redundancy and edge feature computation, resulting in performance that cannot fit the requirements in real-time communication scenarios. 
Therefore, it is still a challenge to design lightweight GNN-based AMR methods that balance high accuracy and ubiquity.

In this paper, we propose a lightweight graph neural network called STF-GCN, which is based on multi-domain feature fusion and constructed with an adaptive correlation-based graph structure. 
Specifically, we first use the time domain as the anchor point, integrate spatial and frequency domain information, and perform fusion embedding of multi-domain features with an encoder that combines CNN and Bi-LSTM. 
The fused multi-domain features are used to construct spatial-temporal-frequency (STF) graph nodes. 
Fused features whose feature dimensions and corresponding channels are sliced into graph nodes. 
Edge features among nodes enable learning about interactions and potential connections across dimensions. 
In addition, we propose an adaptive correlation-based method to generate the adjacency matrix, where the node features are embedded in the matrix with an adaptive correlation function to reduce the complexity of graph structure. 
Subsequently, we propose a staged PoolGAT layer for graph structure feature extraction, which consists of stacked graph convolution layers, a pooling layer, and an attention mechanism that aims to reduce computational complexity while minimizing feature loss. 
Finally, the global average pooled feature vectors are passed through the fully connected layer to obtain the classification results. 

Experimental results illustrate that the overall recognition performance achieved by the STF-GCN on the RML2016.10a, RML2016.10b and RML22 datasets reaches the highest in existing AMR tasks, 64.35\%, 66.04\% and 70.95\%, respectively. The main contributions of this paper are shown below.
\begin{itemize}
\item We propose an adaptive correlation-based GNN framework for multi-domain information fusion, which allows the graph structure to be more adapted to modulated signal features in terms of node features and edge features.
\item We develop a novel method for generating adjacency matrices based on adaptive correlation function, which allows the adjacency matrix to comprehensively represent node relationships while reducing computational complexity.
\item We design a PoolGAT layer that coarsens graph structural features and efficiently aggregates and updates key features of graph nodes and adjacency matrices.
\item Experimental results on RadioML2016 series and RML22 datasets demonstrate that our method outperforms the SOTA DL-AMR methods in all recognition accuracy metrics and allows for comparatively fast inference speed.
\end{itemize}
\section{Related work}
\begin{table*}

\centering
  \caption{Comparison of Related DL-AMR Methods}
\begin{tabular}{|c|c|c|c|c|c|}

\hline
\textbf{Model} &\textbf{Publication}& \textbf{Network Architecture} & \textbf{Sample Format} & \textbf{Dataset}  & \textbf{Performance (OA)} \\ \hline
\multicolumn{1}{|c|}{{ULCNN \cite{Ultra}}}
&2024
& \multicolumn{1}{c|}{CNN}                     
& \multicolumn{1}{c|}{I/Q}                     
& \multicolumn{1}{c|}{RML2016.10a}                     
& 62.47\%   \\ \hline
\multicolumn{1}{|c|}{{PET-CGDNN \cite{PETCGDNN}}}
&2021
& \multicolumn{1}{c|}{CNN \& GRU}                     
& \multicolumn{1}{c|}{I/Q}                     
& \begin{tabular}[c]{@{}c@{}}RML2016.10a \& \\ RML2016.10b\end{tabular}                    
& 60.44\%(a), 63.82\%(b)   \\ \hline
\multicolumn{1}{|c|}{{MSGNet \cite{MSGNET}}}
& 2024
& \multicolumn{1}{c|}{CNN \& GRU}                     
& \multicolumn{1}{c|}{I/Q, A/P \& R/I}                     
& \begin{tabular}[c]{@{}c@{}}RML2016.10a \& \\ RML2016.10b\end{tabular}                    
&  62.92\%(a), 64.95\%(b)   \\ \hline
\multicolumn{1}{|c|}{{AMC-NET \cite{AMCNET}}}
&2023
& \multicolumn{1}{c|}{CNN \& QKV Attention}                     
& \multicolumn{1}{c|}{I/Q}                     
& \begin{tabular}[c]{@{}c@{}}RML2016.10a \& \\ RML2016.10b\end{tabular}                    
&  62.51\%(a), 64.63\%(b)   \\ \hline
\multicolumn{1}{|c|}{{CE-FuFormer \cite{zhaocaidan}}}
&2023
& \multicolumn{1}{c|}{CNN \& Transformer}                                        
& \multicolumn{1}{c|}{I/Q}                     
& \multicolumn{1}{c|}{RML2016.10a}                     
& 63.02\%   \\ \hline
\multicolumn{1}{|c|}{{TSFFN \cite{TSFFN}}}
&2024
& \multicolumn{1}{c|}{CNN \& QKV Attention}                     
& \begin{tabular}[c]{@{}c@{}}I/Q, STFT, PWVD \& \\ Handcrafted Features \end{tabular}                    
& \begin{tabular}[c]{@{}c@{}}RML2016.10a \& \\ RML2016.10b\end{tabular}                    
&  62.70\%(a), 64.56\%(b)   \\ \hline
\multicolumn{1}{|c|}{{DH-TR \cite{DHTR}}}
& 2024
& \multicolumn{1}{c|}{CNN, GRU \& Transformer}                     
& \multicolumn{1}{c|}{I/Q}                     
& \begin{tabular}[c]{@{}c@{}}RML2016.10a \& \\ RML2016.10b\end{tabular}                    
&  62.81\%(a), 65.41\%(b)   \\ \hline\hline\hline
\multicolumn{1}{|c|}{{AvgNet \cite{AvgNet}}}
& 2022
& \multicolumn{1}{c|}{Typical GNN}                     
& \multicolumn{1}{c|}{I/Q}                     
& \begin{tabular}[c]{@{}c@{}}RML2016.10a \& \\ RML2016.10b\end{tabular}                    
&  \begin{tabular}[c]{@{}c@{}}62.93\% (a, 8:2 train/test ratio) \\ 64.58\% (b, 1:1 train/test ratio) \end{tabular}        \\ \hline
\multicolumn{1}{|c|}{{CTGNet \cite{CTGNet}}}
&2023
& \multicolumn{1}{c|}{CNN, Transformer \& GNN}                     
& \multicolumn{1}{c|}{I/Q}                     
& \begin{tabular}[c]{@{}c@{}}RML2016.10a \& \\ RML2016.10b\end{tabular}                    
&  \begin{tabular}[c]{@{}c@{}}62.07\% (a, 8:2 train/test ratio) \\ 64.33\% (b, 6:4 train/test ratio) \end{tabular}        \\ \hline
\multicolumn{1}{|c|}{{DeepSIG \cite{DeepSig}}}
&2023
& \multicolumn{1}{c|}{ResNet18, LSTM, DiffPool}                     
& \multicolumn{1}{c|}{I/Q}                     
& \multicolumn{1}{c|}{RML2016.10a}                     
& 62.19\% (Unknown division ratio)   \\ \hline
\end{tabular}

\label{tab:related} 
\begin{threeparttable}
\begin{tablenotes}
\item[1] OA denotes Overall Accuracy across all SNRs. The default dataset division ratio is \textbf{6:2:2}. 
\item[2] Performance is quoted directly from the compared papers.
\end{tablenotes}
    \end{threeparttable}

\end{table*}
This section focuses on methods closely related to DL-AMR and GNN-based AMR tasks. Table \ref{tab:related} summarizes the most relevant SOTA methods applied to the similar datasets.

\subsection{DL-based AMR Methods}
O'Shea et al. \cite{o2016convolutional} first use CNN for AMR tasks on raw I/Q signal sequences, demonstrating higher performance and stability compared to traditional methods. 
They also proposed the RML2016.10a benchmark dataset for subsequent researchers to test model performance, which led to the widespread use of CNN to extract spatial features from modulated signals. Peng et al. \cite{peng} convert the raw I/Q signals into grid-like topology images and feed them to CNN for training. Guo et al. \cite{Ultra} proposed the ultralight convolutional neural network (ULCNN) for unmanned aircraft vehicle systems. 
To extract temporal domain features of the signal, Rajendran et al. \cite{LSTM2} and Ke et al. \cite{DAE} attempt to model such temporal relationship within modulated signals based on RNN. 
Researchers then coincidentally investigate how to synthesize the strengths of both networks. MCLDNN \cite{MCLDNN} divides the I/Q signal into three branch CNN for spatial feature extraction and then use LSTM to extract temporal features, which is still the most widely used baseline model. CNN-LSTM \cite{CNNLSTM} builds a dual-stream model of CNN and LSTM to extract I/Q and A/P features. Wang et al. \cite{multistream} try to incorporate CBAM attention and multi-head attention based on multi-stream input and CRNN architecture. STARNet \cite{STARnet} parallelizes feature extraction with CNN and GRU extractors and automatically exploits discriminative information based on signal reconstruction performance. As Transformer has made a splash in computer vision field, Cai et al. \cite{TRN} attempt to migrate the Transformer architecture to the AMR task. SigFormer \cite{Sigformer} introduces a pyramid Transformer architecture to encode internal features of modulated signals. Zhang et al. \cite{AMCNET} propose a frequency domain denoising module and replaces the LSTM layer in CRNN with a fusion attention mechanism, after which MCDformer \cite{MCDformer} further improved the performance based on a similar architecture. Feng et al. \cite{TSFFN} propose a method that integrates spectrograms, manually extracted features, and temporal features using QKV attention mechanism to improve signal discrimination. IQFormer \cite{IQFormer} and DH-TR \cite{DHTR} achieve the highest recognition performance on the current RML16 series datasets due to their effective integration of the CNN-RNN-Transformer architecture.

\subsection{GNN-based Methods}
Graph is a data structure that models a set of objects (nodes) and their relationships (edges), which is naturally well-suited for extracting and modeling modulation features. Reference \cite{ModulationRecognitionWithGCN} is the first attempt to obtain the modulation feature information through graph mapping.
AvgNet \cite{AvgNet} uses the convolutional feature between graph nodes to construct the diagonal of the adjacency matrix and creates two adjacency matrices with I/Q signals for training. 
Tonchev et al.\cite{AMR_GCN_TF} introduced a graph convolution-based approach for I/Q channel fusion. 
Inspired by the method of translating constellation diagrams of received signals into images \cite{gnnReview}, Ghasemzadeh et al. \cite{RobustGCNAMR} convert the constellation diagram into a graph structure and constructs the adjacency matrix. 
GGCNN \cite{GGCNN} is designed to mitigate the effects of additive Gaussian white noise (AWGN) by extracting higher-order statistical features and combining them with time-domain features within GNN. 
CTGNet \cite{CTGNet} uses a sliding window to reorganize the I/Q signal into a matrix, which is then fed into a CNN-Transformer network to dynamically map the matrix into the graph structure.
LAGNet\cite{LAGNet} uses an attention mechanism to map the A/P components to the graph structure via a LSTM-based encoder.
\section{Signal model}
\begin{figure*}
    \centering
    \includegraphics[width=0.85\textwidth]{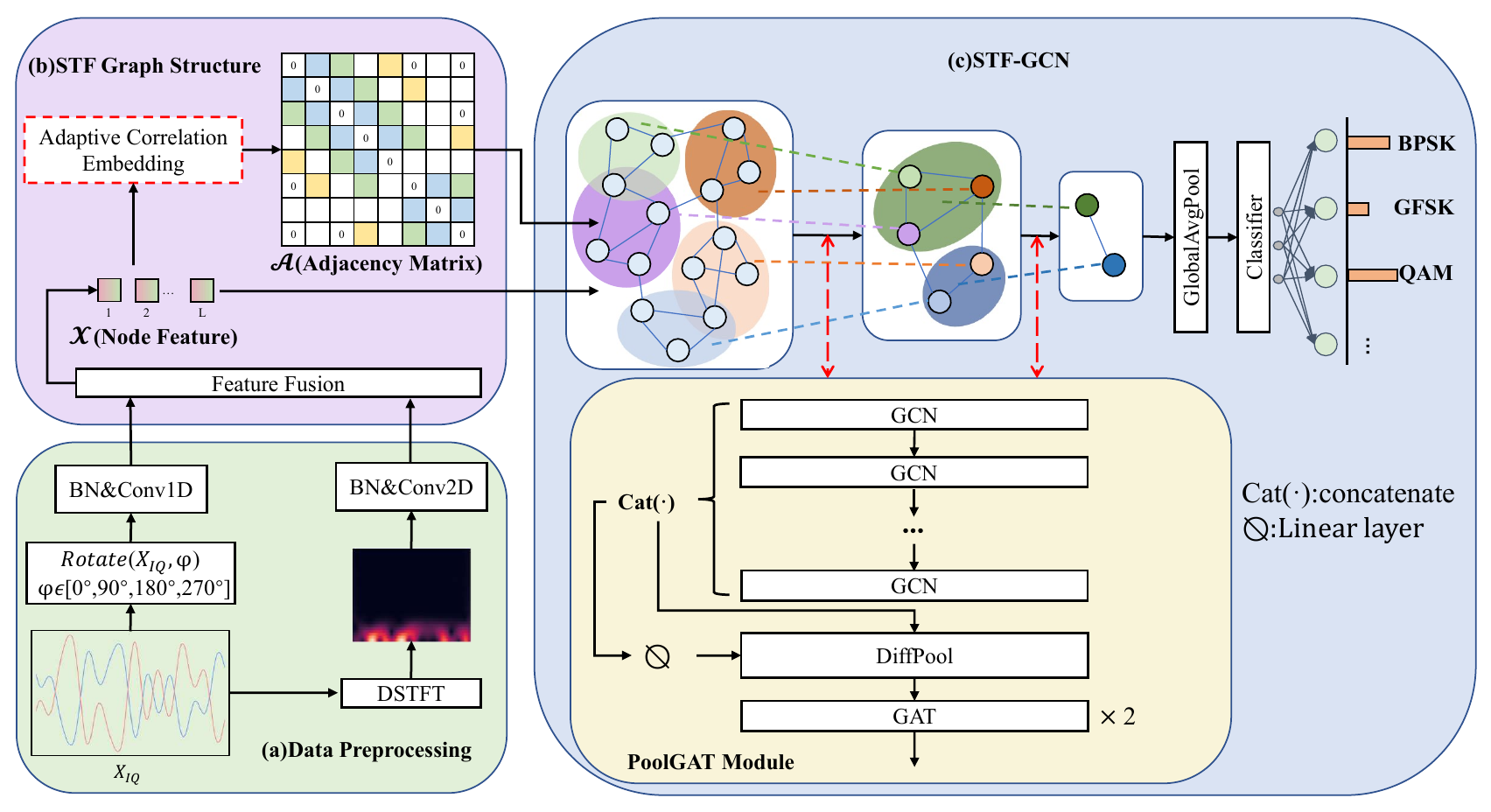}
    \caption{The overall network architecture of the proposed STF-GCN.}
    \label{fig:model}
\end{figure*}
In urban intelligent transportation scenarios, the establishment of wireless channels should focus on multipath propagation, time delay and Doppler effect because of high speed movement and building occlusion.
The expression for the received signal through the channel is given by:
\begin{equation}
\label{eq1}
\begin{aligned}s_{r}(t) = \sum_{l=1}^{L} \eta_l e^{j(2\pi \Delta_lt + \phi_l)} s(t - \kappa_l) + n(t)\end{aligned},
\end{equation}

where $t\in[1,2,\cdots,\Gamma]$, $\Gamma$ denotes the number of signal sampling points, $L$ represents the number of paths in multipath signals, $s_{r}(t)$ is the received signal, $s(t)$ is the original transmitted signal, and $n(t)$ represents the AWGN. Furthermore, $\Delta_l$, $\eta_l$, $\kappa_l$, and $\phi_l$ correspond to the Doppler shift, amplitude attenuation, time delay, and phase shift of the $l$-th path, respectively.

By default, complex signals are transmitted, and thus the received signals are typically represented as I/Q signals:
\begin{equation}
\label{eq2}
s_{IQ}(t)=I(t)+jQ(t),
\end{equation}
where $s_{IQ}(t)$ represents the received I/Q signal, $I(t)$ and $Q(t)$ represent the in-phase and quadrature parts of the received signal, respectively. 
$s_{IQ}$ will be divided into two paths, which will be transmitted through the same channel, so that the received signals can be represented by a two-dimensional sequence as
\begin{align}
\label{eq3}
\begin{aligned}
\textbf{x}_{IQ}
&=\left[\begin{array}{l}
\textbf{x}_I \\
\textbf{x}_Q
\end{array}\right]
\\&=\left[\begin{array}{cccc}
\textbf{x}_I\left(1\right),&\textbf{x}_I\left(2\right), &\cdots,&\textbf{x}_I\left(\Gamma\right)\\
\textbf{x}_Q\left(1\right),&\textbf{x}_Q\left(2\right), &\cdots,&\textbf{x}_Q\left(\Gamma\right)
\end{array}\right].
\end{aligned}
\end{align}
Since the I/Q signal contains the amplitude and phase information of the original signal, it is the most commonly used input for the AMR methods.

\section{Method}
The overview of the proposed architecture is shown in Fig. \ref{fig:model}, there are three main components: data preprocessing, spatial-temporal-frequency (STF) graph structure construction, and STF-GCN combined with classifier.
The modulation recognition process is summarized as follows: The received I/Q signals undergo data augmentation and discrete short-time fourier transform (DSTFT) to obtain spatial and frequency domain information, which are then fused to form the node features of the STF graph structure. 
The adjacency matrix is constructed based on adaptive correlation function with the multi-domain fused node features, defining the edge features of STF graph structure. 
The fused nodes and adjacency matrix are then input into staged PoolGAT layers for training to achieve a attentively coarsened graph structure. 
Finally, these resulting feature vectors are fed into the classifier for classification. Detailed descriptions of these modules are provided in the following subsections.

\subsection{Data Preprocessing}
To enrich the node features of the graph structure, we introduce the discrete short-time fourier transform (DSTFT) to extract frequency domain information from the original signal. 
The time-frequency spectrogram contains the frequency and phase of local sinusoidal regions with window shifts, allowing analysis of instantaneous frequency and amplitude from a time-frequency domain perspective. 
Since we are dealing with discrete signals obtained by sampling continuous signals, DSTFT is applied. The expression for DSTFT is as follows:
\begin{equation}
\label{eq4}
\textbf{x}_{STFT}(m,k)=\sum_{n=-\infty}^\infty \textbf{x}_{IQ}\left(n\right)w\left(n-m\right)e^{-j2\pi\frac kNn},
\end{equation}
where $\textbf{x}_{STFT}(m,k)$ represents the DSTFT result at time index $m$ and frequency index $k$, $w(n-m)$ denotes the window function, and $N$ is the number of DFT points. Notice that the blackman window is employed as the window function in this paper.

The spatial and frequency domain information extracted by the DSTFT is then pre-processed and combined into the time domain.
Specifically, we first applied a random rotation operation to the original I/Q signals as data augmentation. 
This enhances the diversity of the training data and helps the model learn more robust spatial feature representations. 
It is noteworthy that this operation only changes the phase information of the I/Q signal without affecting the frequency domain information, thus maintaining the correspondence between two input features. 
The data augmentation expression is:
\begin{equation}
\label{eq5}
\hat{\textbf{x}}_{IQ} = Rotate((\textbf{x}_{IQ},\varphi)\mid\varphi\in[0^{\circ},90^{\circ},180^{\circ},270^{\circ}]),
\end{equation}
where $\hat{\textbf{x}}_{IQ}$ represents the I/Q signal after data augmentation, $\textbf{x}_{IQ}$ denotes the original I/Q signal, and $\varphi$ is the angle of rotation, selected with equal probability among $[0^{\circ},90^{\circ},180^{\circ},270^{\circ}]$, $Rotate(.) $ represents as follows:
\begin{equation}
\label{eq6}
\hat{\textbf{x}}_{IQ}=\left[\begin{array}{l}
\hat{\textbf{x}}_{I} \\
\hat{\textbf{x}}_{Q}
\end{array}\right]=\left[\begin{array}{cc}
\cos \varphi & -\sin \varphi \\
\sin \varphi & \cos \varphi
\end{array}\right]\textbf{x}_{IQ}.
\end{equation}
After that, we perform feature concatenation of the enhanced spatial domain features and the frequency domain features. 
We begin by standardizing the feature dimensions across spatial and frequency domains using batch normalization and convolution layers.
The spatial domain input is denoted as $\hat{\textbf{x}}_{IQ}\in\mathbb{R}^{2\times L}$, and the frequency domain input as $\textbf{x}_{STFT}\in\mathbb{R}^{1\times f\times L}$, where $f$ represents the number of sinusoidal components obtained from the DSTFT operation. They are pre-processed to obtain the temporal-spatial domain feature $\textbf{X}_{SD}$ and the temporal-frequency domain feature $\textbf{X}_{FD}$. The processing equations are shown below:
\begin{equation}
\label{eq7}
\textbf{X}_{SD}=Conv1D(\hat{\textbf{x}}_{IQ},\mathcal{O}),
\end{equation}
\begin{equation}
\label{eq8}
\textbf{X}_{FD}=Squeeze(Conv2D(\textbf{x}_{STFT},\mathcal{O})),
\end{equation}
where $\mathcal{O}$ represents the number of output channels and $Squeeze(.)$ denotes the dimensionality reduction operation. We convert $\hat{\textbf{x}}_{IQ}$ and $\textbf{x}_{STFT}$ to $\textbf{X}_{SD}\in\mathbb{R}^{\mathcal{O}\times \mathcal{H}}$ and $\textbf{X}_{FD}\in\mathbb{R}^{\mathcal{O}\times \mathcal{H}}$ with the same number of channels through convolution. 
Specifically, the Conv2D convolution kernel size is $(f,1)$, which is designed to compress the frequency domain feature to match the temporal-spatial domain feature dimension. 
As convolution is similar to window shifting, $\textbf{X}_{SD}$ and $\textbf{X}_{FD}$ exhibit temporal correlation between dimensions. This helps to integrate different domain features.

\subsection{Adaptive Correlation-based STF Graph Structure}
\begin{figure*}
    \centering
    \includegraphics[width=0.8\textwidth]{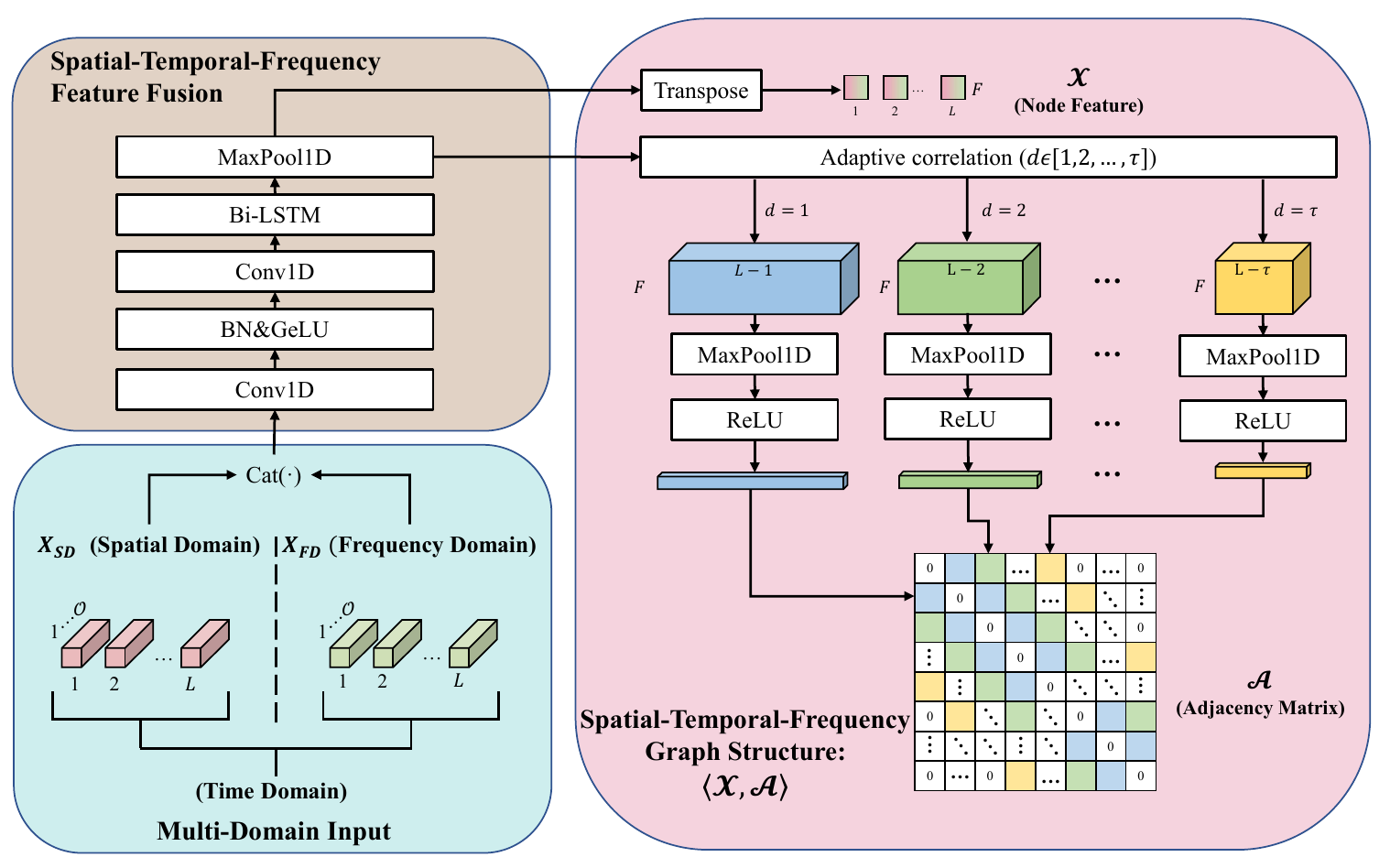}
    \caption{Construction of a graph structure based on the fusion of space-time-frequency features of three domains.}
    \label{fig:adj}
\end{figure*}
Fig. \ref{fig:adj} and Algorithm \ref{alg:graph} illustrate the adaptive correlation-based STF graph structure construction module. This module constructs nodes for the graph structure using fused features and derives edge features from their correlations. These edge features are then adapted to the node features before entering the STF-GCN. There are two sub-modules: the feature fusion module and the adaptive correlation-based adjacency matrix generation module.

The feature fusion module employs a encoder composed of CNN and Bi-LSTM to integrate $\textbf{X}_{SD}$ and $\textbf{X}_{FD}$ into graph node feature $\textbf{X}_{STF}\in\mathbb{R}^{F\times \mathcal{H}}$. 
This ensures the temporal correlation of different domain information within the nodes of a graph structure. 
The advantage of fusion embedding during the pre-training stage is that it avoids the parallel training of multiple graph structures, reducing computational complexity and contributing to the lightweight design of the model overall.
The feature fusion method involves first connecting the features according to their feature dimensions, then applying a 1D convolution with a kernel size of 1 to traverse and fuse features across different channels. This is followed by further processing with another 1D convolution with a kernel size of 1, after normalization and the GeLU activation function. The formula for the CNN part is as
\begin{equation}
\label{eq9}
\textbf{X}_{SF}=Conv1D(Conv1D_{1}(Cat(\textbf{X}_{SD},\textbf{X}_{FD}))),
\end{equation}
where $Cat(.)$ represents the concatenation operation, and $Conv1D_{1}$ includes 1D convolution, normalization, and the GeLU activation function. After obtaining $\textbf{X}_{SF}$, a Bi-LSTM layer is added to enhance the contextual connections of the graph node feature, resulting in $\textbf{X}_{STF}$.

Finally, the second dimension of $\textbf{X}_{STF}$ is downsampled by maxpooling to reduce the complexity. The downsampled feature set $\boldsymbol{\mathcal{X}}=\{\textbf{X}_1,\textbf{X}_2,\cdots,\textbf{X}_{\mathcal{N}}\}\in\mathbb{R}^{\mathcal{N}\times F}$ are used as graph node features, where $\mathcal{N}$ represents the number of nodes, $F$ represents the feature dimensions of each node. Since we chose to slice the nodes based on feature dimensions, the values of $\mathcal{H}$ and $\mathcal{N}$ are the same.

Then we use the node features $\mathcal{X}$ to construct the adjacency matrix $\boldsymbol{\mathcal{A}}\in\mathbb{R}^{\mathcal{N}\times \mathcal{N}}$ of the STF graph structure. Each element in the adjacency matrix of a general graph structure represents the edge weights between different nodes, and for signal graph structures, the adjacency matrix is equivalent to the correlations between different nodes embedded with information about the windowing signal. Multi-domain information after local signal window segmentation has been embedded in node features. In order to explore the correlation between nodes, inspired by the correlation function, we designed and constructed the adjacency matrix based on adaptive correlation. Specifically, the $(d+1)th$ multi-dimensional diagonal correlation feature is first computed as
\begin{equation}
\label{eq11}
\textbf{a}_{d}=\boldsymbol{\mathcal{X}}[0\colon \mathcal{N}-d, :]\odot\boldsymbol{\mathcal{X}}[d\colon \mathcal{N}, :],
\end{equation}
where $\odot$ represents the element-wise multiplication, $\boldsymbol{\mathcal{X}}$ is the node features to extract the correlation among the dimensions within each node feature, the variable $\tau$ is set to control the number of diagonals that capture the potential correlations, so that offset $d\in[1,2,\cdots,\tau]$. It should be noted that the correlation will decrease with increasing node distance because the modulated signal contains timing characteristics. 
The obtained $\textbf{a}_{d}$ is a two-dimensional vector. The length of the first dimension is $\mathcal{N}-d$, which is the length of the $(d+1)$th diagonal of the adjacency matrix, just the right length to be embedded into the adjacency matrix. The second dimension represents the multi-dimensional correlation information. Considering the computational complexity, we performed a series of operations including max pooling layer and ReLU activation function to compress the diagonal features over the second dimension into one-dimension features $\boldsymbol{\mathcal{A}}_{d}\in\mathbb{R}^{(\mathcal{N}-d)}$, which can be expressed by
\begin{equation}
\label{eq12}
\boldsymbol{\mathcal{A}}_{d}=\mathrm{ReLU}(MaxPool1D(\textbf{a}_{d})).
\end{equation}

Then $\boldsymbol{\mathcal{A}}_{d}$ is embedded in $\boldsymbol{\mathcal{A}}$, which can be represented as
\begin{align}
\label{eq13}
&\begin{aligned}
&\boldsymbol{\mathcal{A}} =\text{diag}\_\text{embed}(\boldsymbol{\mathcal{A}}_1, \boldsymbol{\mathcal{A}}_2, \dots, \boldsymbol{\mathcal{A}}_d, \dots, \boldsymbol{\mathcal{A}}_\tau)\\
&\setlength{\arraycolsep}{3pt}
=\begin{bmatrix}
  0 & \mathcal{A}_1^1 & \mathcal{A}_2^1 & \cdots & \mathcal{A}_\tau^1 & 0 & \cdots & 0\\
  \mathcal{A}_1^1 & 0 & \mathcal{A}_1^2 & \mathcal{A}_2^2 & \cdots & \mathcal{A}_\tau^2 & \ddots & \vdots\\
  \mathcal{A}_2^1 & \mathcal{A}_1^2 & 0 & \mathcal{A}_1^3 & \mathcal{A}_2^3 & \ddots & \ddots & 0\\
  \vdots & \mathcal{A}_2^2 & \mathcal{A}_1^3 & 0 & \mathcal{A}_1^4 & \mathcal{A}_2^4 & \ddots & \mathcal{A}_\tau^{L-\tau}\\
  \mathcal{A}_\tau^1 & \vdots & \mathcal{A}_2^3 & \mathcal{A}_1^4 & 0 & \mathcal{A}_1^5 & \ddots & \vdots\\
  0 & \mathcal{A}_\tau^2 & \ddots & \mathcal{A}_2^4 & \mathcal{A}_1^5 & 0 & \ddots & \mathcal{A}_2^{L-2}\\
  \vdots & \ddots & \ddots & \ddots & \ddots & \ddots & \ddots & \mathcal{A}_1^{L-1}\\
  0 & \cdots & 0 & \mathcal{A}_\tau^{L-\tau} & \cdots & \mathcal{A}_2^{L-2} & \mathcal{A}_1^{L-1} & 0
\end{bmatrix},
\end{aligned}
\end{align}
where the function $\text{diag}\_\text{embed}(.)$ is to convert $(\boldsymbol{\mathcal{A}}_1, \boldsymbol{\mathcal{A}}_2, \cdots, \boldsymbol{\mathcal{A}}_d, \cdots, \boldsymbol{\mathcal{A}}_\tau)$ into a diagonal matrix and then embed into $\boldsymbol{\mathcal{A}}$, and $\mathcal{A}_{d}^{i}$ represents the edge weight of node $i$ and $i+d$.

In summary, the constructed adjacency matrix $\boldsymbol{\mathcal{A}}$ not only extracts the correlation between embedded modulation symbols but also adapts to multi-domain information. This adaptability helps the STF-GCN extract key information for discriminating modulation types. 

\begin{algorithm}
    \renewcommand{\algorithmicrequire}{\textbf{Input:}}
	\renewcommand{\algorithmicensure}{\textbf{Output:}}
	\caption{Construction of The Spatial-Temporal-Frequency Graph Structure}
    \label{alg:graph}
    \begin{algorithmic}[1] 
        \REQUIRE raw I/O signal $\textbf{x}_{IQ}$;
	\ENSURE node features $\boldsymbol{\mathcal{X}}$, adjacency matrix $\boldsymbol{\mathcal{A}}$;

        \STATE Generate $\hat{\textbf{x}}_{IQ}$, $\textbf{x}_{STFT}$ according to Eq. (\ref{eq4})~(\ref{eq5});
        \STATE Generate $\textbf{X}_{SD}$, $\textbf{X}_{FD}$ according to Eq. (\ref{eq7})~(\ref{eq8});
        \STATE (Feature Fusion)Generate $\textbf{X}_{STF}$ according to Eq. (\ref{eq9});
        \STATE Generate $\boldsymbol{\mathcal{X}}$ through Bi-LSTM, max pooling layer and transpose operation;
        
        \FORALL {$d=1,2,\cdots,\tau$}
            \STATE Calculate $d$th multidimensional diagonal according to Eq. (\ref{eq11});
            \STATE Calculate $d$th one-dimensional diagonal according to Eq. (\ref{eq12});
        \ENDFOR
        
        \STATE Embed $(\boldsymbol{\mathcal{A}}_1, \boldsymbol{\mathcal{A}}_2, \cdots, \boldsymbol{\mathcal{A}}_d, \cdots, \boldsymbol{\mathcal{A}}_\tau)$ into $\boldsymbol{\mathcal{A}}$ according to Eq. (\ref{eq13});
        
        \STATE \textbf{return} $\boldsymbol{\mathcal{X}}$, $\boldsymbol{\mathcal{A}}$.
    \end{algorithmic}
\end{algorithm}

\subsection{STF-GCN combined with classifier}
In order to extract features from the constructed STF graph structure, we develop a STF-GCN based on staged coarsened graph convolution network layers, PoolGAT, which is inspired by Diffpool \cite{Diffpool} and graph attention mechanism \cite{GAT}. 
We enable nodes in each PoolGAT layer to learn distinguishable cluster assignments, mapping them to a set of clusters through Diffpool. It helps to combine node features and adjacency matrice to reduce the number of nodes gradually, which simplifies the signal graph structure to reduce the number of subsequent training parameters and the computational effort in forward inference. In addition, we enable the coarsened graph structure to focus the key edge features among these clusters through an attention mechanism, thus forming an attentively coarsened input to the next PoolGAT layer.

A PoolGAT layer consists of three parts: stacked GCN layers, coarsening of the graph structure, and an attention mechanism. Specifically,the node features transmission mechanism for neighboring GCN layers is as follows:
\begin{equation}
\label{eq14}
\boldsymbol{\mathcal{X}}^{\psi_{p+1}}=\widehat{\textbf{D}}^{-1/2}(\boldsymbol{\mathcal{A}}^{\psi}+\textbf{I}) \widehat{\textbf{D}}^{-1/2} \boldsymbol{\mathcal{X}}^{\psi_{p}} \mathbf{\Theta},
\end{equation}
where $\psi\in[1,\cdots,\Psi]$, $p\in[1,\cdots,g-1]$, $\Psi$ and $g$ are the number of PoolGAT layers and the number of stacked GCN layers set, respectively, $\textbf{I}$ represents the self-loops operation, $\widehat{D}_{ii}=\sum_{j} \widehat{A}_{ij}^{\psi}$ is to act as a normalizer, and $\mathbf{\Theta}$ represents the trainable parameters. By stacking $g$-layer GCNs, the assignment matrix $\textbf{S}$ and the embedding matrix $\textbf{Z}$ can be obtained via
\begin{equation}
\label{eq15}
\textbf{S}^\psi=GCN_{g,linear}(\boldsymbol{\mathcal{A}}^\psi,\boldsymbol{\mathcal{X}}^\psi),
\end{equation}
\begin{equation}
\label{eq16}
\textbf{Z}^\psi=GCN_{g,embed}(\boldsymbol{\mathcal{A}}^\psi,\boldsymbol{\mathcal{X}}^\psi),
\end{equation}
where $GCN_{g,linear}(.)$ represents the linear layer applied to node features after stacked $g$-layer GCN, $GCN_{g,embed}(.)$ represents foward directly after stacked $g$-layer GCN.

In the coarsening of the graph structure, the aggregated nodes $\widehat{\boldsymbol{\mathcal{X}}}^{\psi}$ and  the adjacency matrix $\boldsymbol{\mathcal{A}}^{\psi+1}$ are updated as follows:
\begin{equation}
\label{eq17}
\widehat{\boldsymbol{\mathcal{X}}}^{\psi}=(\textbf{S}^\psi)^T \textbf{Z}^\psi\in\mathbb{R}^{\mathcal{N}_{\psi+1}\times F}
\end{equation}
\begin{equation}
\label{eq18}
\boldsymbol{\mathcal{A}}^{\psi+1}=(\textbf{S}^\psi)^T \boldsymbol{\mathcal{A}}^\psi\textbf{S}^\psi\in\mathbb{R}^{\mathcal{N}_{\psi+1}\times\mathcal{N}_{\psi+1}}
\end{equation}
\begin{equation}
\label{eq19}
\mathcal{N}_{\psi+1}=\frac{\mathcal{N}_\psi}{\mathcal{D}},
\end{equation}
where $\widehat{\boldsymbol{\mathcal{X}}}^{\psi}=\{\hat{\textbf{x}}_1^\psi,\hat{\textbf{x}}_2^\psi, \cdots,\hat{\textbf{x}}_{\mathcal{N}_{\psi+1}}^\psi\}$, $\mathcal{N}_{\psi+1}$ denotes the number of nodes after coarsening, $\mathcal{N}_\psi$ represents the number of nodes in the $\psi$th layer, $\mathcal{D}$ is the coarsening factor of PoolGAT.

Additionally, we introduce the graph attention mechanism after graph structure coarsening. This allows nodes to focus on directly related nodes, based on whether edge feature weights between nodes are greater than zero, without relying on prior knowledge of the entire graph structure. This can effectively aggregate the feature information of local nodes. For the feature $\textbf{x}_i^{\psi}$ of the $i$th node and  $\textbf{x}_j^{\psi}$ of the $j$th node that have a direct relationship, there is a corresponding attention parameter $\alpha_{ij}$ as follows:
\begin{equation}
\label{eq20}
\alpha_{ij}=\frac{\exp\left(\text{LeakyReLU}\left(\boldsymbol{\beta}^T[\mathbf{W} \hat{\textbf{x}}_i^{\psi}\|\mathbf{W} \hat{\textbf{x}}_j^{\psi}]\right)\right)}{\sum_{k\in\mathcal{M}_i}\exp\left(\text{LeakyReLU}\left(\boldsymbol{\beta}^T[\mathbf{W} \hat{\textbf{x}}_i^{\psi}\|\mathbf{W} \hat{\textbf{x}}_k^{\psi}]\right)\right)}
\end{equation}
where $\boldsymbol{\beta}$ and $\mathbf{W}$ represent the trainable parameters, $\|$ means the connection of the weighted nodes, and $\mathcal{M}_i$ represents the set of nodes that are directly related to the $i$th node. In addition, we normalize the per-attention parameter so that we can get the feature expression of the $i$th node after the attention mechanism as:
\begin{equation}
\label{eq21}
\tilde{\textbf{x}}_i^{\psi}=\sum_{k\in\mathcal{M}_i} \alpha_{ik}\hat{\textbf{x}}_i^{\psi},
\end{equation}
where the set of the $(\psi+1)th$ layer nodes input $\boldsymbol{\mathcal{X}}^{\psi+1} =\{\tilde{\textbf{x}}_1^{\psi},\tilde{\textbf{x}}_2^{\psi},\cdots,\tilde{\textbf{x}}_{\mathcal{N}_\psi}^{\psi+1}\}$. 

After the $\Psi$-layer PoolGAT, a separate GCN layer is added to merge the node features with the adjacency matrix extracted from the last PoolGAT layer. Finally, the graph structure of the last PoolGAT layer is downsampled with the global average pooling operation, and the resulting feature vector is classified according to a linear layer to obtain the probability distribution $p$, and the modulation type index $\hat{p}$ of the largest probability distribution is taken as the final classification result.
The loss function used in the model is the cross entropy loss as $\mathcal{L}_{CE}$, denoted as
\begin{equation}
\label{eq22}
\mathcal{L}_{CE}(\hat{p},y)=-\sum_{i=1}^\zeta y_ilog(\hat{p}_i)
\end{equation}
where $\zeta$ is the total number of categories, $y$ represents the true label of the sample, and $\hat{p}$ denotes the predicted label. 
\begin{algorithm}
\renewcommand{\algorithmicrequire}{\textbf{Input:}}
\renewcommand{\algorithmicensure}{\textbf{Output:}}
\caption{The Forward Propagation of STF-GCN}
\label{alg:stf-gcn}
\begin{algorithmic}[1]
\REQUIRE Time domain signal: $\textbf{x}_{IQ}$;
\ENSURE Predicted label: $\hat{p}$;
\STATE set hyper-parameters:Optimizer: AdamW, Initial learning rate: $lr$, Scheduler of update $lr$: scheduler, Training epoch: $T$, Number of feature nodes: $\mathcal{N}$, Dimension of the node feature: $F$, Out channels for preprocessing: $\mathcal{O}$, Number of diagonals of the $\boldsymbol{\mathcal{A}}$: $\tau$, Number of GCN layers in PoolGAT: $g$, Number of PoolGAT layers: $\Psi$, Coarsening factor: $\mathcal{D}$;
\FOR{$epoch$ = $1$ to $T$}
    \STATE Construct $G<\boldsymbol{\mathcal{X}}$, $\boldsymbol{\mathcal{A}}>$ according to Algorithm 1;
    \FOR{$\psi$ = $1$ to $\Psi$}
        \STATE Compute $\textbf{S}^\psi$ and $\textbf{Z}^\psi$ according to Eq. (\ref{eq15}),(\ref{eq16});
        \STATE Update $\widehat{\boldsymbol{\mathcal{X}}}^{\psi}$ and $\boldsymbol{\mathcal{A}}^{\psi+1}$ according to Eq. (\ref{eq17})-(\ref{eq19});
        \STATE Calculate $\boldsymbol{\mathcal{X}}^{\psi+1}$ via the graph attention mechanism  according to Eq. (\ref{eq20}),(\ref{eq21}); 
    \ENDFOR
    \STATE Obtain the feature vectors by a graph convolution layer and a global average pooling;
    \STATE Calculate predict label $\hat{p}$ by fully connected layer classifier;
    \STATE Calculate $\mathcal{L}_{CE}(\hat{p},y)_{epoch}$ according to Eq. (\ref{eq22});
    \STATE $\hat{\theta_G},\hat{\theta_P}$ $\leftarrow$ AdamW($\mathcal{L}_{CE}(\hat{p},y)_{epoch}$, scheduler($lr$)).
    \STATE \% $\hat{\theta_G}$ and $\hat{\theta_P}$ represent parameters of graph structure and PoolGAT, respectively.
\ENDFOR
\STATE \textbf{return} $\hat{p}$
\end{algorithmic}
\end{algorithm}
\subsection{The Forward Propagation of STF-GCN}
The forward propagation of STF-GCN is shown in Algorithm \ref{alg:stf-gcn}. The STF graph structure $G<\boldsymbol{\mathcal{X}}$, $\boldsymbol{\mathcal{A}}>$ will be passed through $\Psi$-layer PoolGAT. Then the probability distribution $p$ is  obtained by the classifier. And the loss $\mathcal{L}_{CE}(\hat{p}, y)$ is calculated and the trainable parameters are updated.

\section{EXPERIMENTS AND RESULTS}
\subsection{RadioML Dataset}
In order to further validate the effectiveness and multi-scenario generalization capability of our approach in urban transportation systems, the public source datasets RML16 series \cite{RML} and  RML22 \cite{RML22} have been carefully selected to evaluate the proposed method. Table \ref{tab:dataset} summarizes the key differences between the two models in terms of their application scenarios.
\begin{itemize}
    \item \textbf{RML16 (RML2016.10a \& RML2016.10b)}: RML16 uses the Rician fading channel, which consists of a direct path and multiple scattering paths. It is designed to simulate line-of-sight transmission (LOS) scenarios, such as microwave links and satellite communications.
    \item \textbf{RML22}: RML22 uses the ETU70 model, which is a pure scattering multipath channel model assuming that the signal does not have a significant direct path. It is used to evaluate the performance of mobile communication systems (e.g., LTE or 5G NR) in urban environments under channel conditions with high delay extension and high travel speeds.
\end{itemize}
\begin{table*}
    \centering
    \caption{Summary of Used Datasets}
    \label{tab:dataset}
    \begin{tabular}{p{0.07\textwidth} p{0.14\textwidth}p{0.28\textwidth} p{0.35\textwidth}}
        \toprule
        \textbf{Details} & \textbf{Description} & \textbf{RadioML2016.10a, RadioML2016.10b} & \textbf{RML22}\\
        \midrule
        \textbf{Channel} & \textbf{Fading} & Rician fading model, &3GPP fading model ETU70,\\
        
        &&Filter tap magnitudes = [0, -0.97, -5.23] dB, &Filter tap magnitudes = [-1, -1, 0, 0, 0, -3, -5, -7] dB,\\
       && Filter tap delays=[0, 4.5, 8.5] ns,  &Filter tap delays = [0, .05, .12, .2, .23, .5, 1.6, 2.3, 5] \\
       &&Num. of taps = 8,& ns, Num. of taps = 8,\\
       &&Max. freq. dev (Doppler) = 1 Hz, &Max. freq. dev (Doppler) = 70 Hz,\\
       
       &&Num. of sinusoids = 8, K-factor = 4 &     Num. of sinusoids = 8 
        \\
        \cline{2-4}
        
        & \textbf{Clock effect} & LO and SRO max. freq. deviation: 500 Hz, 500 Hz & XO, LO and SRO max. freq. deviation: 5 Hz, 500 Hz, 50 Hz \\
        &&LO and SRO std dev per sample = $10^{-2}$, $10^{-2}$ & XO, LO and SRO std dev per sample = $10^{-4}$, $10^{-2}$, $10^{-3}$ \\
        &&&XO to LO scaling = 100, XO to clock scaling = 10 \\
        \midrule
        \textbf{Signal} & \textbf{SNR Range} & -20 dB to 18 dB (in steps of 2 dB)& -20 dB to 20 dB (in steps of 2 dB)\\
        & \textbf{Number of Samples} & 220,000 (a), 1,200,000 (b) & 462,000 \\
        & \textbf{Dimension} & 2×128 & 2×128 \\
        & \textbf{Duration} & 0.64 ms & 4.5 ms \\
        & \textbf{Samples per Symbol} & 8 & 2 \\
        & \textbf{Sample Rate} & 200 kHz & 30 kHz \\    
        & \textbf{Center Frequency}&-& 1 GHz \\
        & \textbf{Clock Rate}&-& 100 MHZ  \\
        & \textbf{Division Ratio} & 6:2:2 & 6:2:2 \\
        \midrule
        \textbf{Modulation} & \textbf{Modulation Types} & QPSK, BPSK, 8PSK, 64QAM, 16QAM, BFSK, CPFSK, PAM4, WBFM, AM-DSB, AM-SSB (Only RML2016.10a) & QPSK, BPSK, 8PSK, 64QAM, 16QAM, BFSK, CPFSK, PAM4, WBFM, AM-DSB, AM-SSB \\  
        \bottomrule
    \end{tabular}
\end{table*}

\subsection{Implementation Details}
\begin{table}[ht]
\centering
\caption{IMPLEMENTATION DETAILS}
\label{table:implement detail}
\begin{tabular}{@{}lll@{}}
\toprule
Hyper-parameters & Description                 & Values       \\ \midrule
Optim            & Optimizer                   & AdamW         \\
$lr$             & Initial learning rate  & $1 \times 10^{-3}$ \\
scheduler        & Scheduler of $lr$ update &ReduceLROnPlateau\\
$T$              & Training epoch                       & 100          \\
$\Phi$           & Threshold value of epochs   & 20          \\
$\mathcal{B}$  & Batch size            & 256           \\
$N$              & Number of DFT points & 128\\
$\mathcal{N}$ & Number of feature nodes        & 128           \\
$F$     & Dimension of the node feature     & 16         \\
$\mathcal{O}$      &Out channels for preprocessing    &16\\
$\tau$             &Number of diagonals of the $\boldsymbol{\mathcal{A}}$    &11\\
$g$             &Number of GCN layers    &4\\
$\Psi$             &Number of PoolGAT layers    &2\\
$\mathcal{D}$    &Coarsening factor          &4\\
\bottomrule
\end{tabular}
\end{table}

To rigorously evaluate the recognition performance of our proposed method, we randomly select the training, validation, and test sets from each modulation type across different SNRs in a 6:2:2 ratio. The hyper-parameters in detail for all experiments are summarized in Table \ref{table:implement detail}, where the choice of parameters is based on the best results from the validation set, and the training framework used was Pytorch (v2.1.0 with Python 3.11). The model is performed on a platform with the NVIDIA GeForce RTX 4070 Laptop GPU and Intel(R) Core(TM) i9-13900H.

\subsection{Comparisons With SOTA Deep Models}
\begin{table*}[h]
  \centering
  \caption{Comparison of model performance on three datasets (A: RadioML2016.10a, B: RadioML2016.10b, C: RML22)}
    \begin{tabular}{cccccccccc}
    \hline
    \hline
    \multirow{2}{*}{Model} & \multirow{2}{*}{Dataset} & \multirow{2}{*}{Parameters} & \multirow{2}{*}{Highest Accuracy} & \multicolumn{2}{c}{SNR(dB)} & \multirow{2}{*}{OverAll} & \multirow{2}{*}{Macro-F1} & \multirow{2}{*}{Kappa}\\
        \cline{5-6}
          & \multicolumn{1}{c}{} &       &       &$\leq 0dB$       &$\geq 0dB$       &       \\
    \hline
    \multirow{3}[0]{*}{MCLDNN\cite{MCLDNN}} & A  &   \multirow{3}[0]{*}{0.41M} & 92.82\% & 37.87\% & 91.67\% & 62.21\%  & 0.6443& 0.5843\\
          & B   &  & 93.64\% & 40.71\% & 93.19\% & 64.42\% & 0.6497&0.6046 \\
        & C   &  & 89.71\% & 42.21\% & 86.28\% & 63.64\% & 0.6265 & 0.5979 \\
    \hline
    \multirow{3}[0]{*}{AMC-NET\cite{AMCNET}} & A   &  \multirow{3}[0]{*}{0.47M}  & 92.82\% & 38.56\% & 91.32\% & 62.40\% & 0.6483 & 0.5885 \\
          & B   & & 93.86\% & 41.92\% & 93.34\% & 65.14\% & 0.6487 & 0.6081 \\
          & C   & & 98.1\% & 45.52\% & 93.54\% & 68.90\% & 0.6894 & 0.6571  \\
    \hline
    \multirow{3}[0]{*}{FEA-T\cite{FEA-T}} & A     & \multirow{3}[0]{*}{0.17M} & 89.86\% & 37.20\% & 88.43\% & 60.41\% & 0.6210 & 0.5606  \\
      & B     &  & 93.77\% & 40.58\% & 93.02\% & 64.29\% & 0.6528 & 0.6053 \\
      & C   &  & 89.9\% & 38.29\% & 85.42\% & 60.69\% & 0.6001 & 0.5675 \\
    \hline
    \multirow{3}[0]{*}{STARNet\cite{STARnet}} & A     & \multirow{3}[0]{*}{\textbf{0.02M}} & 93.46\% & 39.84\% & 92.82\% & 63.64\% & - & -  \\
      & B     &  & - & - & - & - & - & - \\
      & C     &  & - & - & - & - & - & - \\
    \hline
    \multirow{3}[0]{*}{MSGNET\cite{MSGNET}} & A     & \multirow{3}[0]{*}{0.15M} & - & - & 92.42\% & 62.92\%   \\
      & B     &  & - & - & - & 64.95\% & - & - \\
      & C     &  & - & - & - & - & - & - \\
    \hline
    \multirow{3}[0]{*}{MCDformer\cite{MCDformer}} & A     & \multirow{3}[0]{*}{0.54M} & 92.68\% & 38.78\% & 91.29\% & 62.53\%& 0.6452 & 0.5879  \\
      & B     &  & 93.93\% & 41.73\% & 93.46\% & 65.08\% & 0.6599 & 0.6120 \\
      & C     &  & 98.07\% & 46.11\% & 93.56\% & 69.22\% & 0.6898 & 0.6614 \\
    \hline
        \multirow{3}[0]{*}{AvgNet\cite{AvgNet}} & A     &\multirow{3}[0]{*}{0.46M} & 92.00\% & 33.80\% & 91.33\% & 62.57\% & 0.6460 & 0.5882 \\
          & B     & & 93.97\% & 40.40\% & 93.26\% & 64.29\% & 0.6452 & 0.6032 \\
          & C     &  & 92.46\% & 44.23\% & 88.51\% & 65.83\% & 0.6547 & 0.6241 \\
    \hline
    \multicolumn{1}{c}{\multirow{3}[1]{*}{STF-GCN(ours)}} & A     & \multirow{3}[0]{*}{0.28M}  & \textbf{93.68\%} & \textbf{40.69\%} & \textbf{93.13\%} & \textbf{64.35\%} & \textbf{0.6668} & \textbf{0.6079} \\
          & B     &  & \textbf{94.48\%} & \textbf{43.39}\% & \textbf{94.51\%} & \textbf{66.04}\% & \textbf{0.6633} & \textbf{0.6227} \\
          & C     &  & \textbf{99.23\%} & \textbf{47.60\%} & \textbf{95.59\%} & \textbf{70.95\%} & \textbf{0.7067} & \textbf{0.6808} \\
    \hline
    \hline
    \end{tabular}%
  \label{tab:bigfinal}%
\end{table*}

\begin{figure*}[h]
  \centering
  \includegraphics[width=0.32\textwidth]{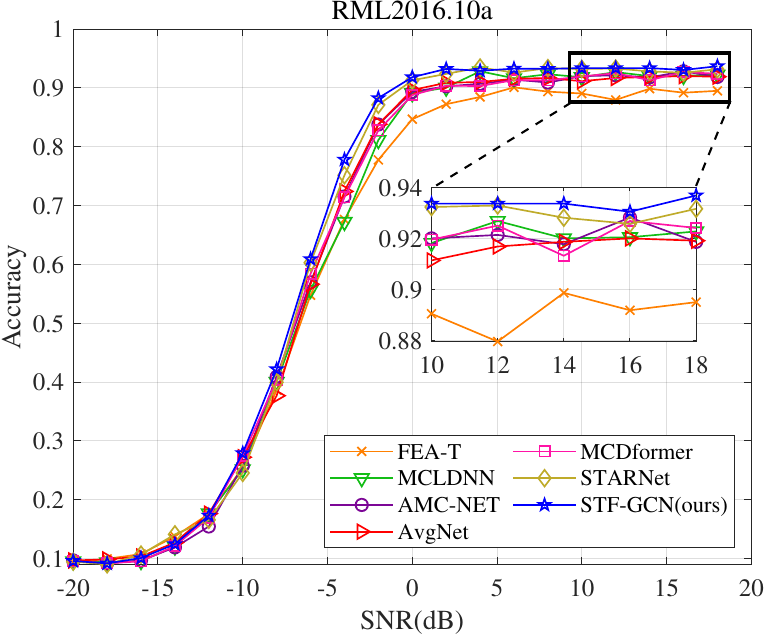}
  \includegraphics[width=0.32\textwidth]{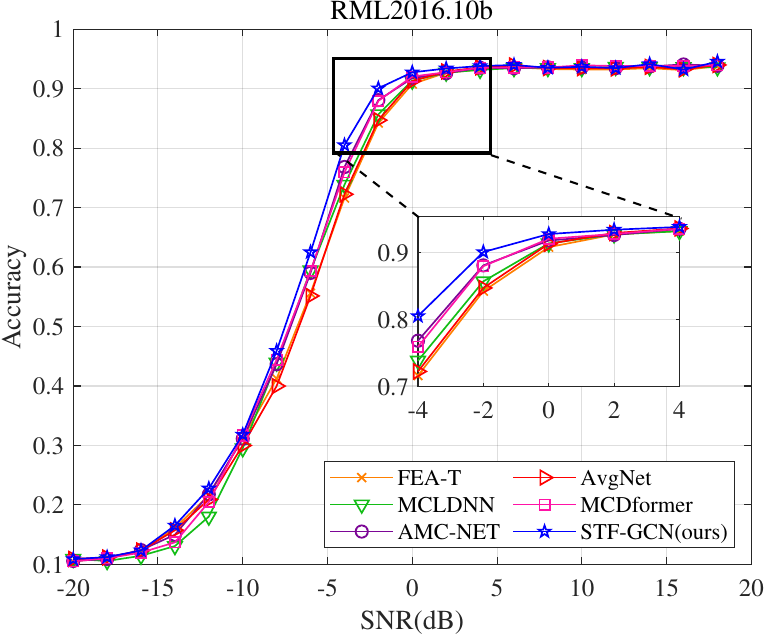}
  \includegraphics[width=0.32\textwidth]{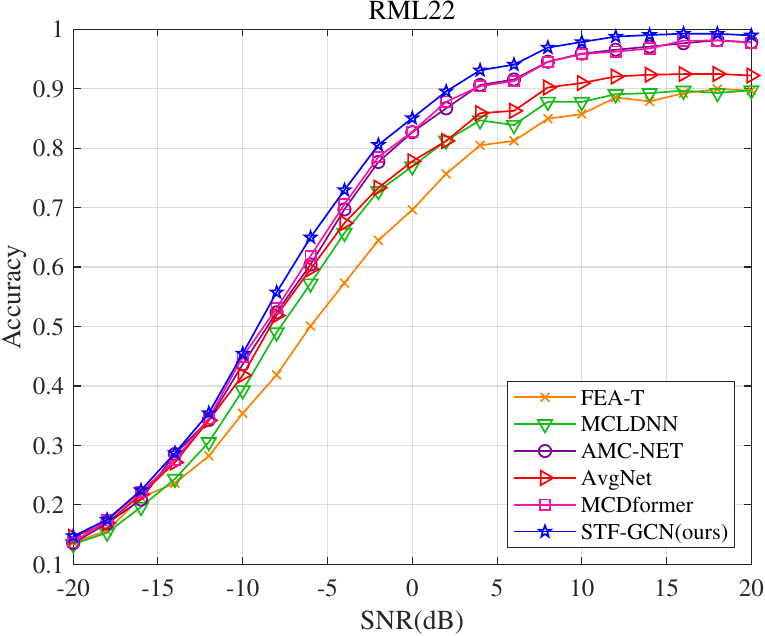}
  \caption{Recognition performance comparison between STF-GCN and other methods on RadioML2016.10a, RadioML2016.10b and RML22 datasets.}
  \label{fig:cp}
\end{figure*}
MCLDNN\cite{MCLDNN}, AMC-NET\cite{AMCNET}, FEA-T\cite{FEA-T}, STARNet\cite{STARnet}, MSGNET\cite{MSGNET}, MCDformer \cite{MCDformer} and AvgNet\cite{AvgNet} are used to compare the performance of our proposed model, STF-GCN, on RadioML2016.10a, RadioML2016.10b and RML22 datasets. Table \ref{tab:bigfinal} displays the performance comparison results, including the number of parameters, highest accuracy, recognition accuracy at different SNRs, overall performance, Macro-F1 score and Kappa score. As illustrated in the results, the proposed STF-GCN outperforms all other models on both three datasets, achieving overall accuracies of 64.35\%, 66.04\%, and 
70.95\%, respectively. Compared to the SOTA models, the overall accuracies improved by 1.78\%, 0.9\% and 1.73\% respectively, achieving the highest average recognition accuracies reported to now for both three datasets. The recognition performance of SOTA models is significantly impacted by low SNR conditions due to noise interference. However, our model achieves average recognition accuracies that surpass those of SOTA models at low SNR ranging from -20 to 0dB by 0.85\%, 1.47\% and  2.08\%, respectively. This demonstrates that our model exhibits greater adaptability and robustness in low SNRs. In addition, we achieve the best performance in terms of both recognition accuracy, maximum accuracy at high SNR, Macro-F1 score and Kappa score, which further reveals that the STF-GCN is a model with strong capabilities.

Fig. \ref{fig:cp} shows the comparison of the average recognition accuracies for a more comprehensive analysis over different SNRs, ranging from -10dB to 18dB. The experimental results clearly show that the proposed STF-GCN has achieved the best recognition behavior at low SNRs, especially ranging from -10dB to 2dB. In high SNR environments, modulated signals are easier to be learned. The overall accuracies of SOTA models are able to exceed 90\% when the SNR is greater than 4dB on RML2016 series and 8dB on RML22. However, when the SNR is greater than 0dB, -2dB and 4dB, respectively, the overall accuracy of the STF-GCN already exceeds 90\% and finally achieves the best performance of 93.68\%, 94.48\% and 99.23 at 18dB, 18dB and 16dB, respectively.

To further explore the computational complexity of the proposed model, we compared the inference time on GPU and CPU and Flops of STF-GCN with FEA-T, AMC-NET, MCDformer and the GNN-based SOTA model AvgNet on dataset RadioML2016.10a and the results are shown in Table \ref{tab:ab4}. It can be seen that MCDformer is optimized for CUDA and thus achieves the best performance on CUDA, however, its high complexity makes it unsatisfactory on CPU. FEA-T uses recognition performance in exchange for lower Flops to achieve optimal CPU inference time. For GNN, the weight computation of the adjacency matrix is unavoidable, but the model itself has low Flops and parameters, so there is not a major gap between the CUDA and CPU performance. STF-GCN is able to achieve a performance improvement of nearly 1.79\% with less than half of the Flops and 63.43\% of the CPU inference time compared to AvgNet, the SOTA model of GNN for AMR tasks, which argues for the deployment possibility of the proposed method.

\begin{table}
\centering
\caption{Comparison of computational complexity and inference time on RML2016.10a.}

\begin{tabular}{ccccc}
\toprule
 \multirow{3}*{Model} &  \multirow{3}*{FLOPs} & \multirow{3}*{\shortstack{Inference time \\on CUDA\\(ms/sample)}} &\multirow{3}*{\shortstack{Inference time \\on CPU\\(ms/sample)}} &\multirow{3}*{OverAll}\\
 \\
 \\
\midrule
FEA-T &\textbf{6.65M} &0.17 &\textbf{0.43} &60.41\% \\
AMC-Net &29.83M &0.27 &0.72 &62.40\% \\
MCDformer &75.02M  & \textbf{0.10}& 1.48 & 62.53\%\\
AvgNet &36.78M &0.58 &1.34 &62.56\%\\
\bottomrule 
\textbf{STF-GCN} &17.41M &0.47 &0.85 &\textbf{64.35\%}\\
\bottomrule
\end{tabular}
\label{tab:ab4}
\end{table}

\subsection{Ablation Study}
\begin{figure}
    \centering
    \includegraphics[width=0.45\textwidth]{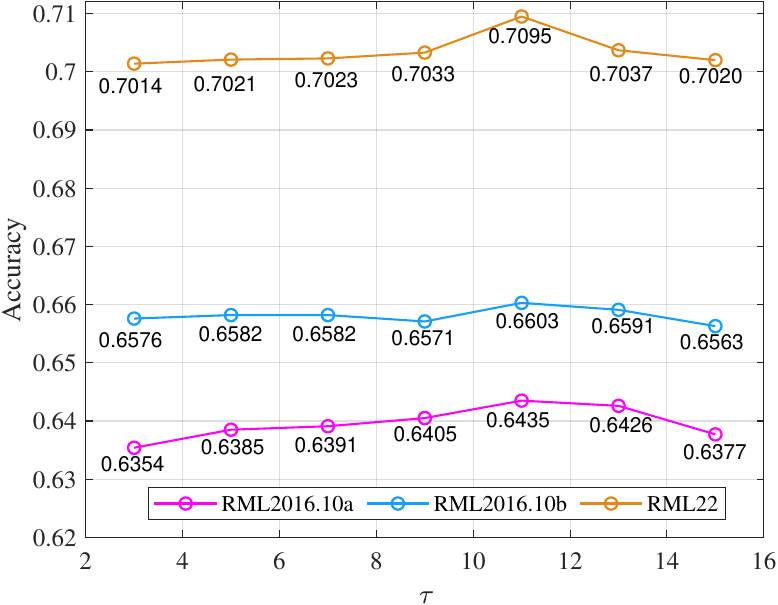}
    \caption{Classification performance for different $\tau$ values on three datasets.}
    \label{fig:tau}
\end{figure}
\begin{table}
\centering
\caption{The input ablation results for graph structure construction}
\begin{tabular}{clcc}
\hline
Datasets                      & \multicolumn{1}{c}{Model} & Overall               \\
\hline
\multirow{4}{*}{RML2016.10a} & w/o Rotation                         & 63.48\%          \\
                             & w/o DSTFT                             & 63.96\%          \\
                             & w/o I/Q signal              & 52.49\%          \\
                             & \textbf{STF-GCN}                 & \textbf{64.35\%} \\
\hline
\multirow{4}{*}{RML22} & w/o Rotation                        & 70.31\%          \\
                             & w/o DSTFT                              & 70.15\%          \\
                             & w/o I/Q signal                & 58.19\%          \\
                             & \textbf{STF-GCN}                 & \textbf{70.95\%}\\
\hline
\end{tabular}
\label{tab:ab1}
\end{table}

\begin{table}
\centering
\caption{The input ablation results for adjacency matrix generation}
\begin{tabular}{clccc}
\hline
& Method & Parameters & FLOPs & Overall               \\
\hline
& Distance-based & \textbf{0.28M} & \textbf{17.41M}  & 63.21\% \\
& KNN & \textbf{0.28M} & \textbf{17.41M}  & 57.56\%\\
& AVG & 0.40M   & 19.78M  & 63.62\%   \\
\hline
& \textbf{STF-GCN(ours)} & \textbf{0.28M} & \textbf{17.41M} & \textbf{64.35\%}\\
\hline
\end{tabular}
\label{tab:adj}
\end{table}

\begin{figure}[b]
    \centering
    \includegraphics[width=0.45\textwidth]{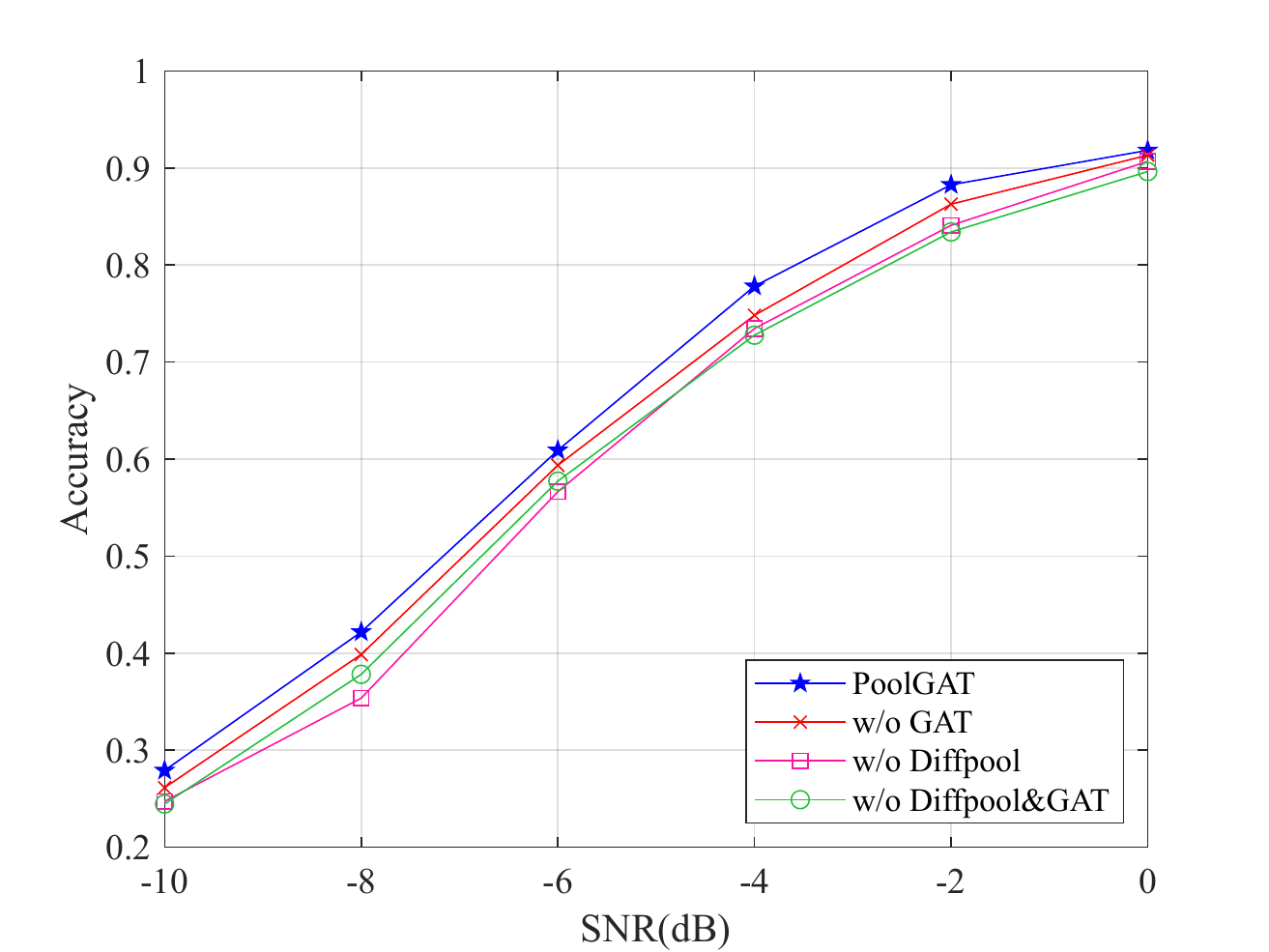}
    \caption{Results of the ablation of the classification performance for the PoolGAT module based on RML2016.10a.}
    \label{fig:poolgat}
\end{figure}
Since $\boldsymbol{\mathcal{A}}$ represents the edge weights of the graph structure, the quality of the graph structure will be directly affected by its sparsity. Fig.\ref{fig:tau} demonstrates that an exploration of the effect of the number of $\boldsymbol{\mathcal{A}}$ diagonal strips $\tau$ on the recognition accuracies was conducted on both three datasets. The experimental results found that when $\tau$ is limited from $3$ to $15$, the model performance can be maintained above 63.54\%, 65.63\% and 70.95\% on three datasets, and both of them reach the highest recognition accuracy at $\tau=11$. This is because both datasets have the same sampling rate and the samples consist of 128 samples, and $\tau=11$ can contain the most relevant mapping of modulation symbols.

We also designed ablation experiments on the model inputs, adjacency matrix generation methods and PoolGAT layer components.
In order to explore the impact of multi-domain inputs for the graph structure construction, Table \ref{tab:ab1} presents experiments investigating the ablation of rotation augmentation, I/Q signal, and DSTFT input on overall recognition performance across two types of fading channels. The experimental results reveal that the accuracy of both channel types decreases by more than 10\% when the I/Q signal input is removed, highlighting that spatial domain information is crucial for AMR tasks. In the case of removing the rotation augmentation or the DSTFT input, there was also a decrease in overall accuracies for both datasets, which proves the effectiveness of the data augmentation and the frequency domain input, and also shows that the multi-domain fusion plays a role in increasing the effect of recognition performance.

Table \ref{tab:adj} compares the proposed adaptive correlation-based method with Distance-based, KNN\cite{knn} and AVG\cite{AvgNet}, to demonstrate the effective contribution of our method to graph structure construction. The combined performance of our method in terms of computational complexity and recognition accuracy is significantly better than the other three methods. The proposed adaptive correlation-based approach does not require the introduction of learnable parameters for the CNN as in the case of AVG, which reduces the complexity of the construction, and we achieve far better recognition performance than other methods.
\begin{table}[h]
\centering
\caption{The ablation results of the PoolGAT layer on RML2016.10a}
\begin{tabular}{clcccc}
\hline
\multicolumn{1}{c}{Model} & Parameters & FLOPs & Overall\\
\hline
 w/o Diffpool & \textbf{0.19M} & 24.67M & 63.18\% &         \\
w/o GAT & 0.27M & \textbf{17.37M} & 63.60\%  &        \\
w/o Diffpool\&GAT & 0.19M & 24.41M & 62.77\%  &        \\
\hline
\textbf{STF-GCN} & 0.28M & 17.41M & \textbf{64.35\%} \\
\hline
\end{tabular}
\label{tab:poolgat}
\end{table}

Fig. \ref{fig:poolgat} and Table \ref{tab:poolgat} demonstrates the ablation experiments of the proposed PoolGAT module with SNR and complexity perspectives. We set up three ablation experiments that include removing the attention mechanism, removing the Diffpool, and leaving only retaining the stacked GCN layers. The experimental results found that if either the attention mechanism or the Diffpool layer is removed, the performance of the model will be degraded, proving the effectiveness of the coarsening operation and the attention mechanism. And without the Diffpool layer, Flops would go up by 41.7\%. When the attention mechanism and Diffpool are completely removed, the performance of the model decreases more significantly, indicating that the two components together have a stronger ability to extract robust signal features and better support the extraction of key features when the graph structure is coarsened. The result also proves that we have achieved a balance between speed-up and recognition performance at low SNRs.

\begin{figure}
	\centering
	\subfloat[RML2016.10a(-4dB,2dB and 8dB)]{
		\includegraphics[width=0.15\textwidth]{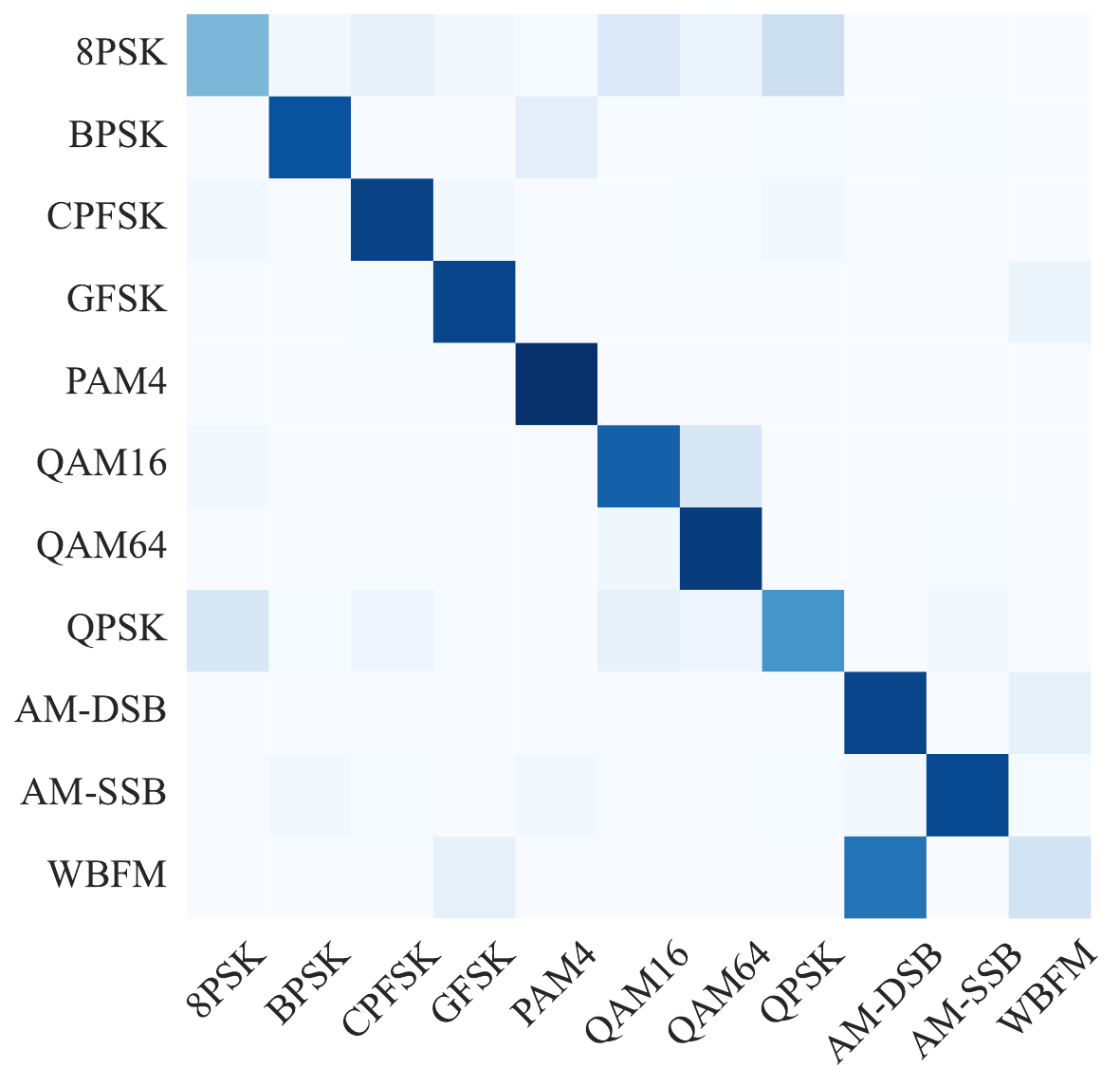}
		\includegraphics[width=0.15\textwidth]{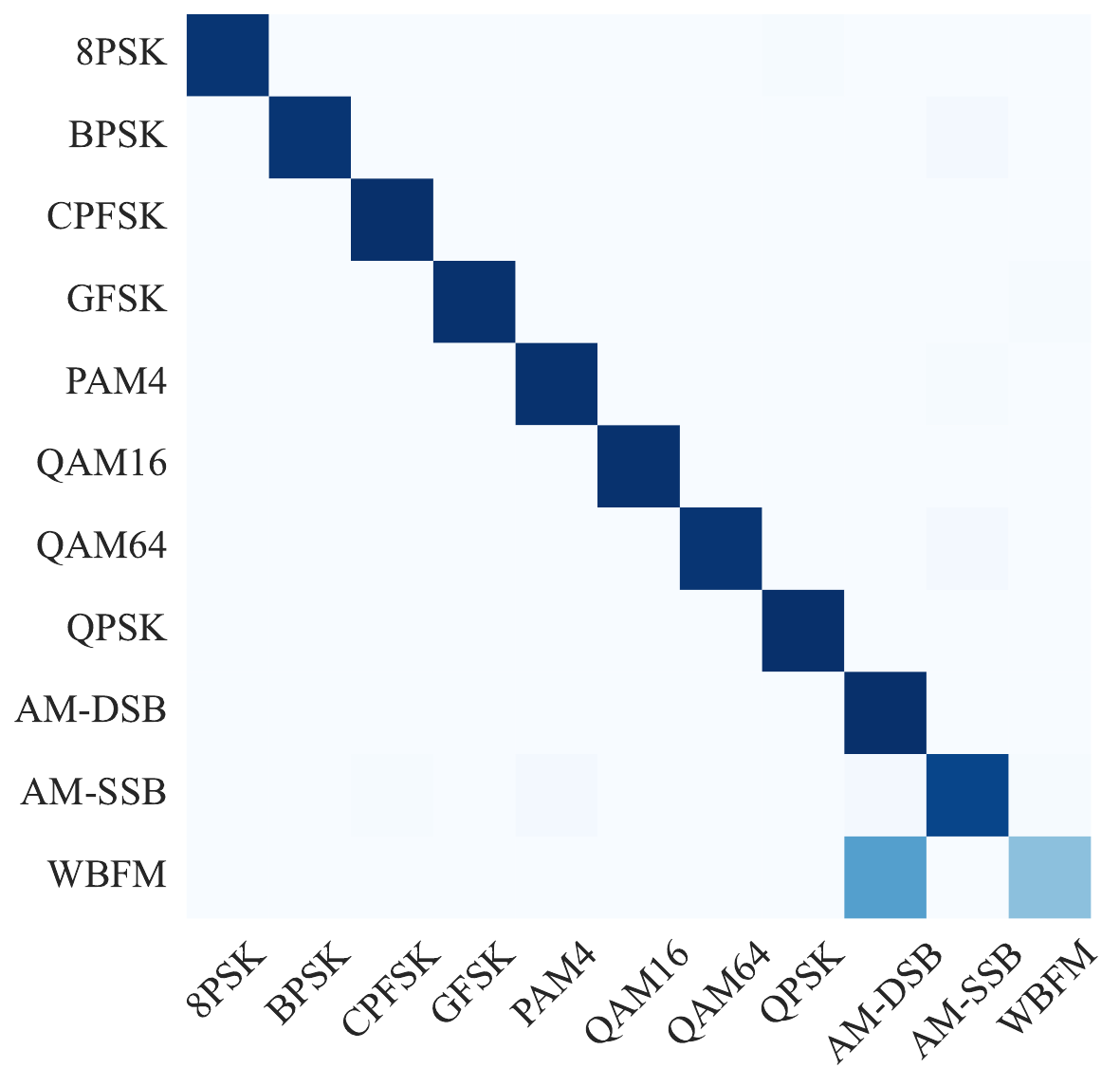}
            \includegraphics[width=0.15\textwidth]{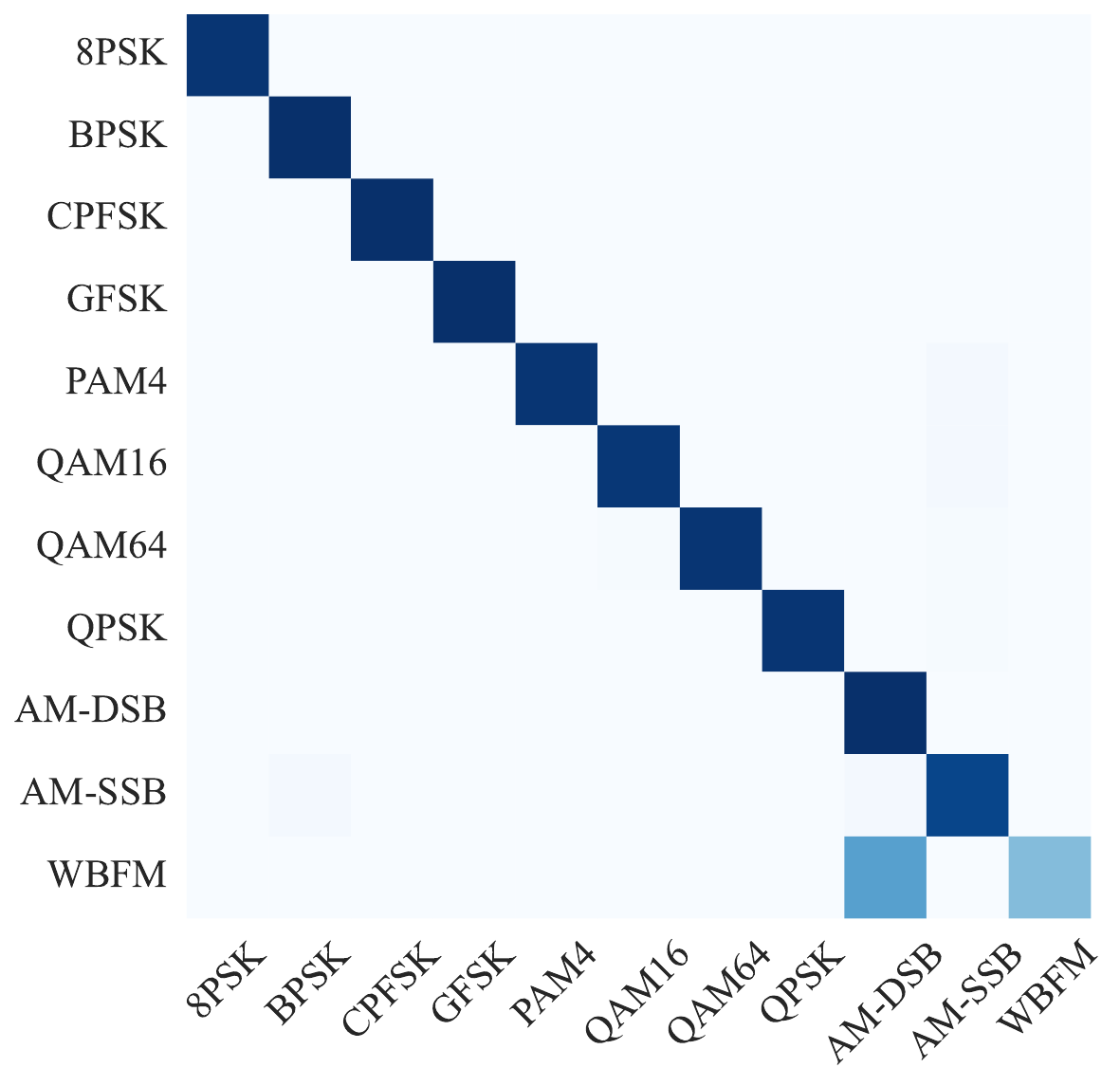}
            }
        \\
        \subfloat[RML22(-4dB,2dB and 8dB)]{
            \includegraphics[width=0.15\textwidth]{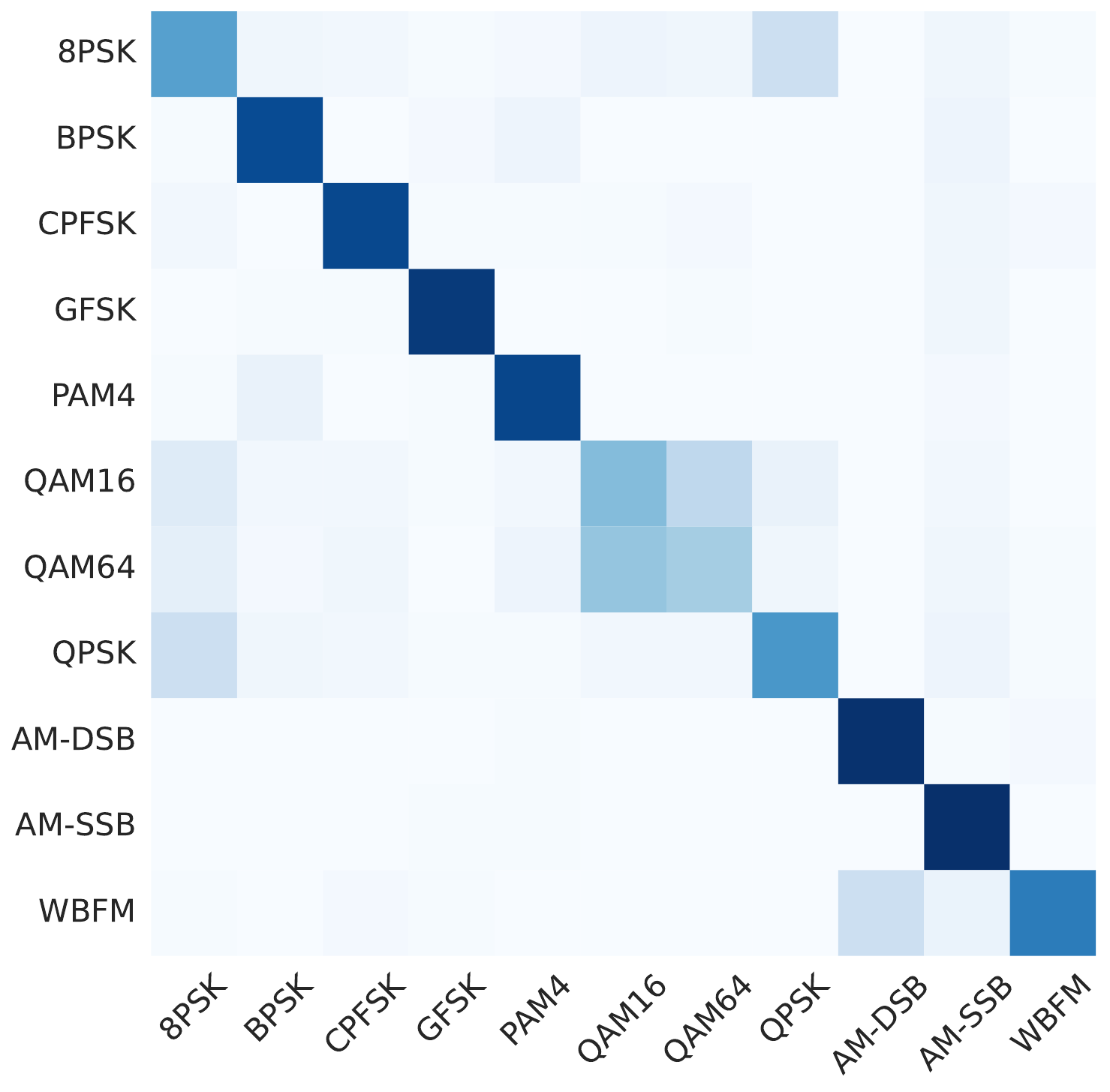}
		\includegraphics[width=0.15\textwidth]{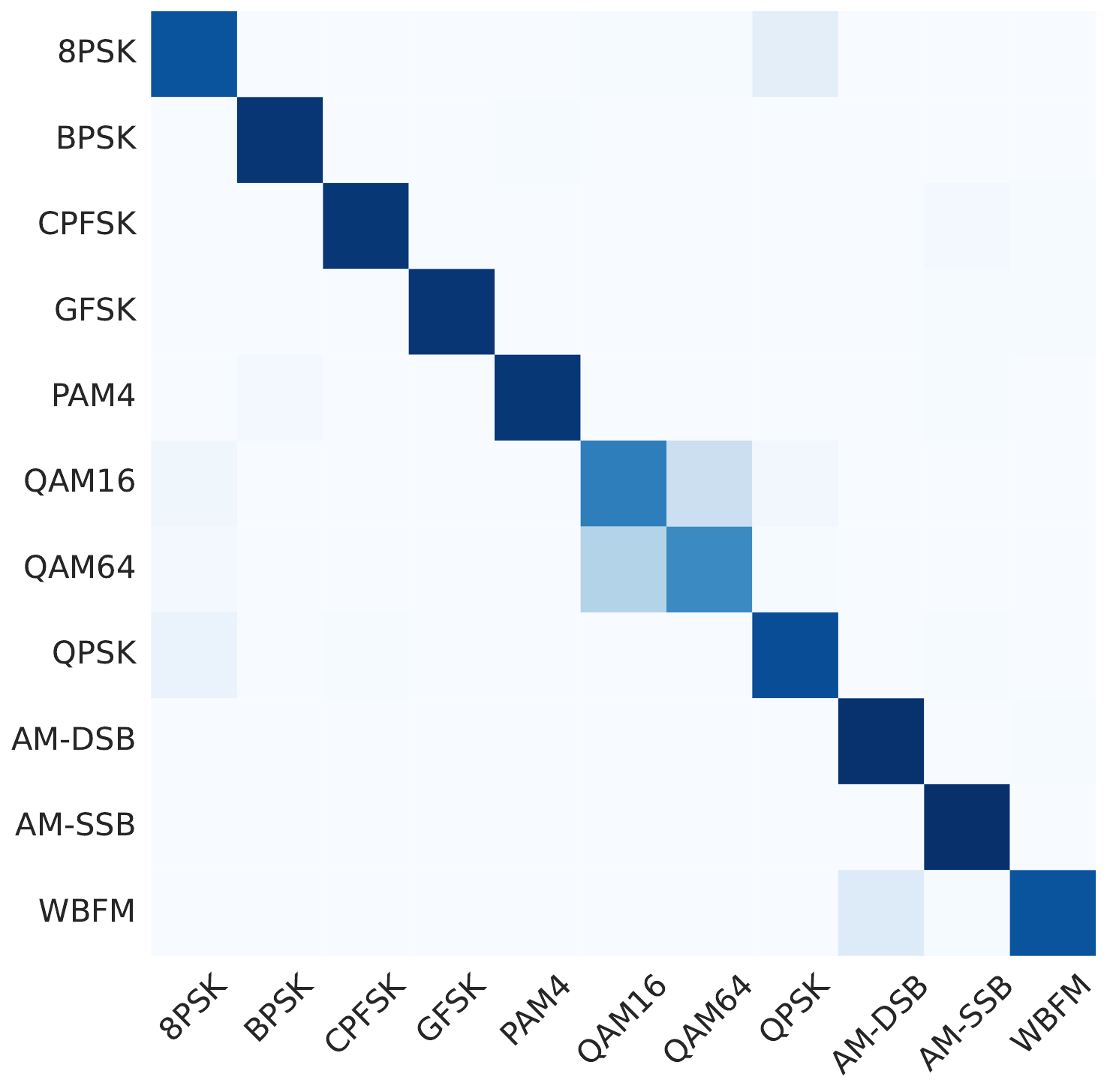}
            \includegraphics[width=0.15\textwidth]{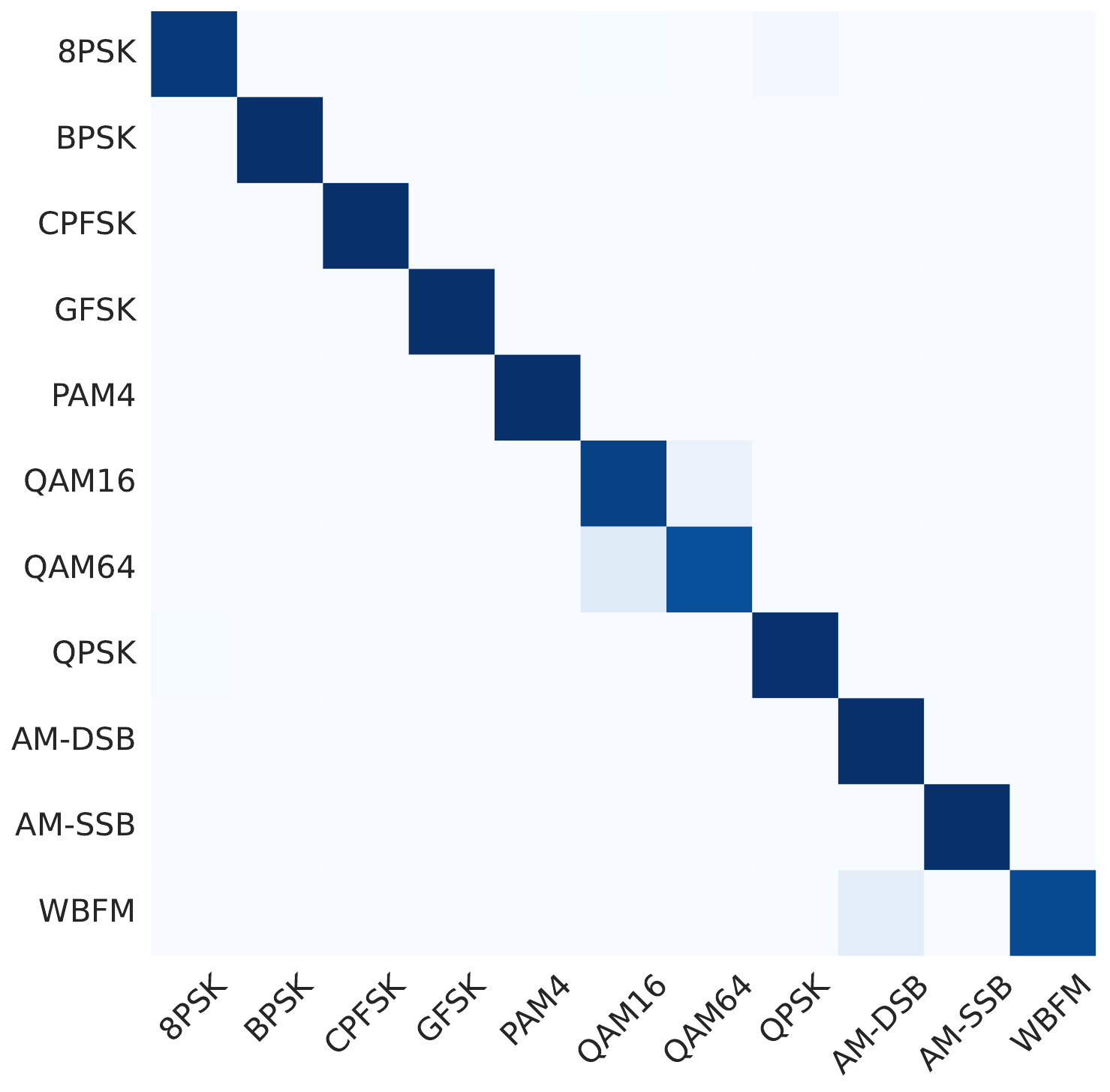}
            }  
        \caption{Normalized confusion matrix at -4dB, 2dB and 8dB SNR on RML2016.10a and RML22.}
	\label{fig:cm} 
\end{figure}

\subsection{Visualization Analysis}
We show the confusion matrices for the two datasets at -4dB, 2dB, and 8dB in Fig. \ref{fig:cm}, from which we can observe the detailed classification results for each modulation type under two different channels. In the RML2016.10a dataset for the small fading channel simulation, at -4dB, most modulation types have only a small error except the PSK modulation type which still has a large error and is recognized as a QAM modulation type. At 2dB and 8dB, the recognition accuracies of all modulation types except WBFM are close to 100\%, and the recognition is very stable and robust. The reason for the lower recognition rate of WBFM is that there are a large number of silence samples in both WBFM and AM-DSB, which leads to some of the WBFM signals to be misinterpreted as AM-DSB signals.
In the RML22 dataset, the multipath effects and delays in urban channels exacerbate intersymbol interference, which significantly affects QAM schemes with higher symbol densities. This is reflected in the need for a higher SNR to properly resolve the confusion between QAM16 and QAM64 compared to the case in the LOS channels.

\begin{figure*} [t!]
	\centering
	\subfloat[BPSK]{
		\includegraphics[height=0.16\textwidth]{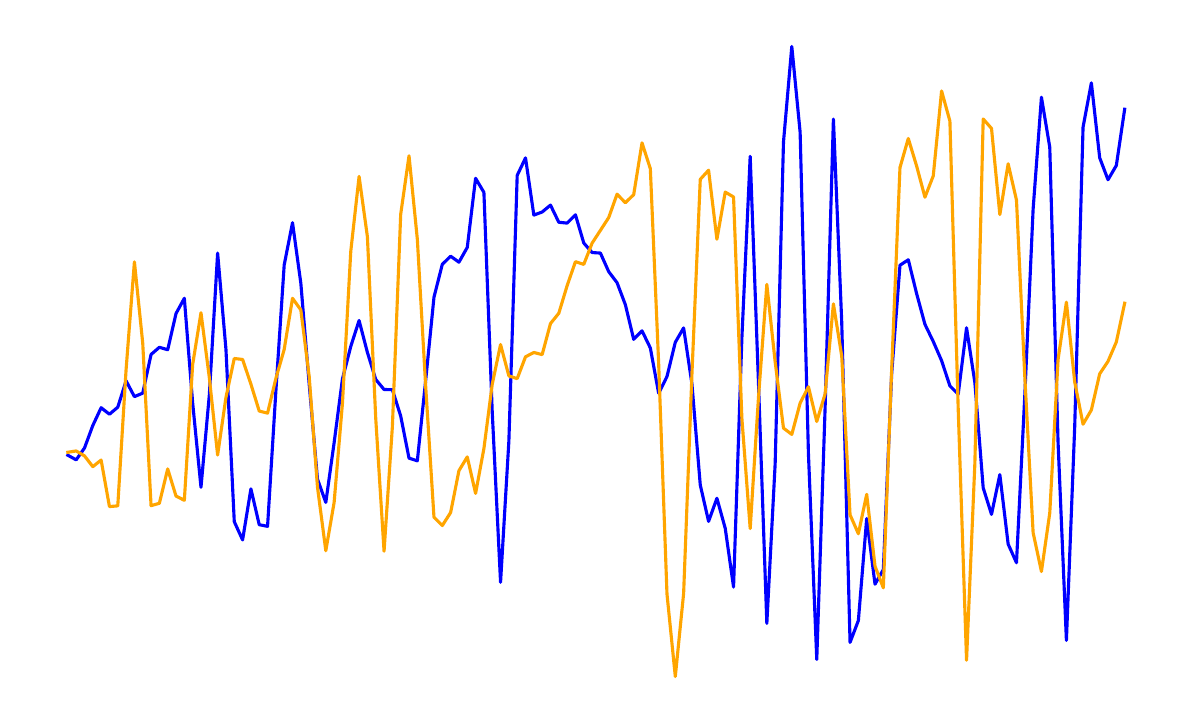}
		\includegraphics[height=0.16\textwidth]{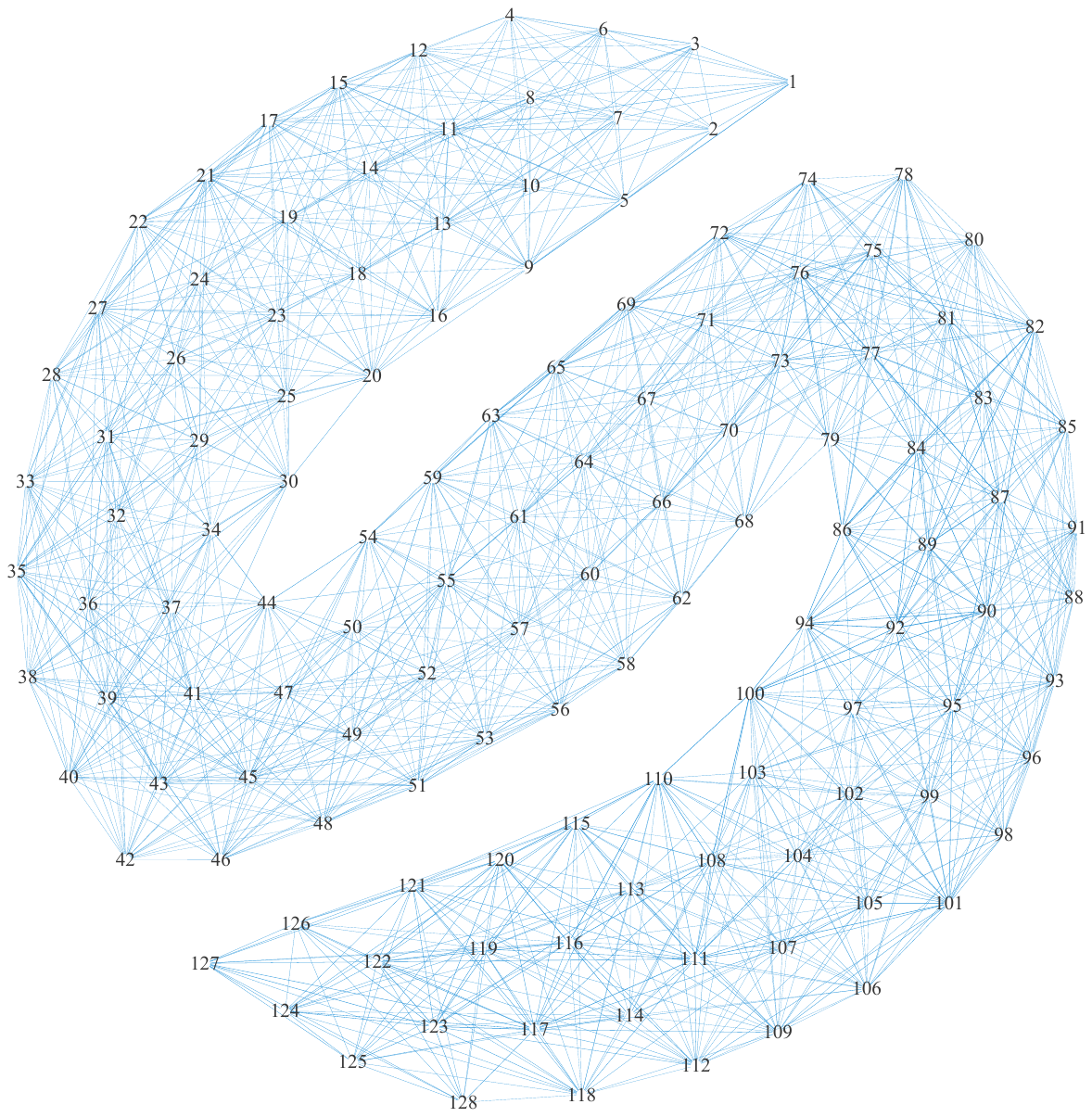}
            \includegraphics[height=0.16\textwidth]{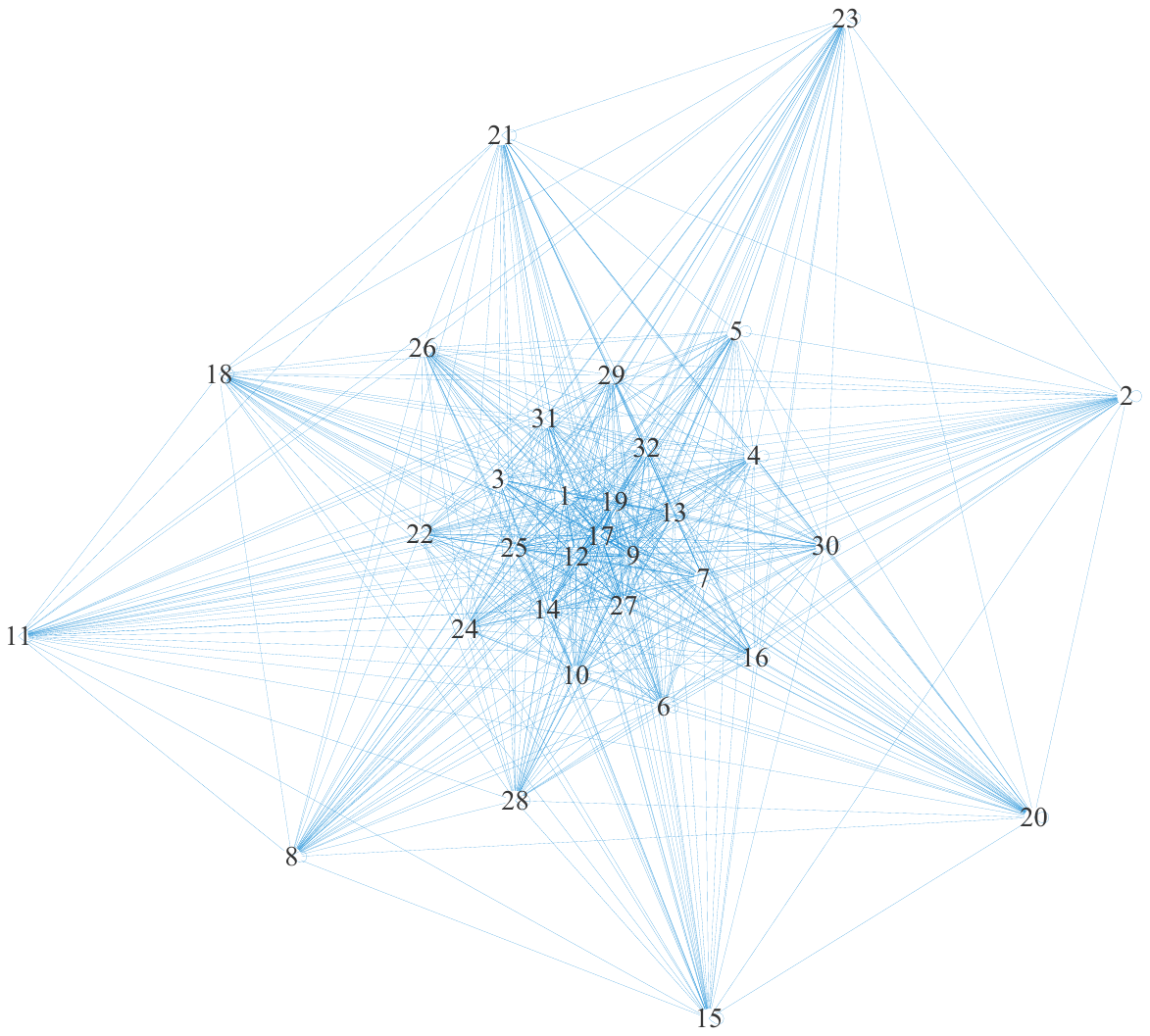}
            }
        \subfloat[QPSK]{
		\includegraphics[height=0.16\textwidth]{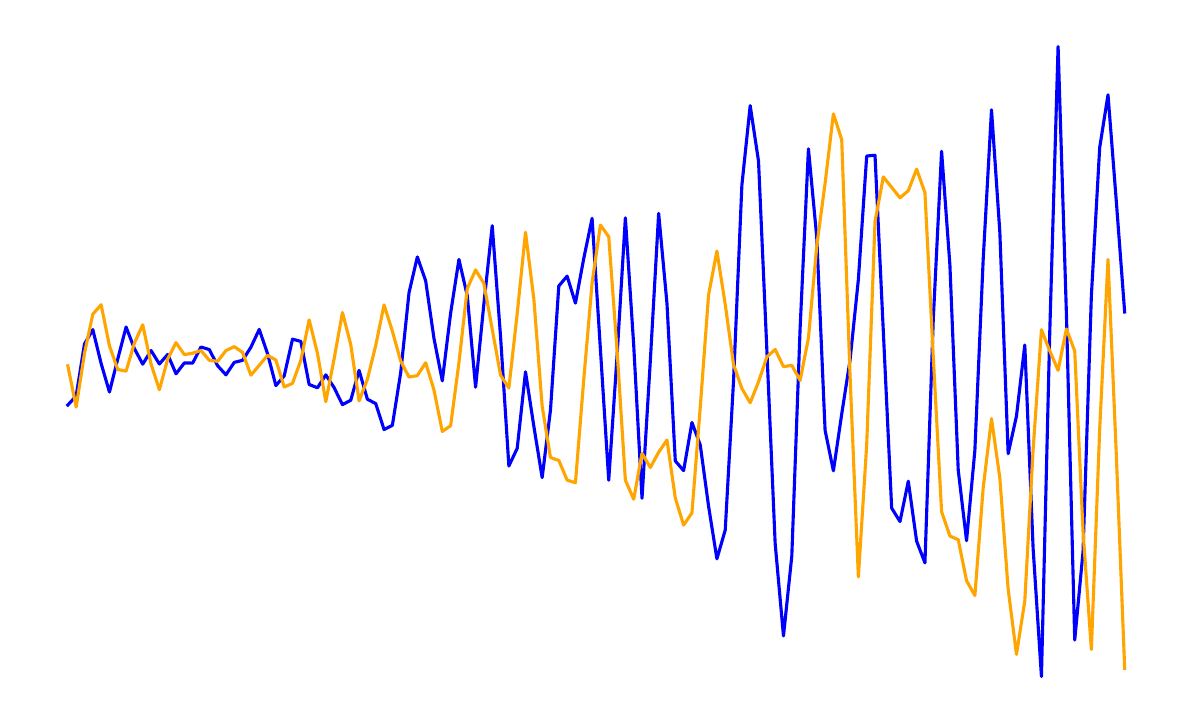}
		\includegraphics[height=0.16\textwidth]{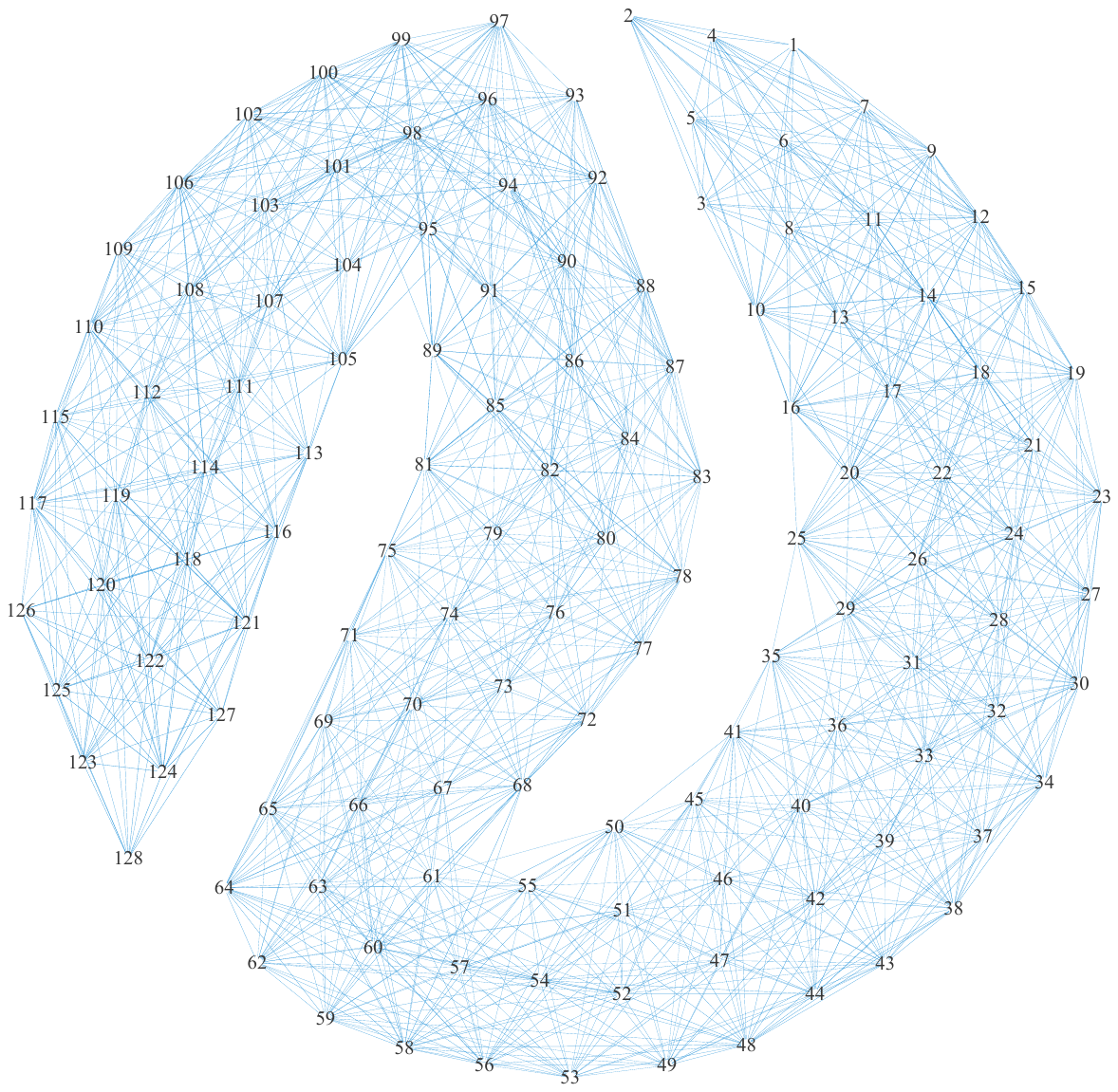}
            \includegraphics[height=0.16\textwidth]{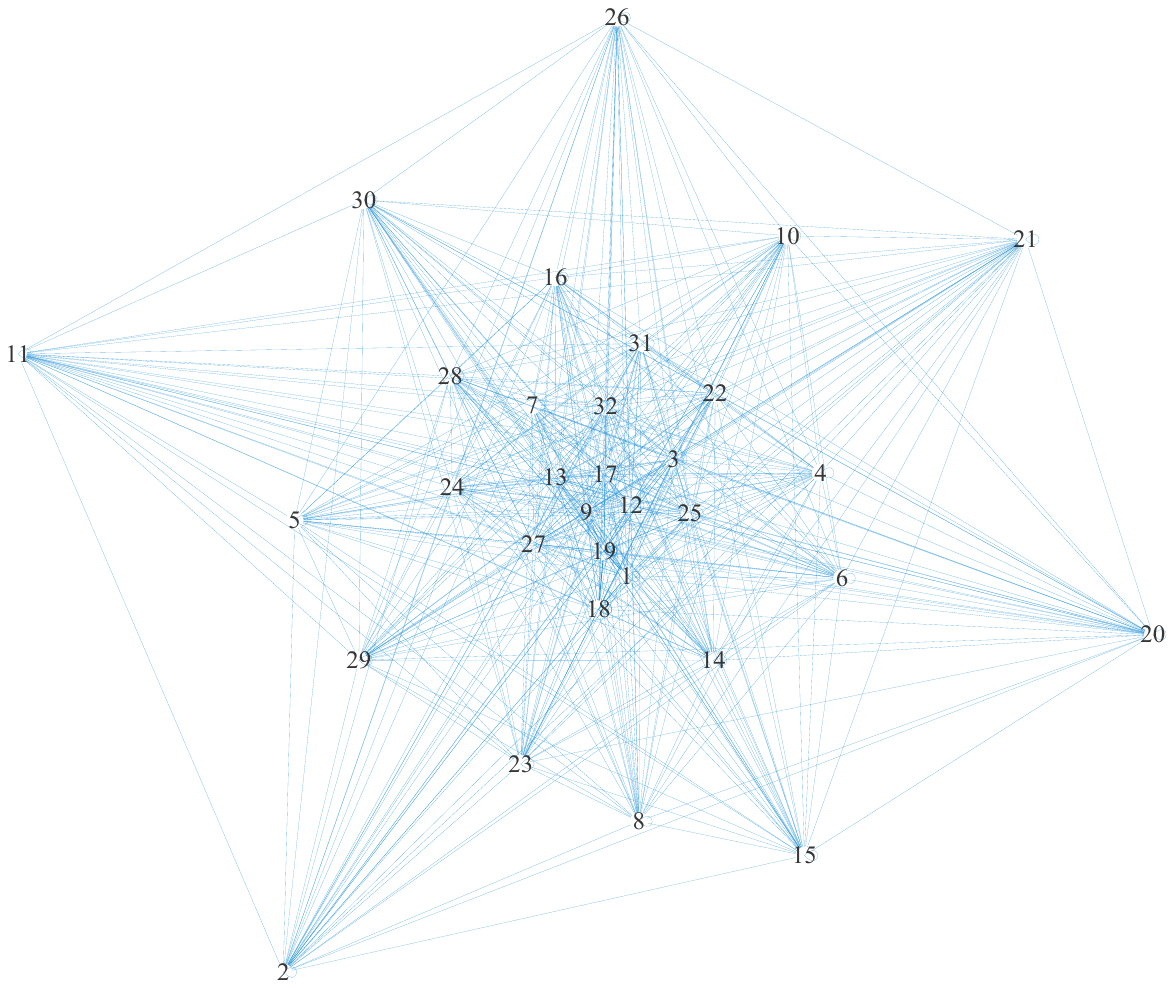}
            }    
 \\
 	\subfloat[8PSK]{
		\includegraphics[height=0.16\textwidth]{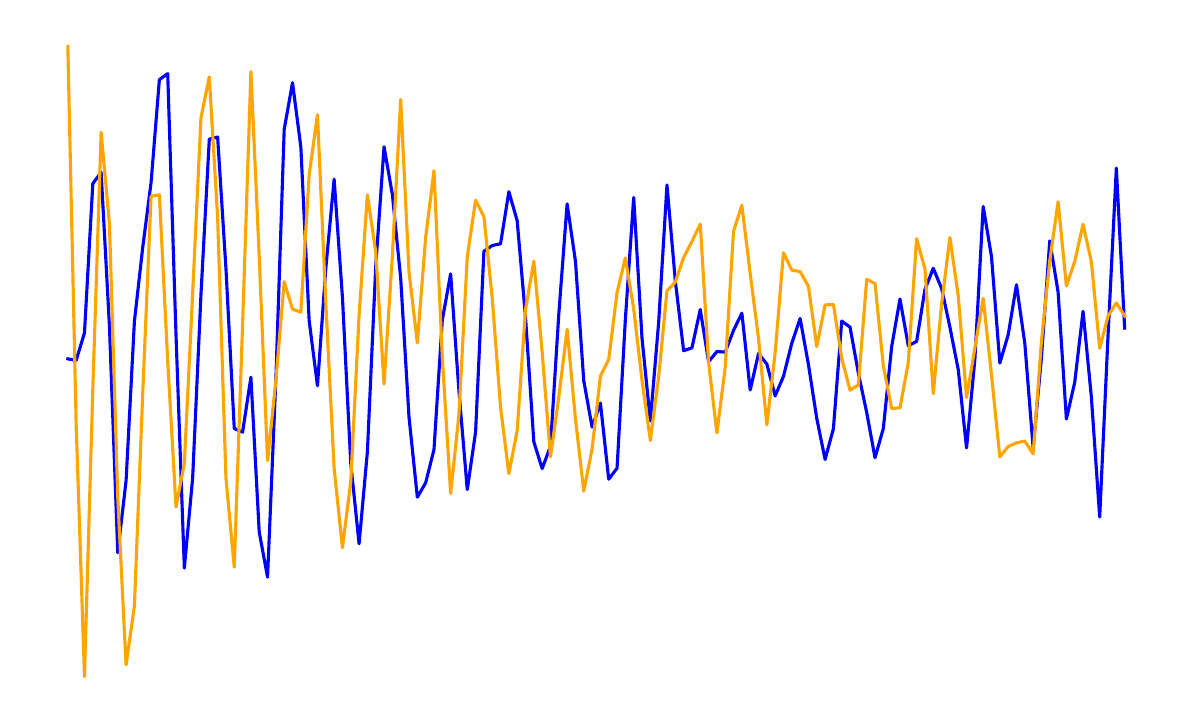}
		\includegraphics[height=0.16\textwidth]{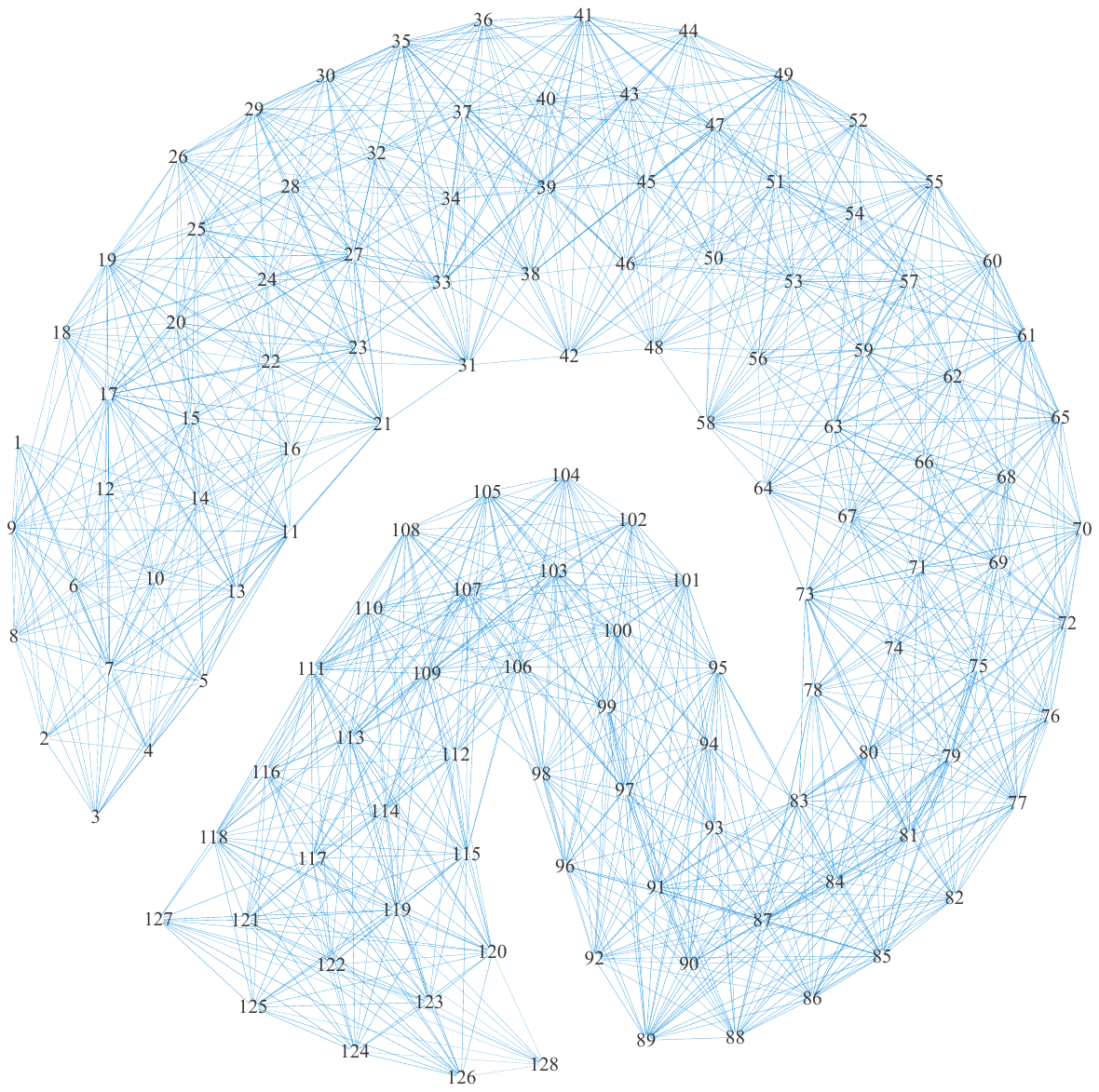}
            \includegraphics[height=0.16\textwidth]{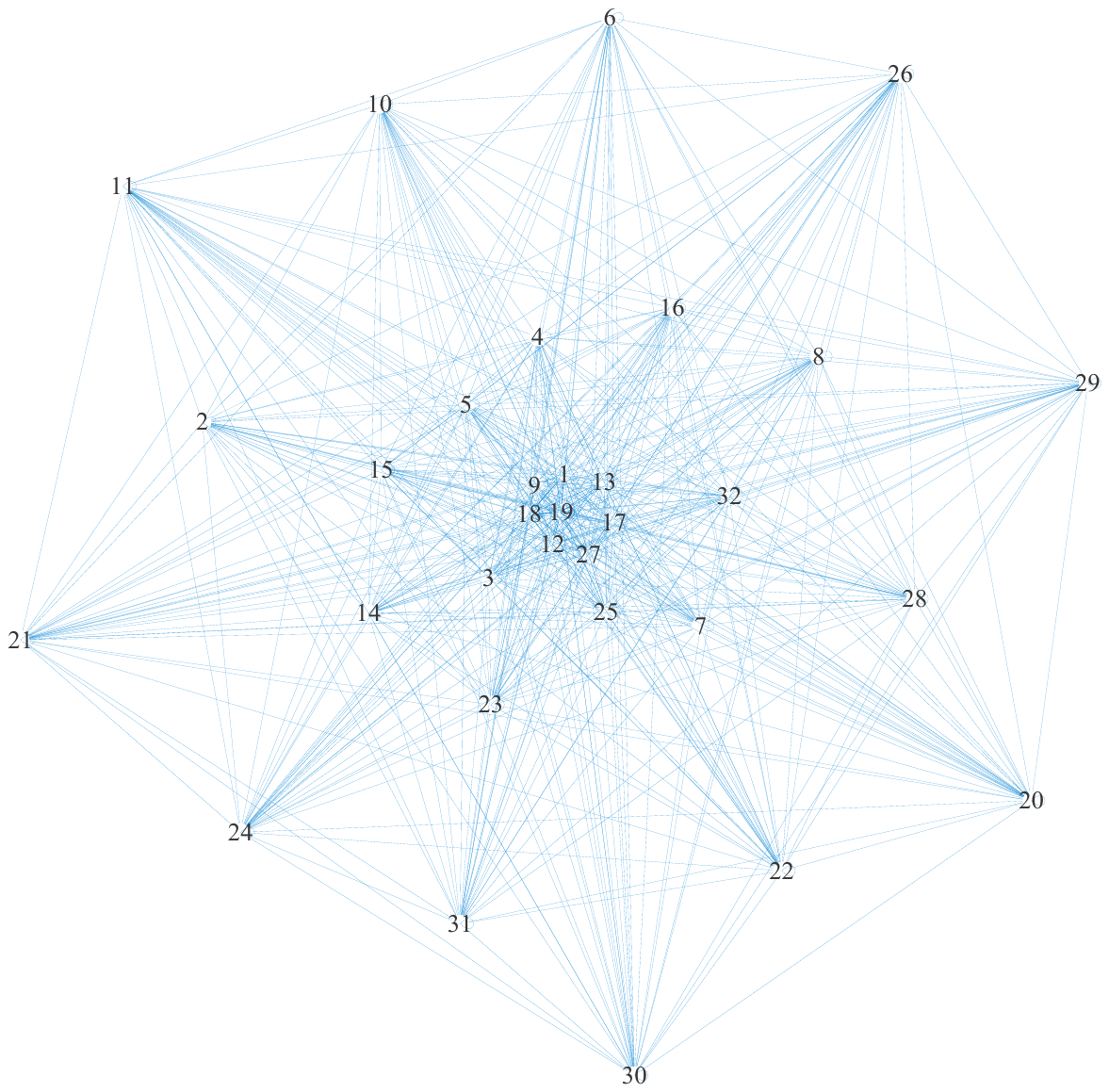}
            }
        \subfloat[CPFSK]{
		\includegraphics[height=0.16\textwidth]{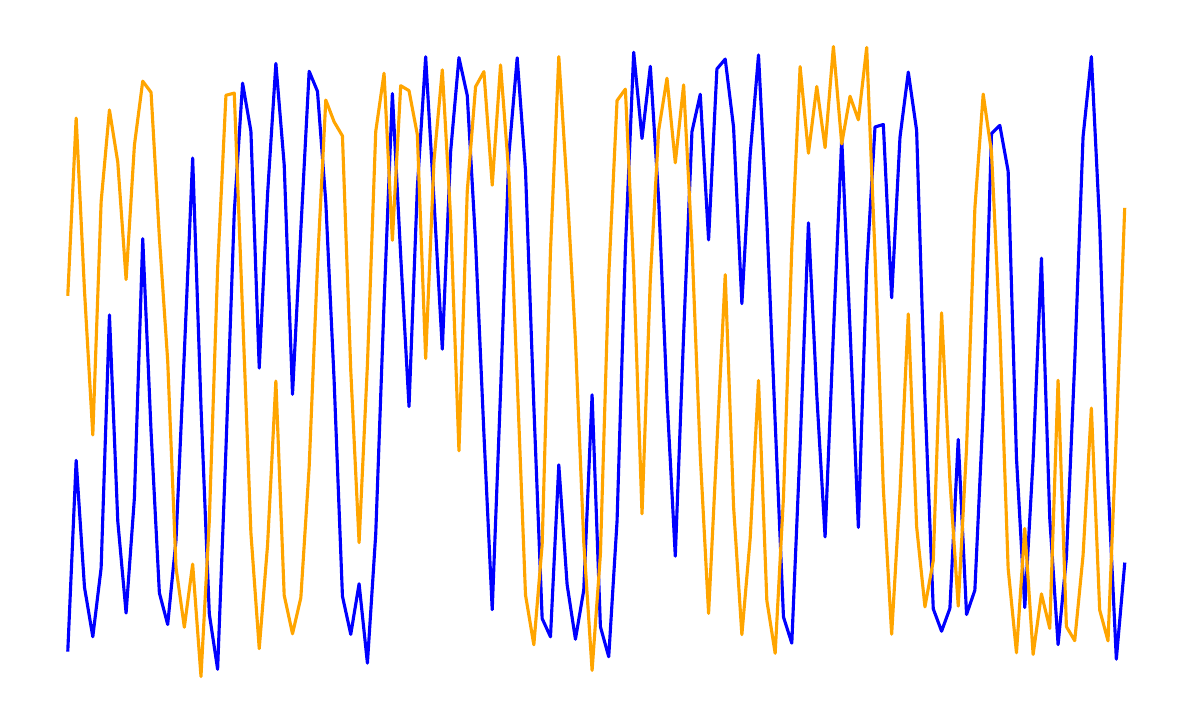}
		\includegraphics[height=0.16\textwidth]{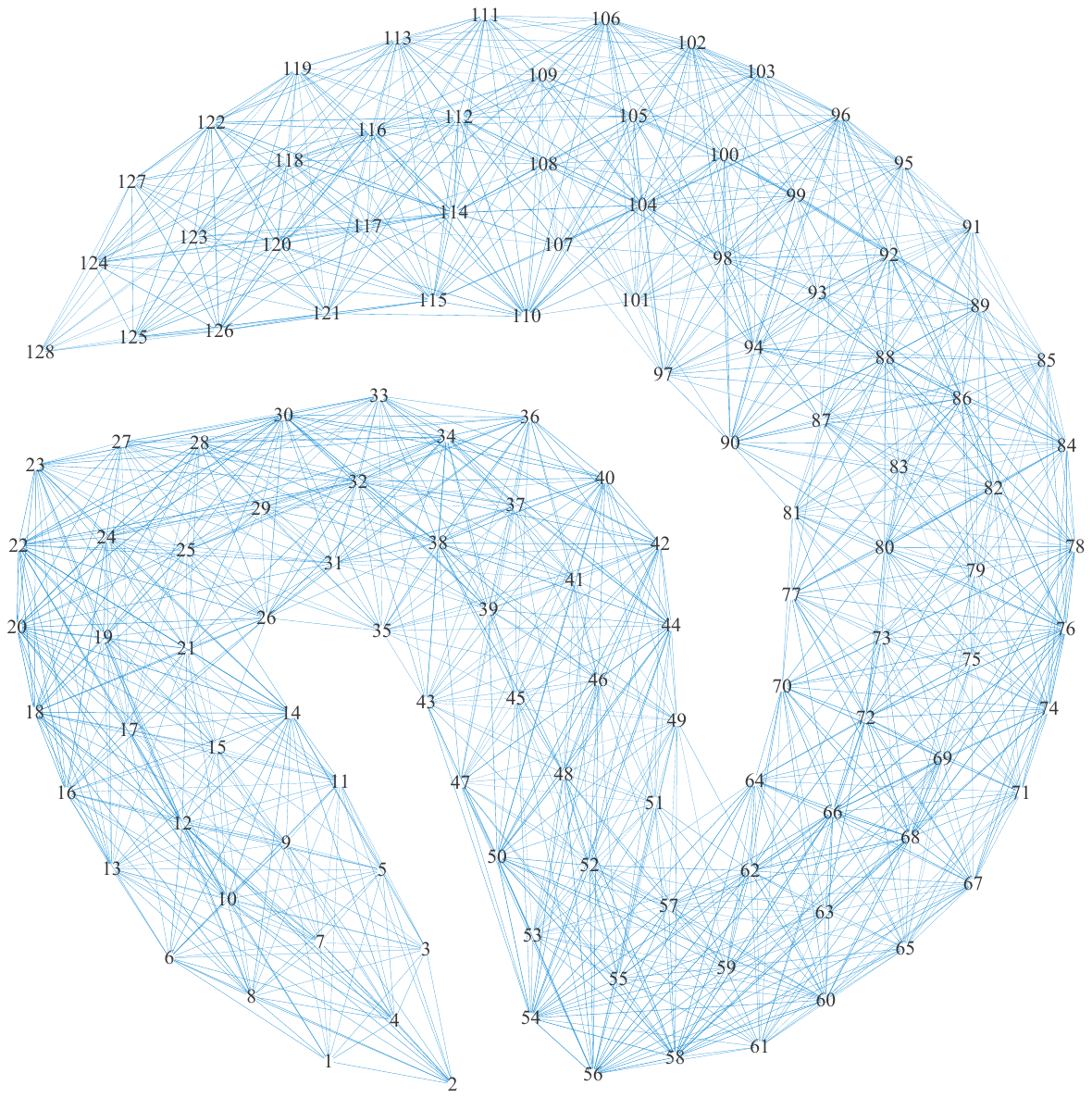}
            \includegraphics[height=0.16\textwidth]{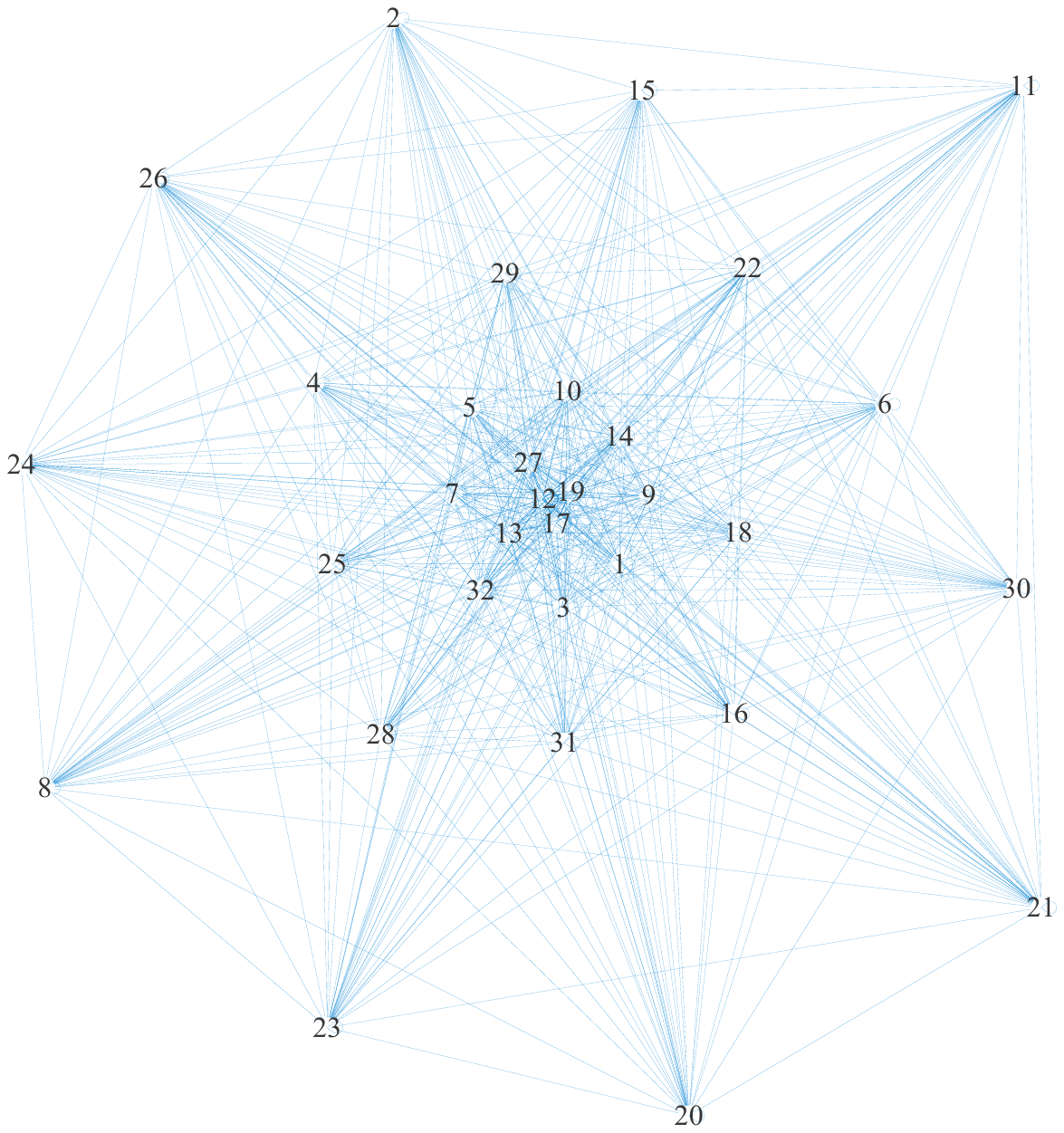}
            }    
 \\
  	\subfloat[GFSK]{
		\includegraphics[height=0.16\textwidth]{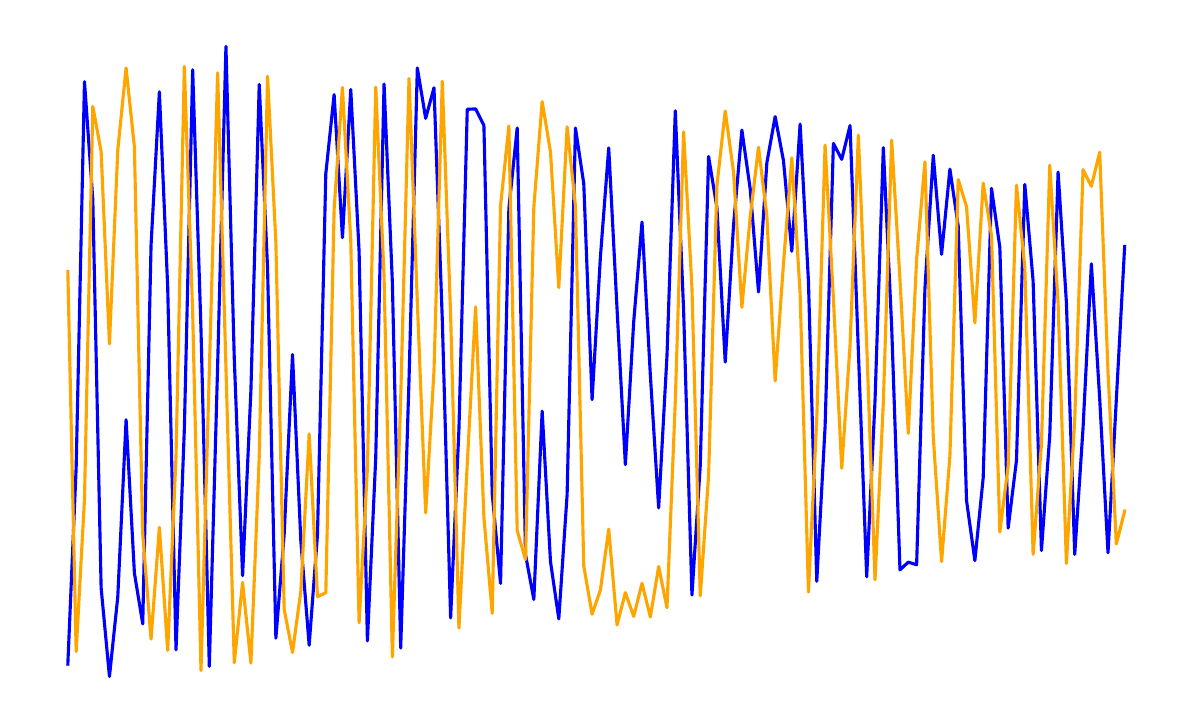}
		\includegraphics[height=0.16\textwidth]{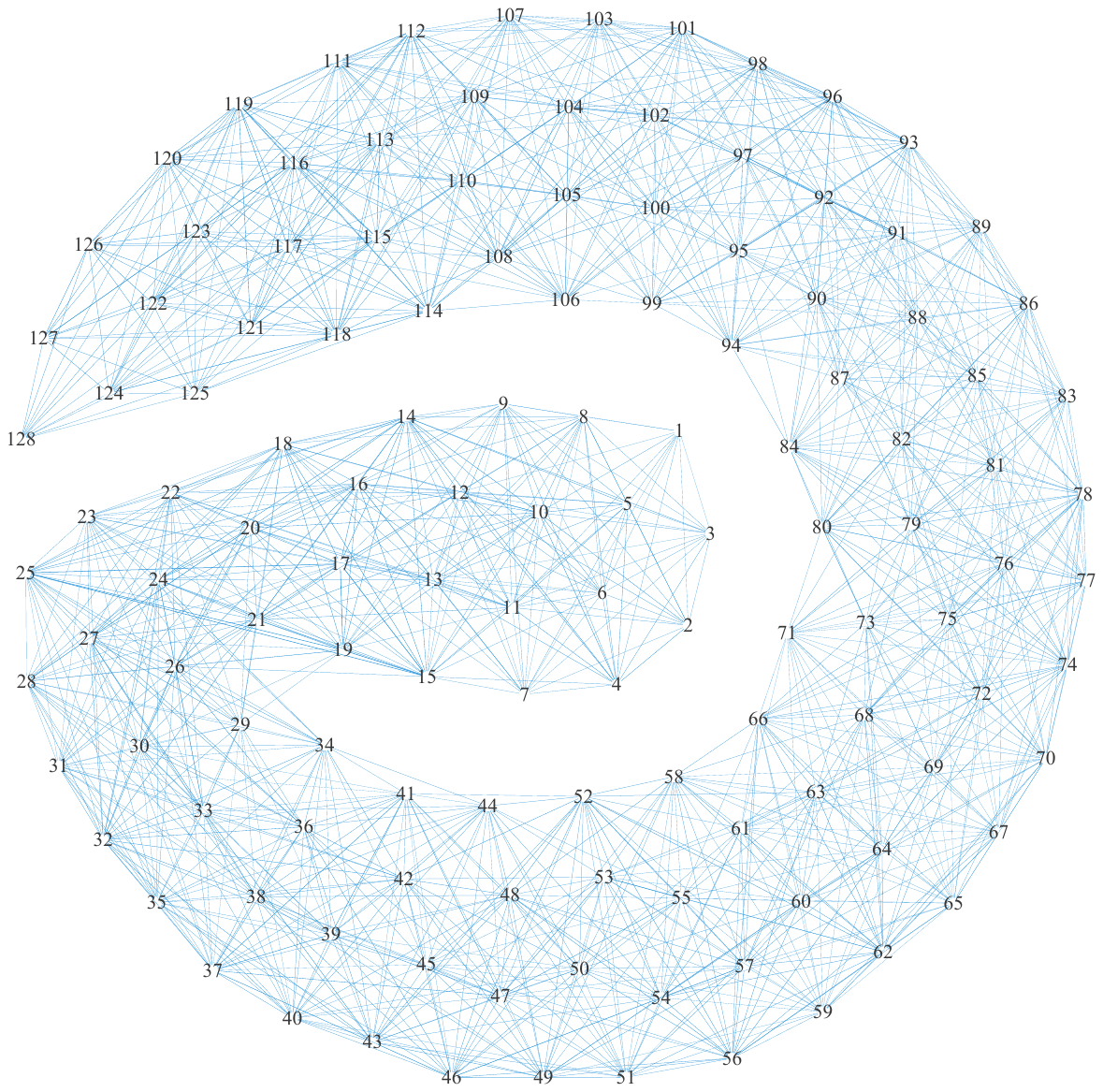}
            \includegraphics[height=0.16\textwidth]{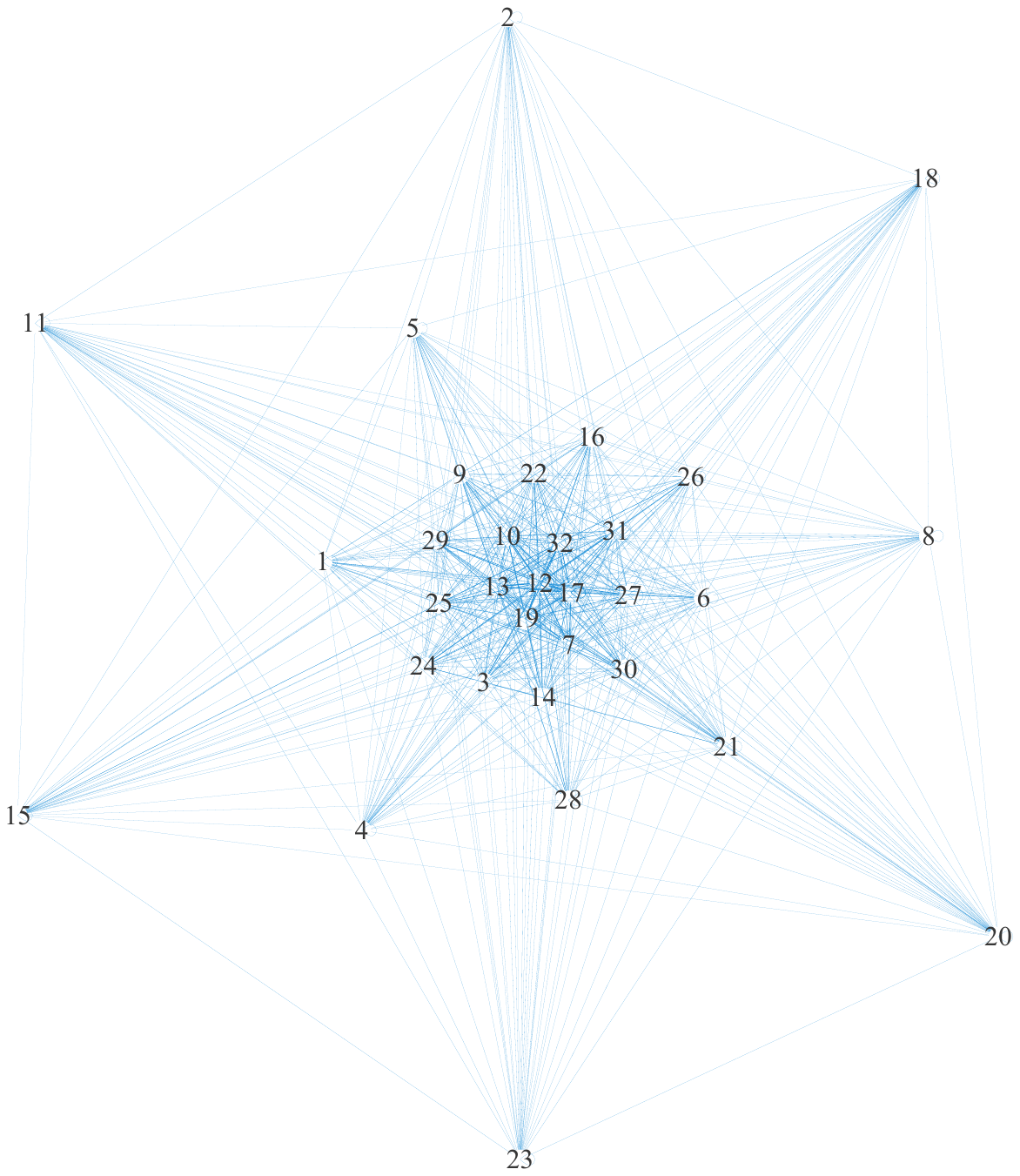}
            }
        \subfloat[PAM4]{
		\includegraphics[height=0.16\textwidth]{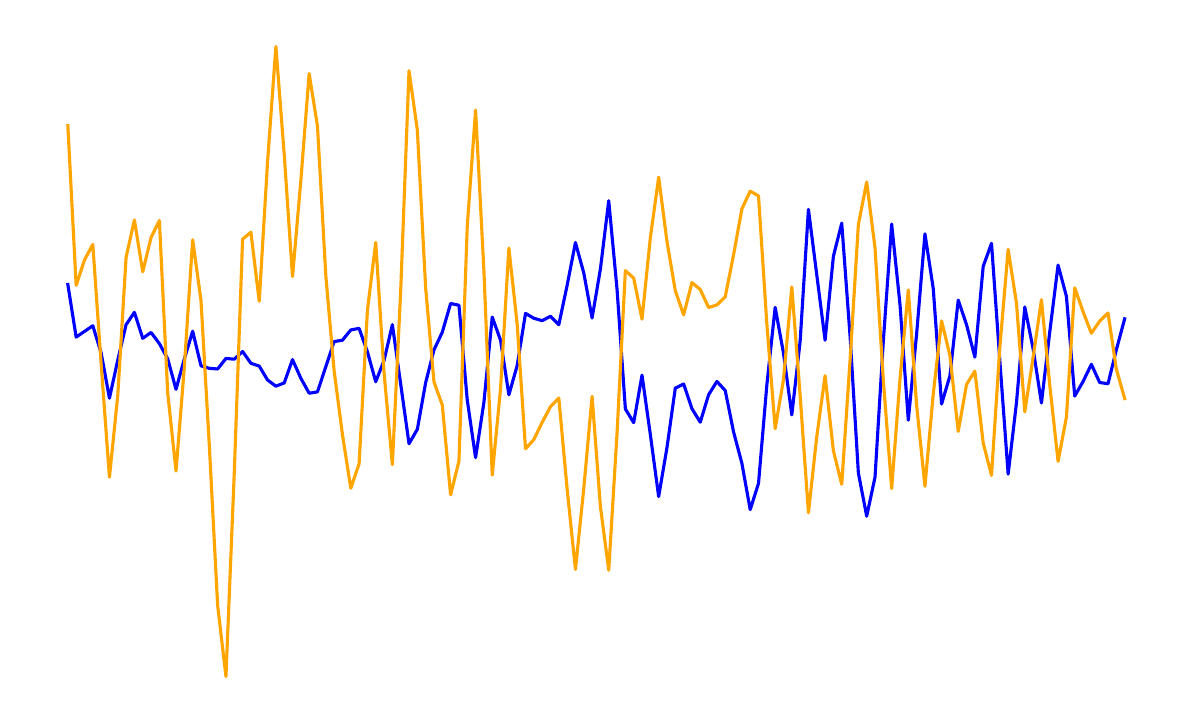}
		\includegraphics[height=0.16\textwidth]{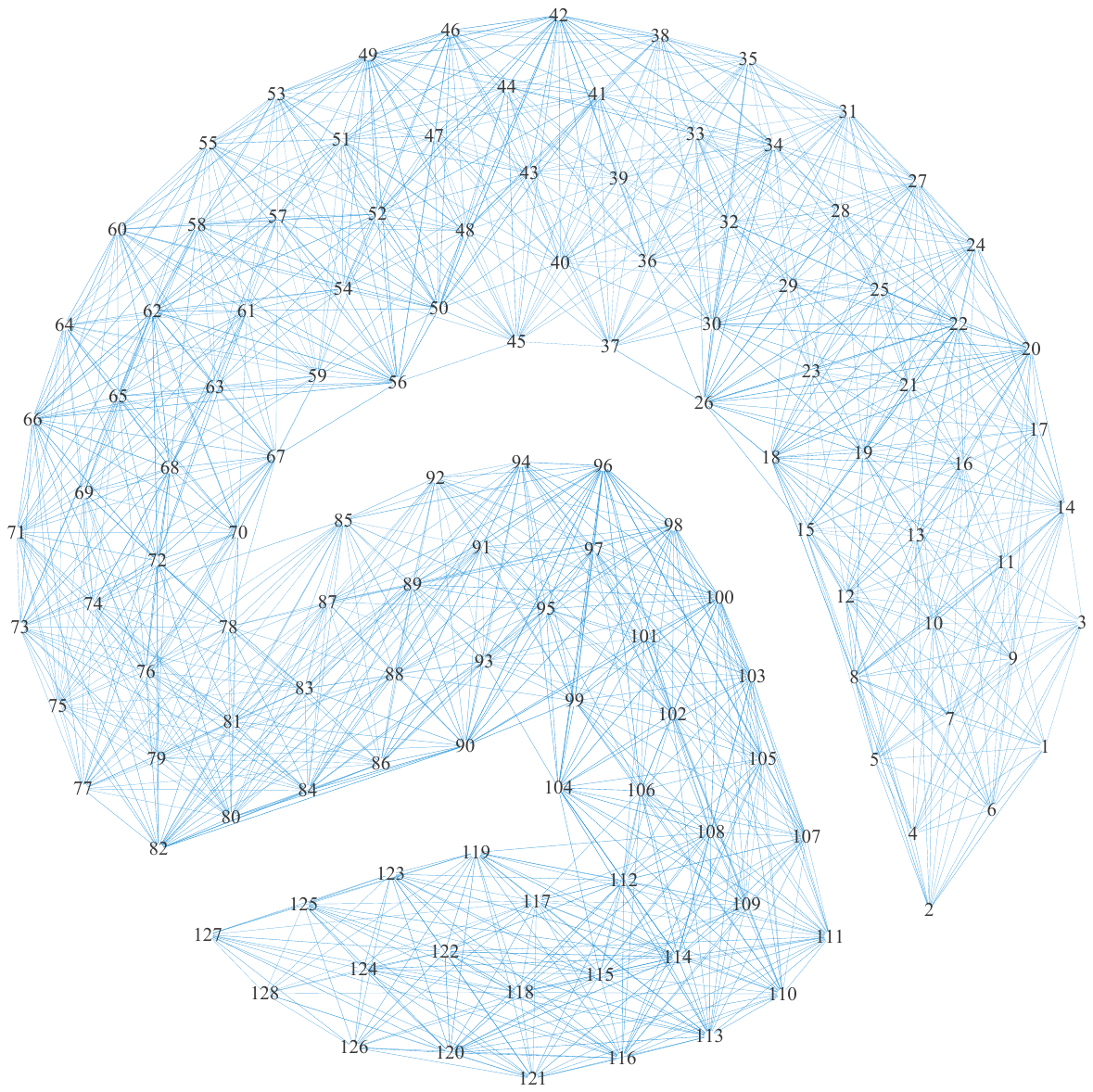}
            \includegraphics[height=0.16\textwidth]{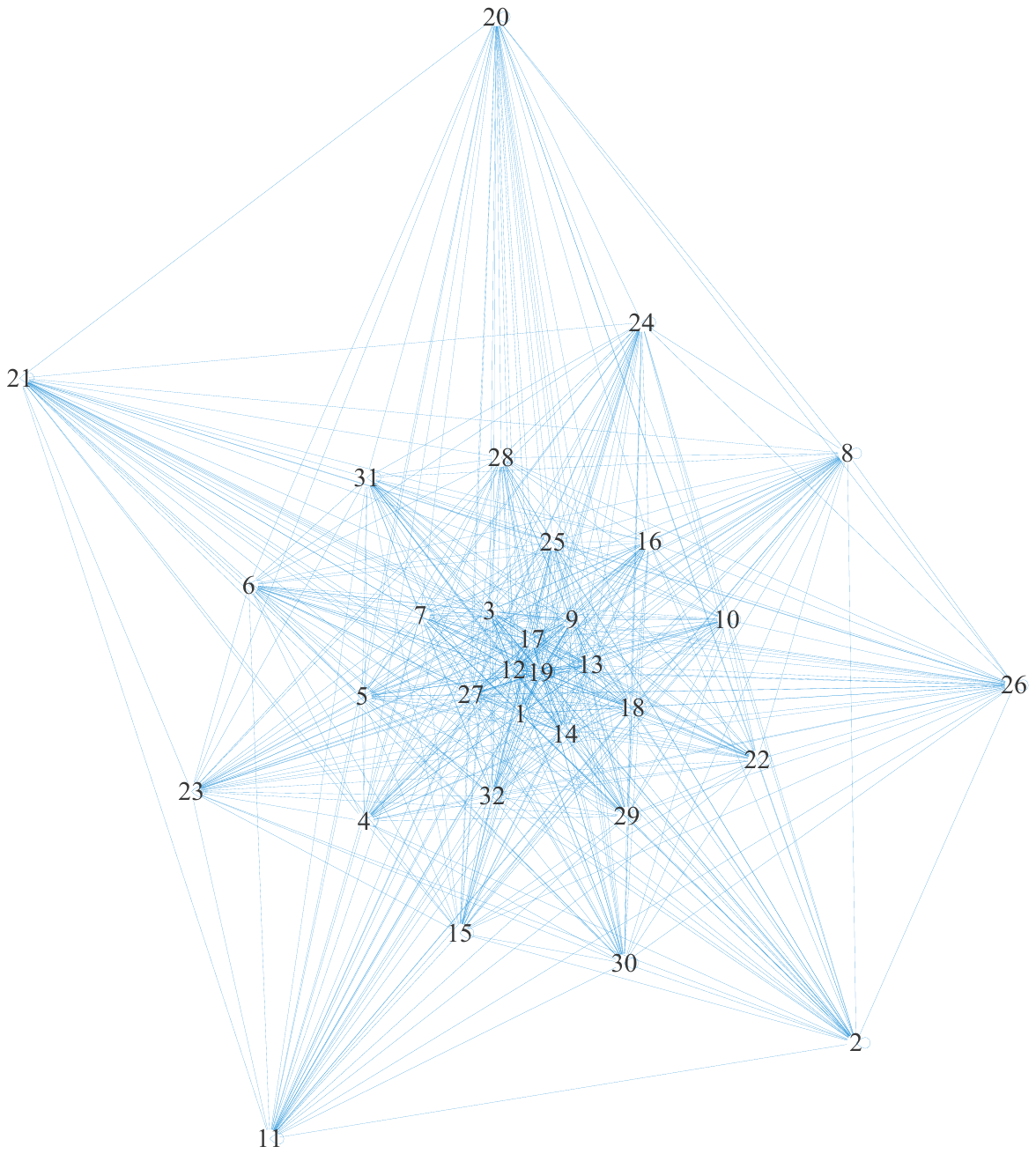}
            }
    \\
        \subfloat[QAM16]{
		\includegraphics[height=0.16\textwidth]{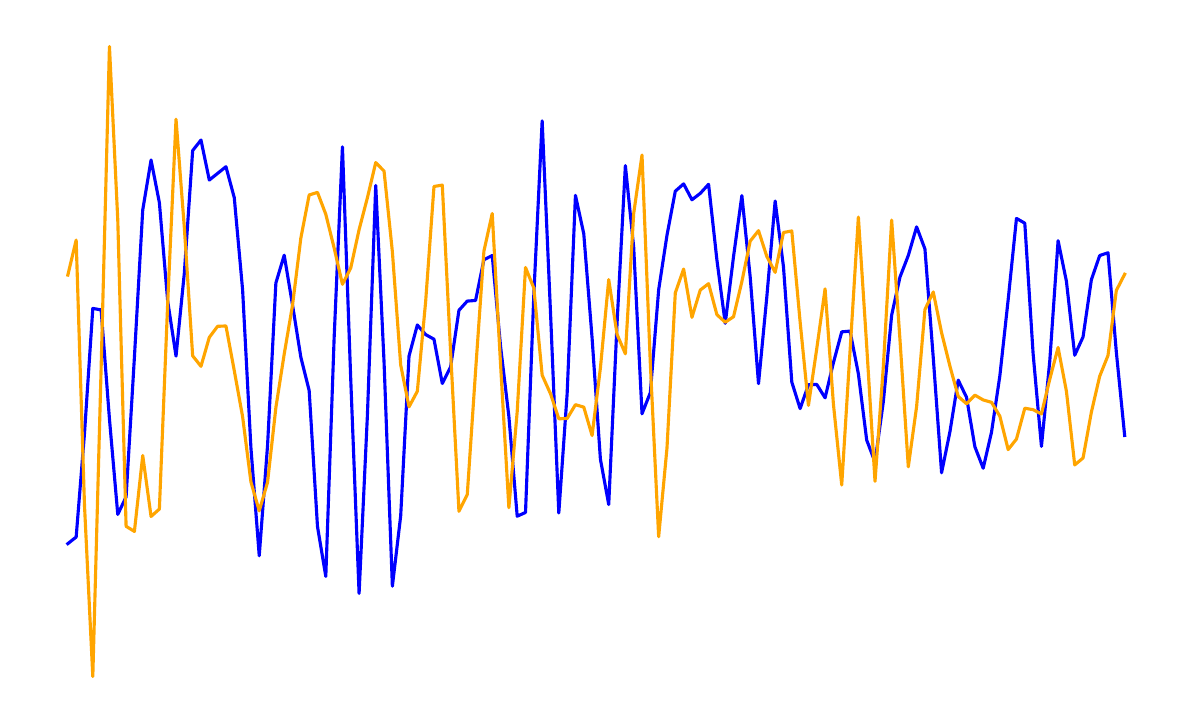}
		\includegraphics[height=0.16\textwidth]{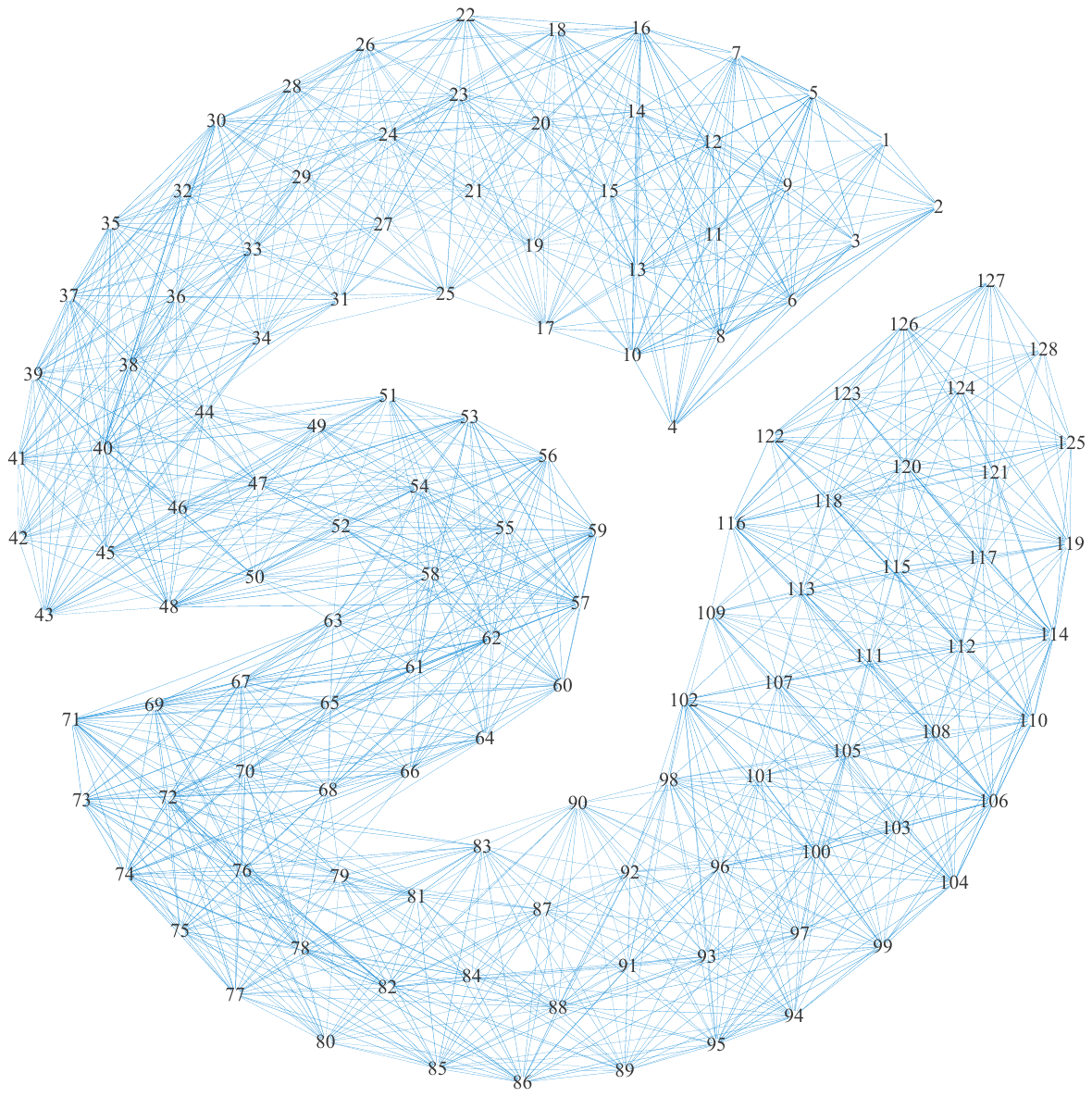}
            \includegraphics[height=0.16\textwidth]{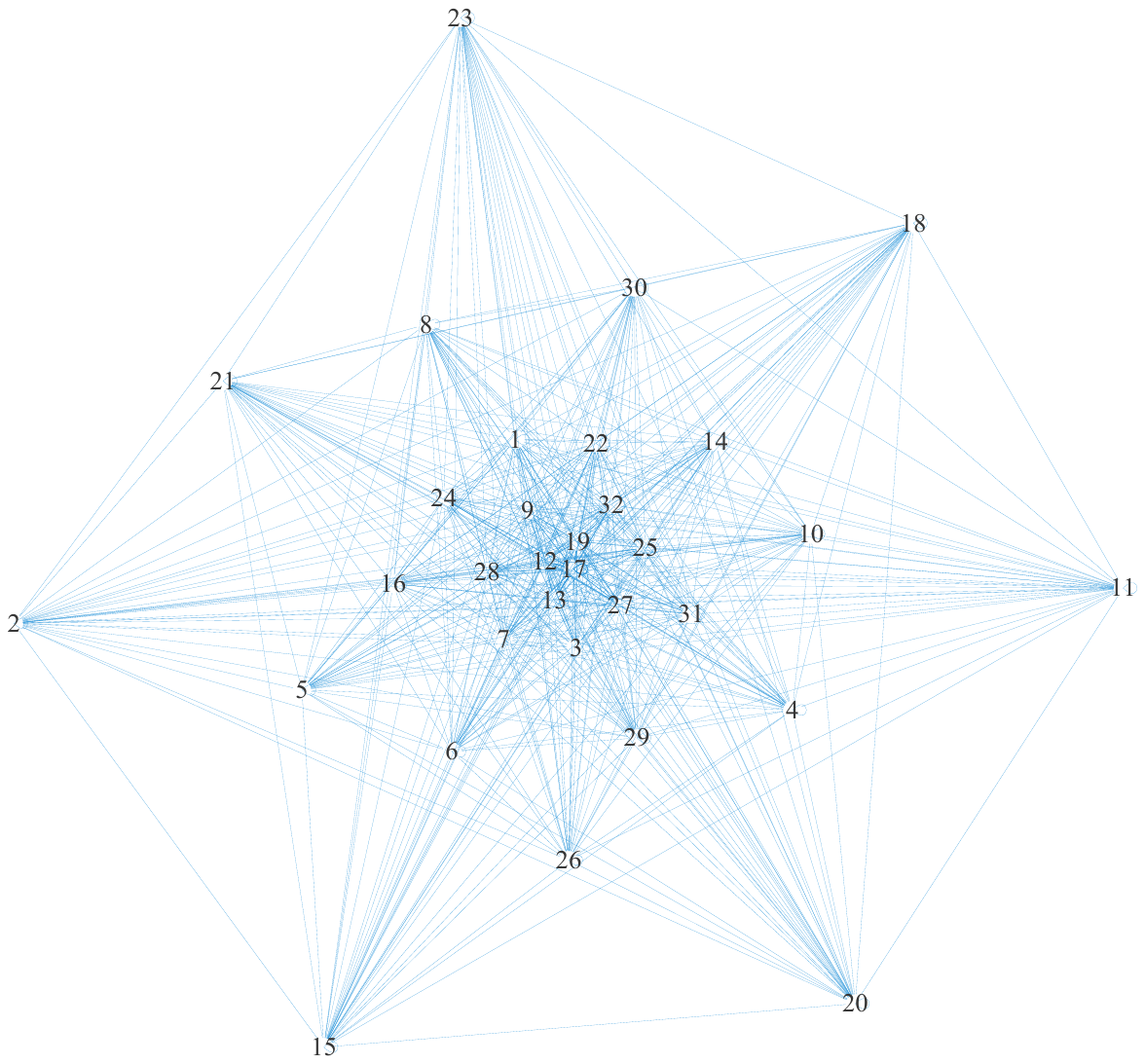}
            }
        \subfloat[QAM64]{
		\includegraphics[height=0.16\textwidth]{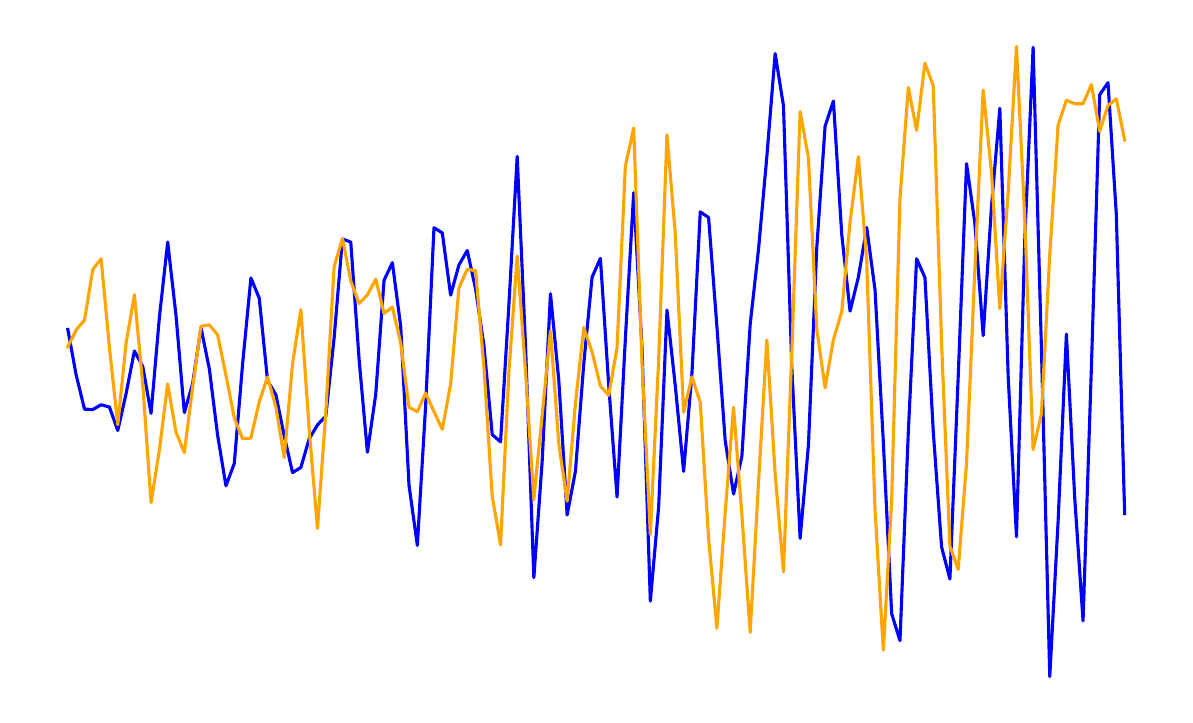}
		\includegraphics[height=0.16\textwidth]{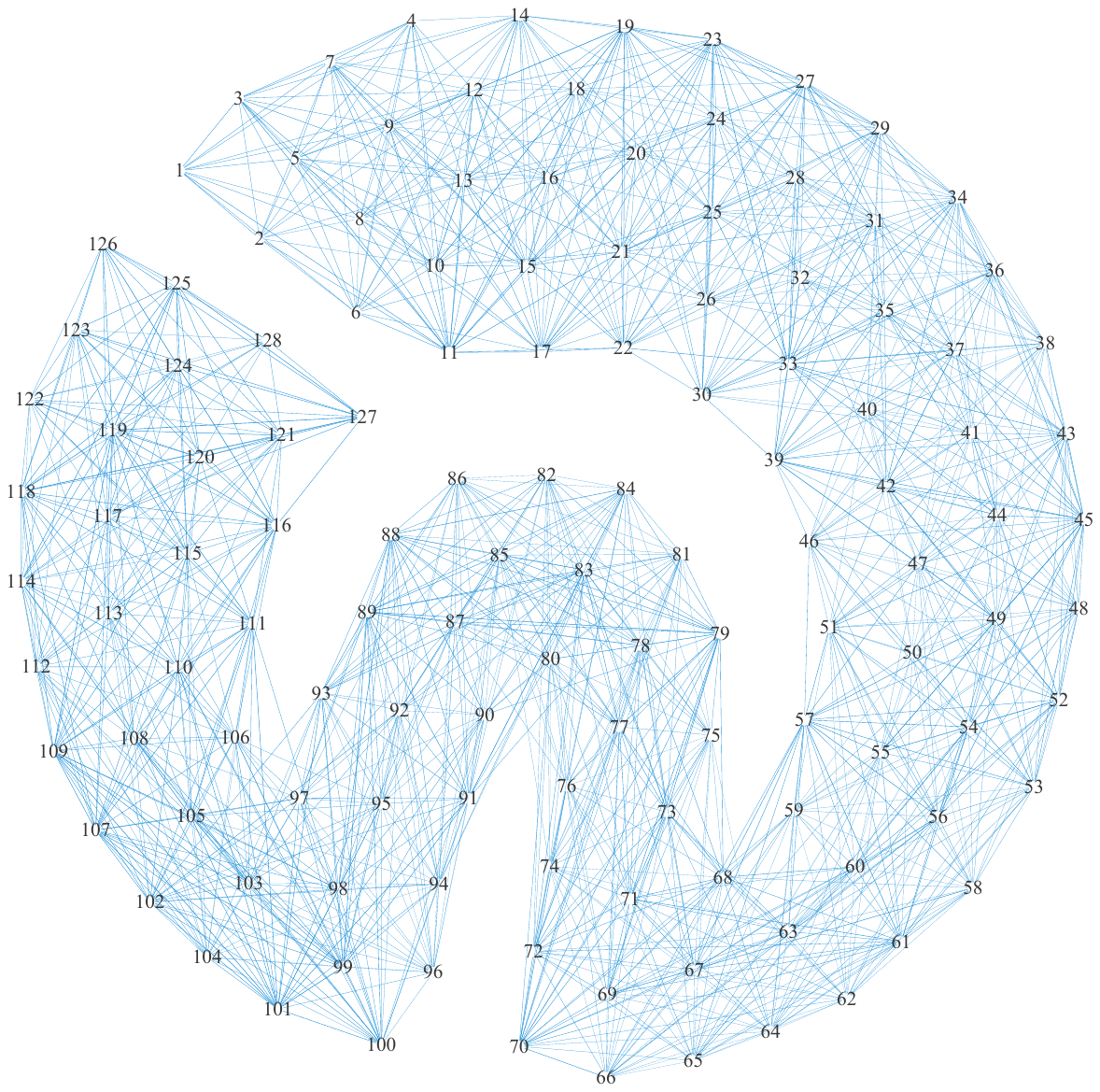}
            \includegraphics[height=0.16\textwidth]{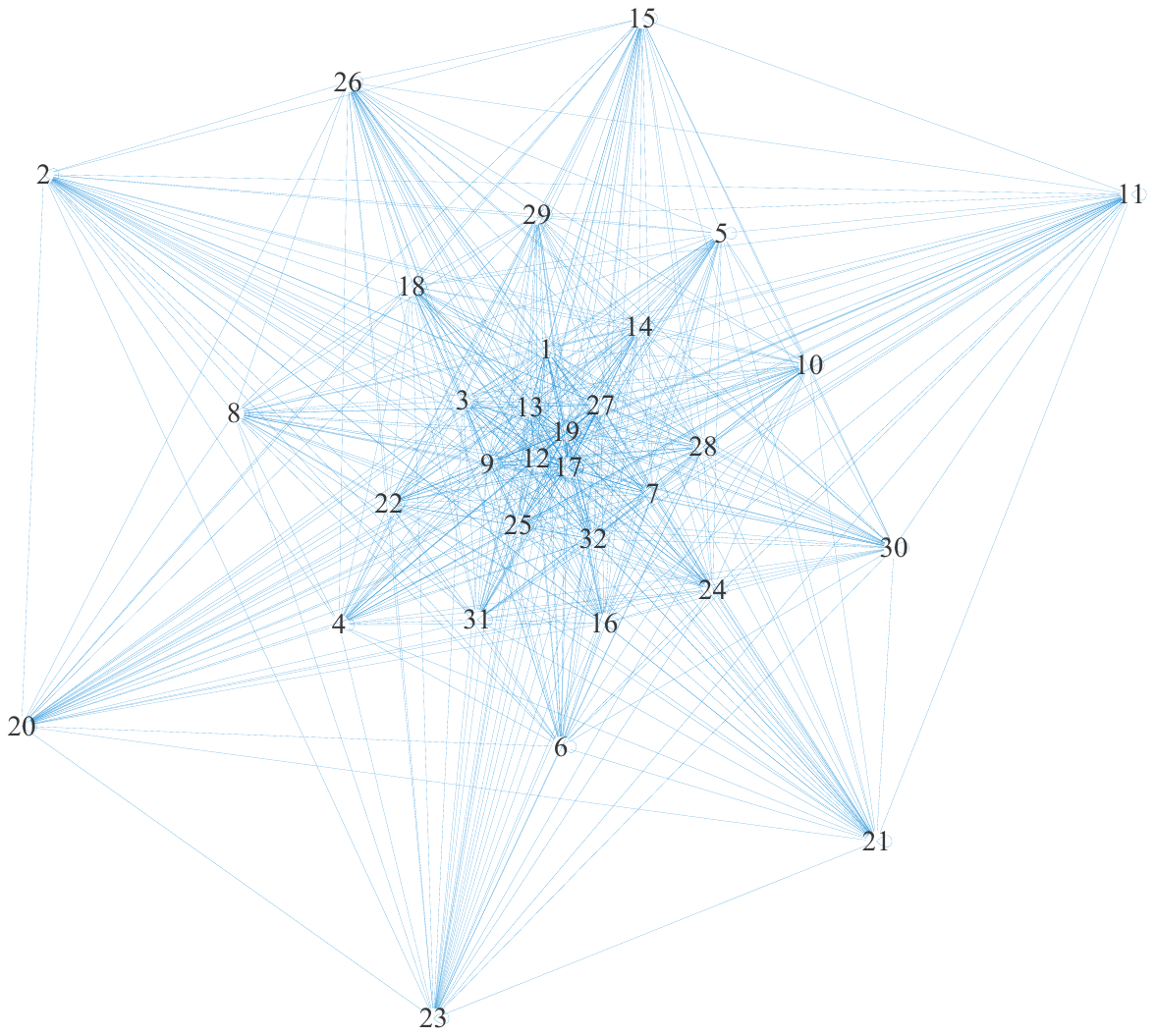}
            }    
 \\
  	\subfloat[WBFM]{
		\includegraphics[height=0.16\textwidth]{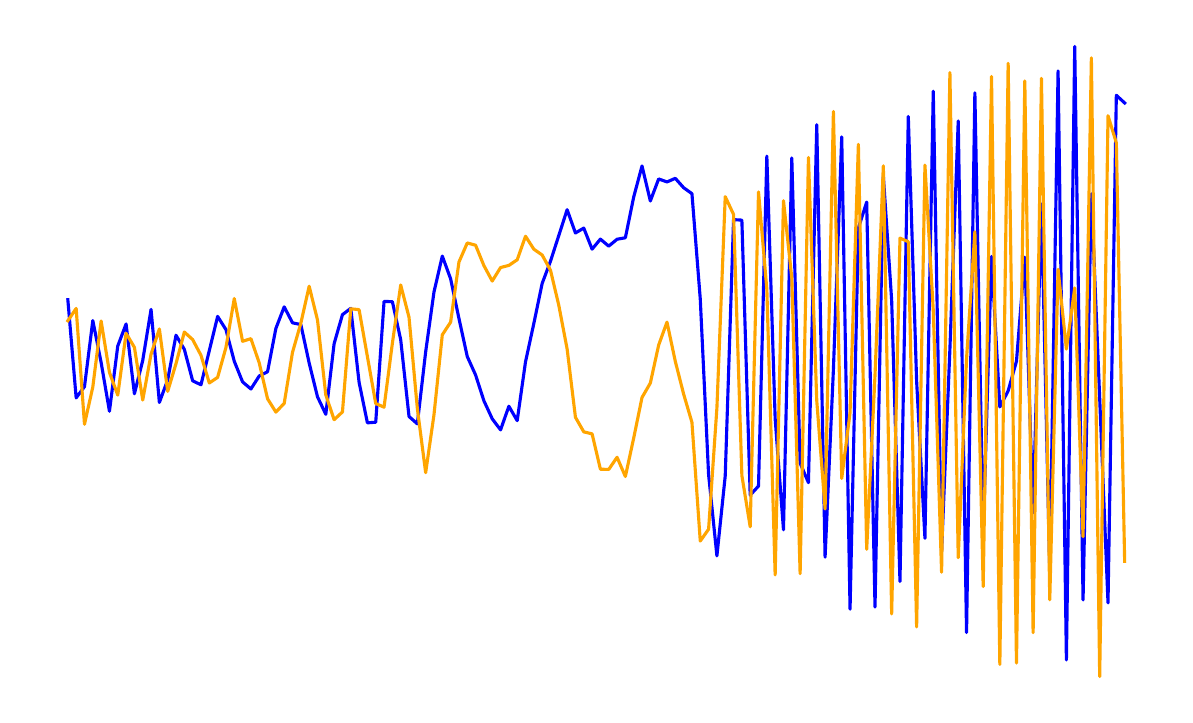}
		\includegraphics[height=0.16\textwidth]{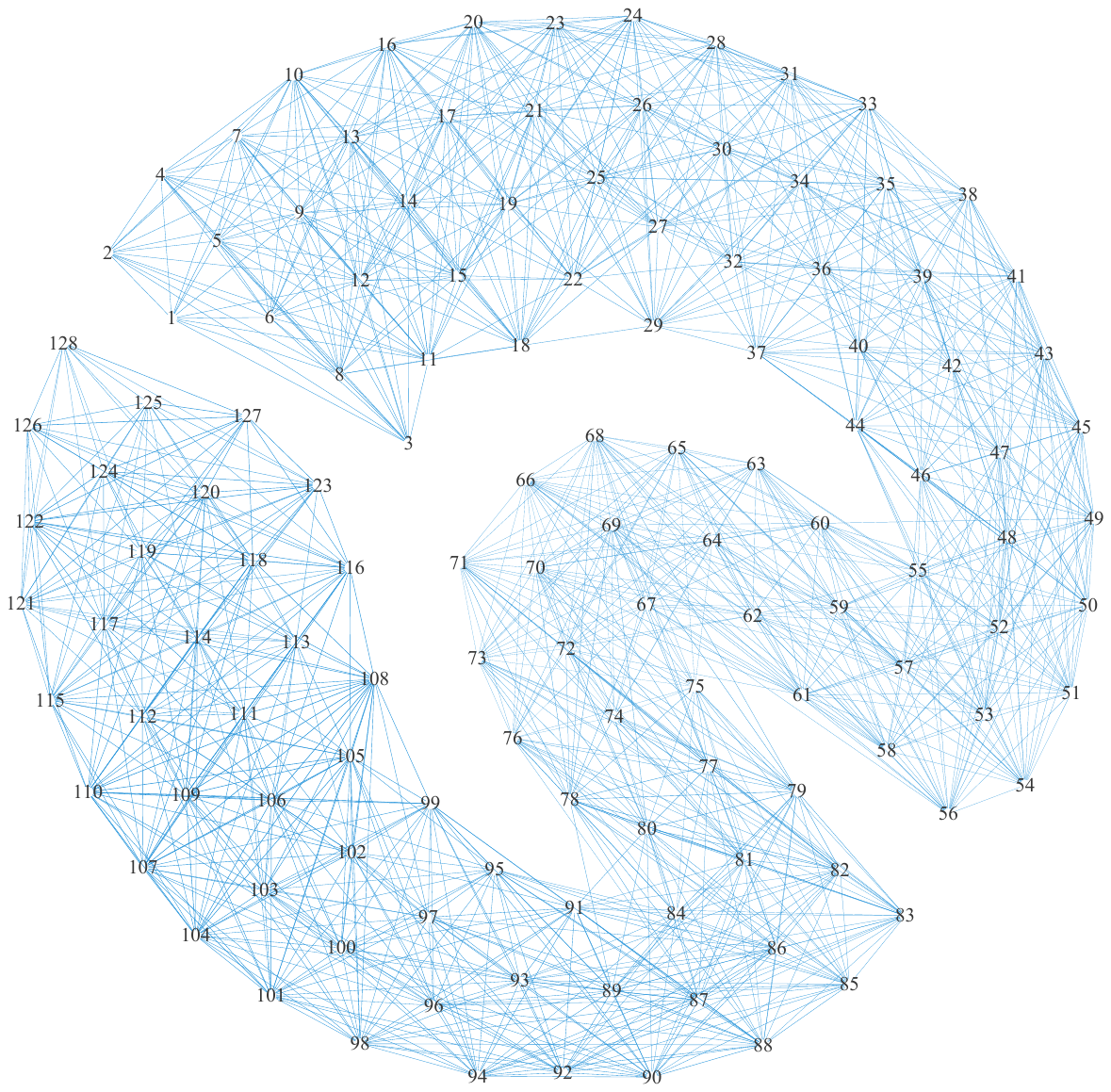}
            \includegraphics[height=0.16\textwidth]{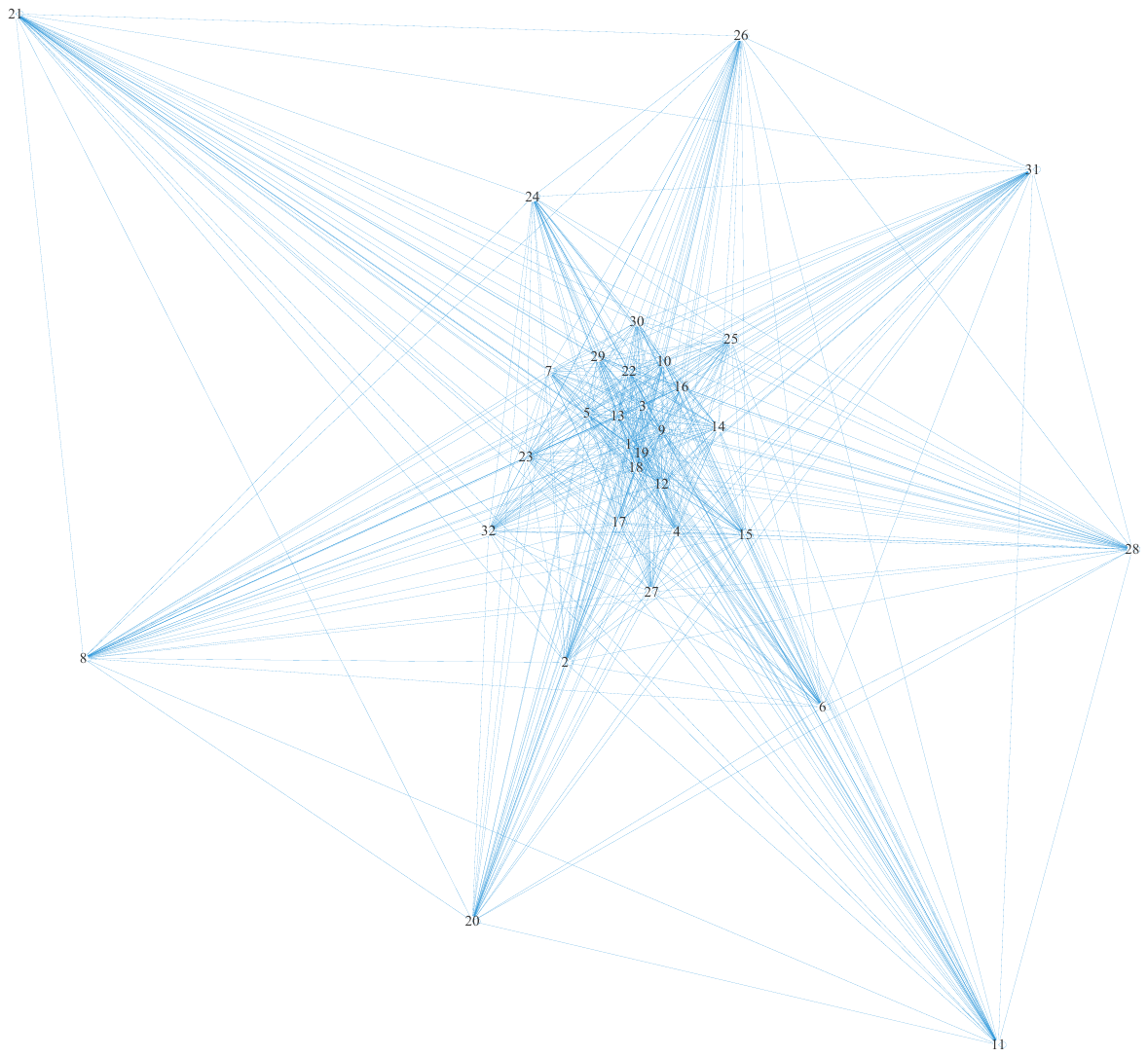}
            }
        \subfloat[AM-DSB]{
		\includegraphics[height=0.16\textwidth]{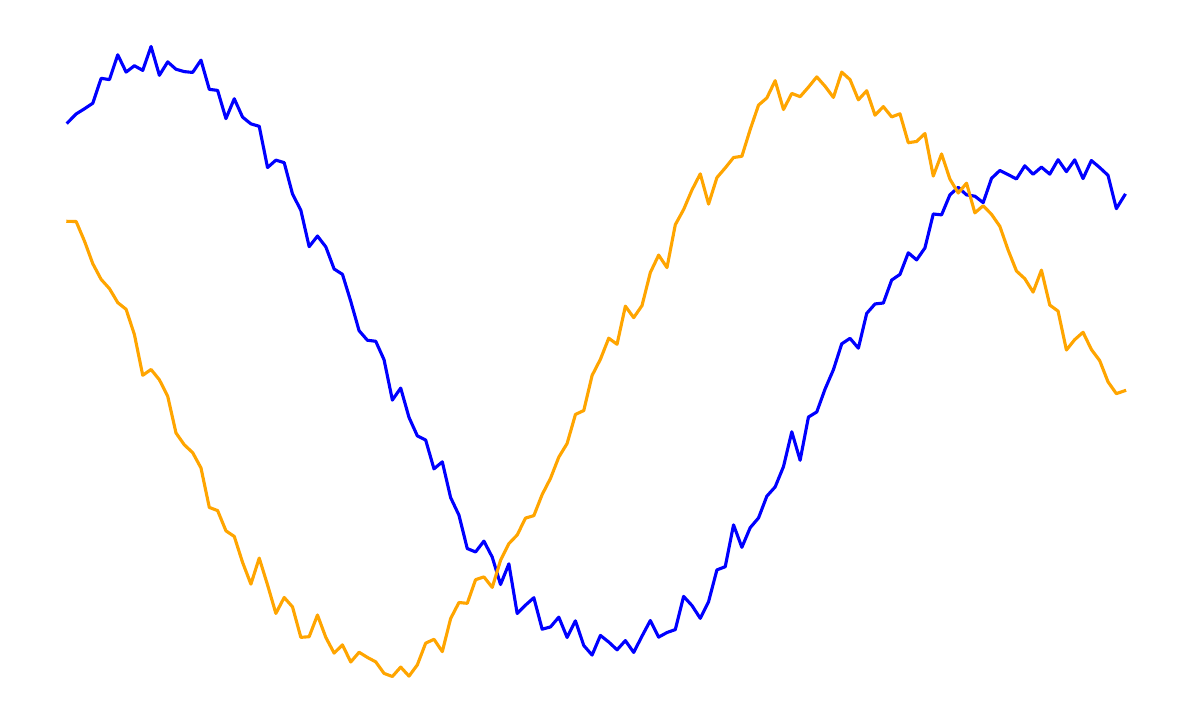}
		\includegraphics[height=0.16\textwidth]{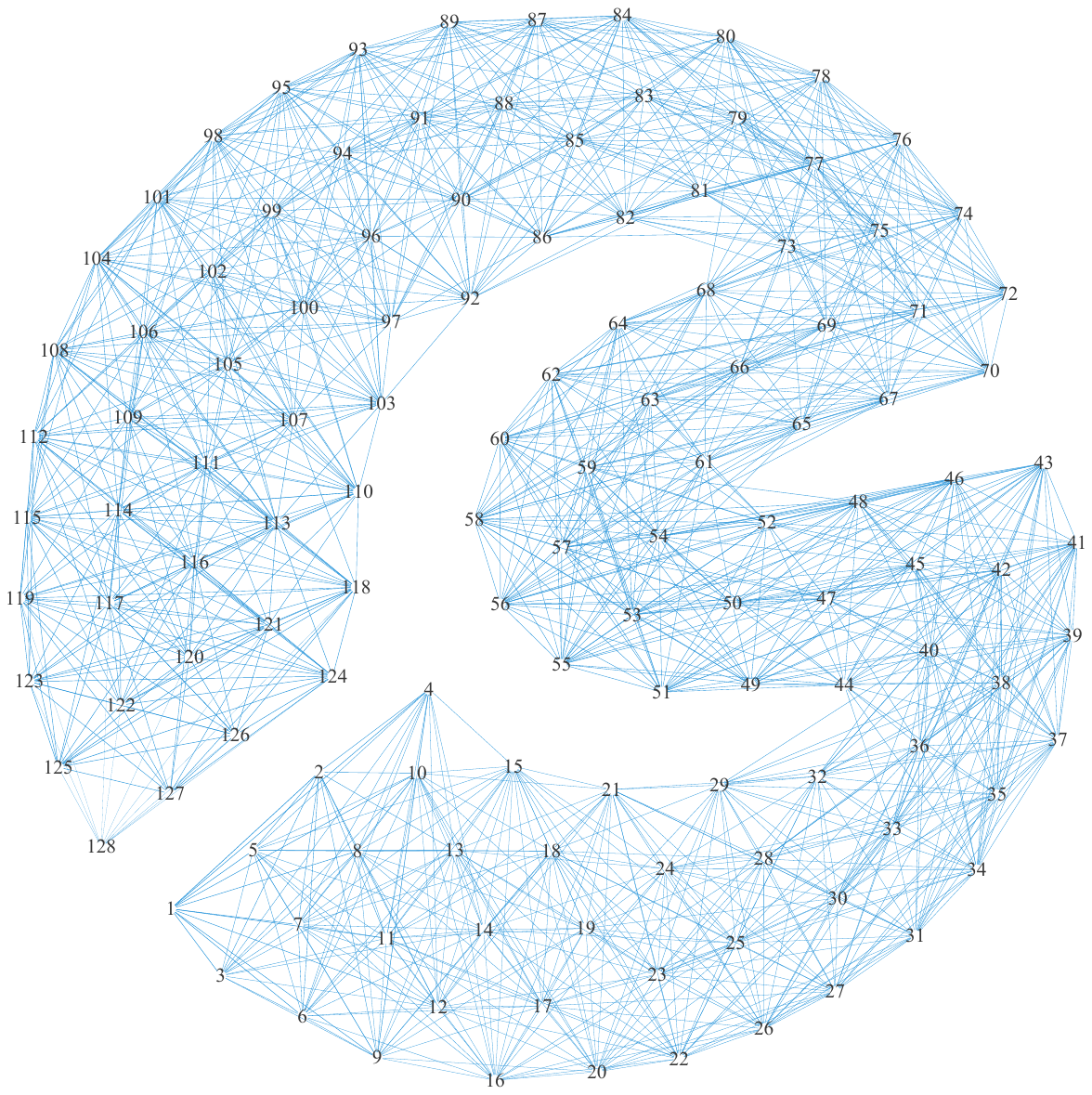}
            \includegraphics[height=0.16\textwidth]{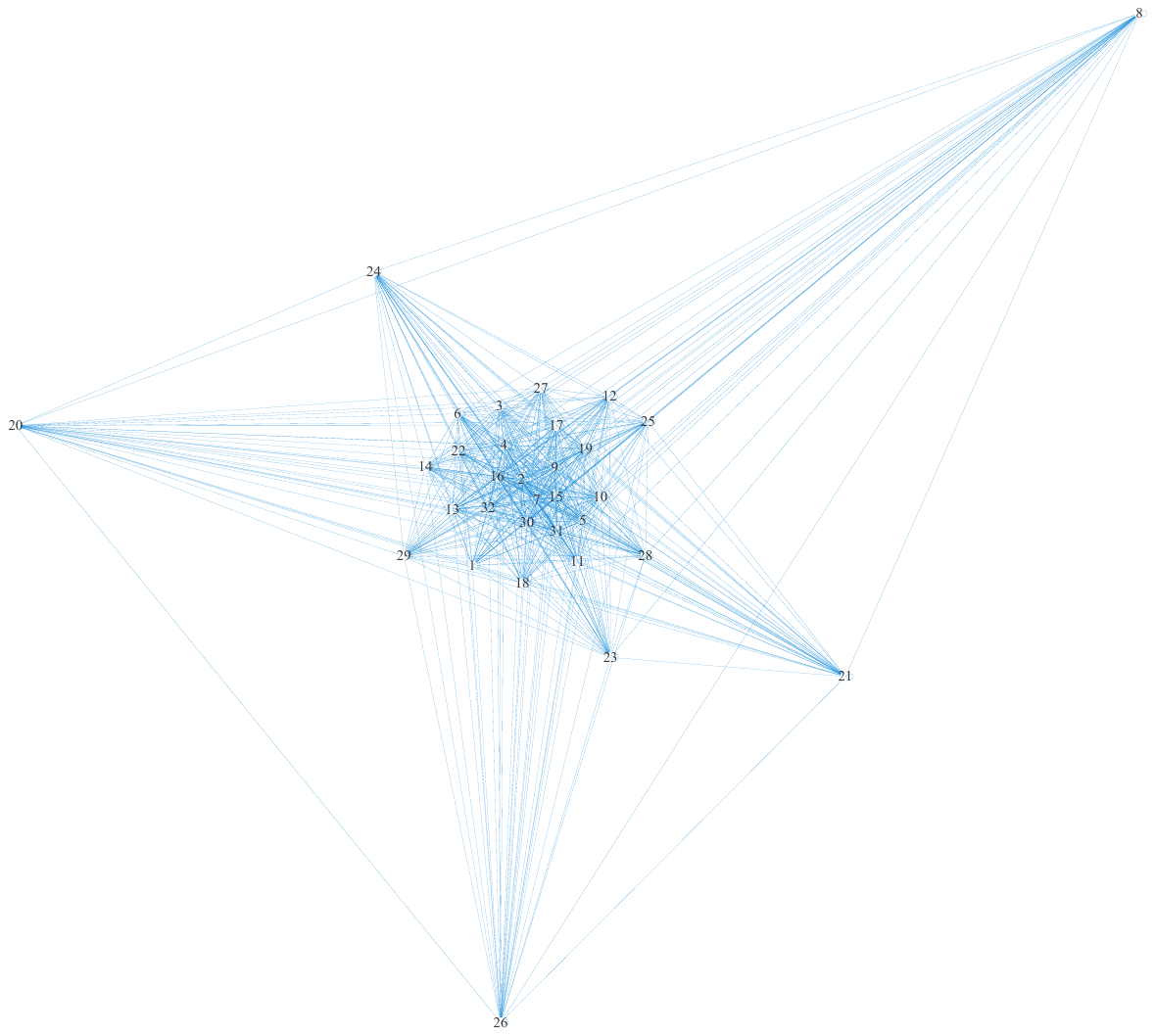}
            }    
 \\
  	\subfloat[AM-SSB]{
		\includegraphics[height=0.16\textwidth]{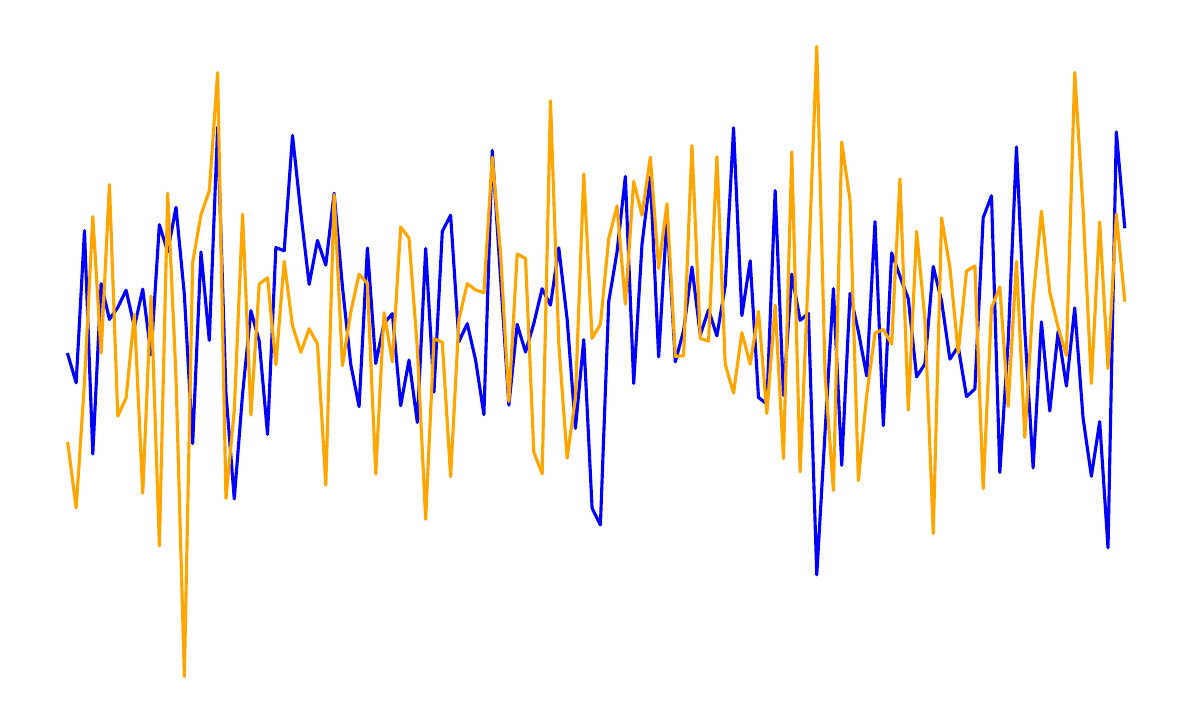}
		\includegraphics[height=0.16\textwidth]{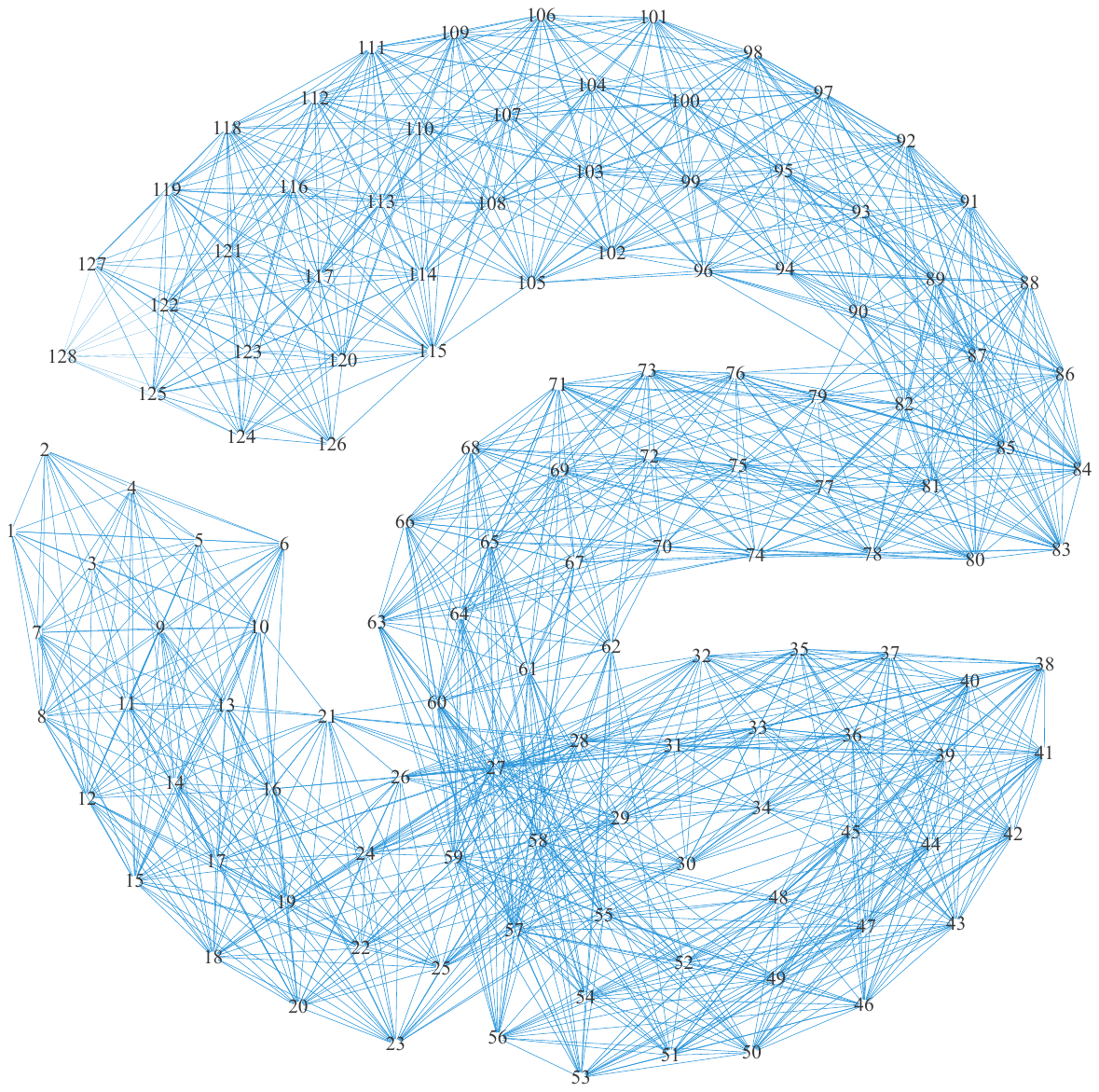}
            \includegraphics[height=0.16\textwidth]{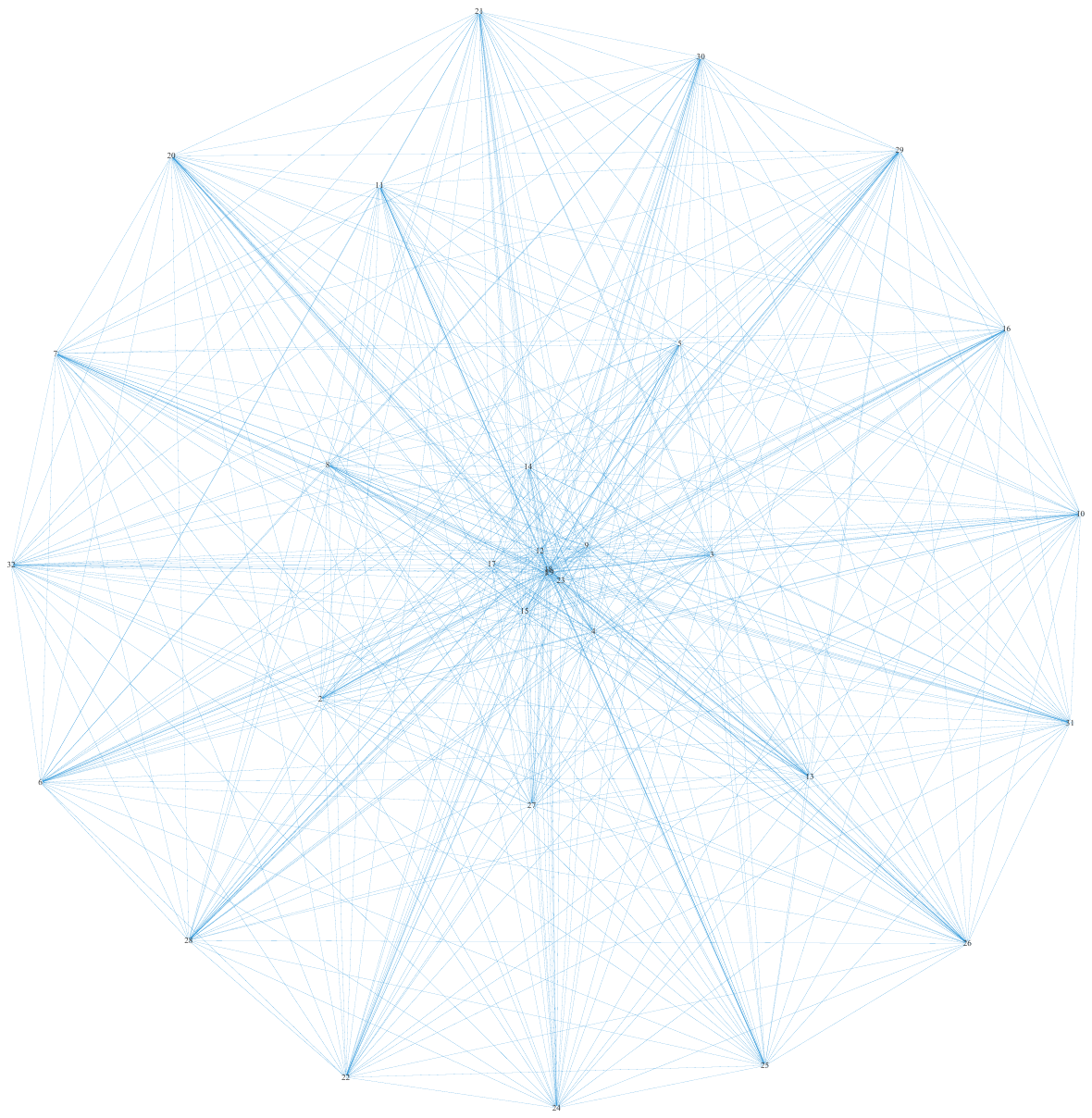}
            }
	\caption{The mapped graphs in STF-GCN for RML22. The pictures for each modulation type are, from left to right, the 16dB SNR I/Q signal, the corresponding initial STF graphical structure, and the attentively coarsened graphical structure after processing by the first PoolGAT layer.}
	\label{fig:graph} 
\end{figure*}
Fig. \ref{fig:graph} represents multi-domain fusion graph structures drawn with open source software Gephi \cite{Gephi}. The visualizations for each modulation type are from left to right: the I/Q signal, the corresponding initial STF graph structure, and the attentively coarsened graphical structure after processing by the first PoolGAT layer. The thickness of the connection lines on the graph structure reflects the size of the edge weights between the nodes. It is observed that there is a significant difference in the distribution and weights of the edges owned by each node on the original graph structure mapped by different modulation types, and thus the generated graph structure can have large differences. In addition, we can find that there are dense node clusters in the central region of the graph structure obtained by the PoolGAT layer through the coarsening and attention mechanism. Furthermore, there are large differences in the composition of node clusters of different modulation types, which proves that the PoolGAT layer can extract the node and edge features of the graph structure well and effectively while coarsening the graph structure at the same time. From the visualization results, the distribution and density of node clusters suggest that our method effectively extracts and coarsens local feature aggregation from the signal after urban channel fading, thereby ensuring the model's robust performance.
\begin{figure*}
	\centering
	\subfloat[RML2016.10a]{
		\includegraphics[width=0.32\textwidth]{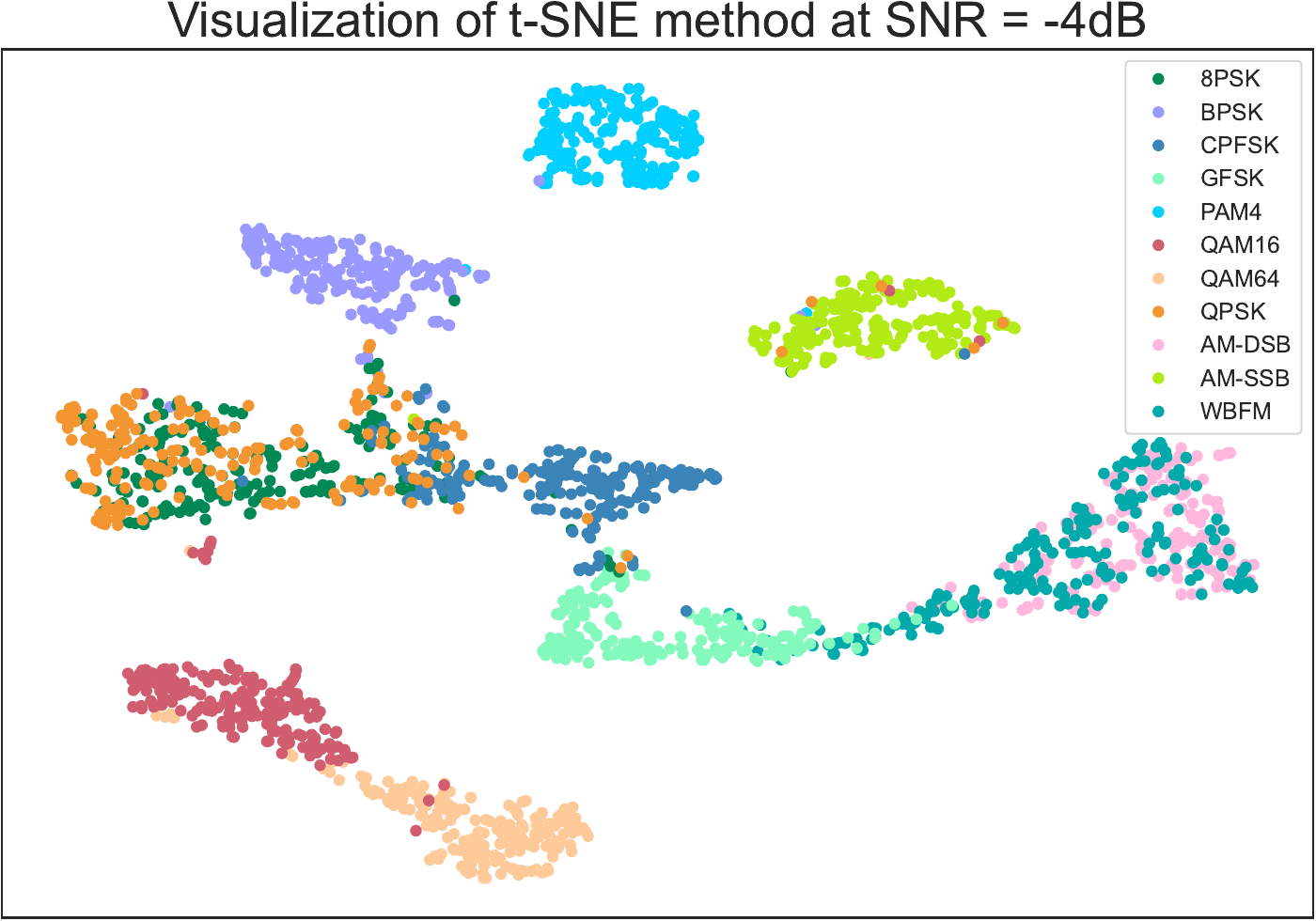}
		\includegraphics[width=0.32\textwidth]{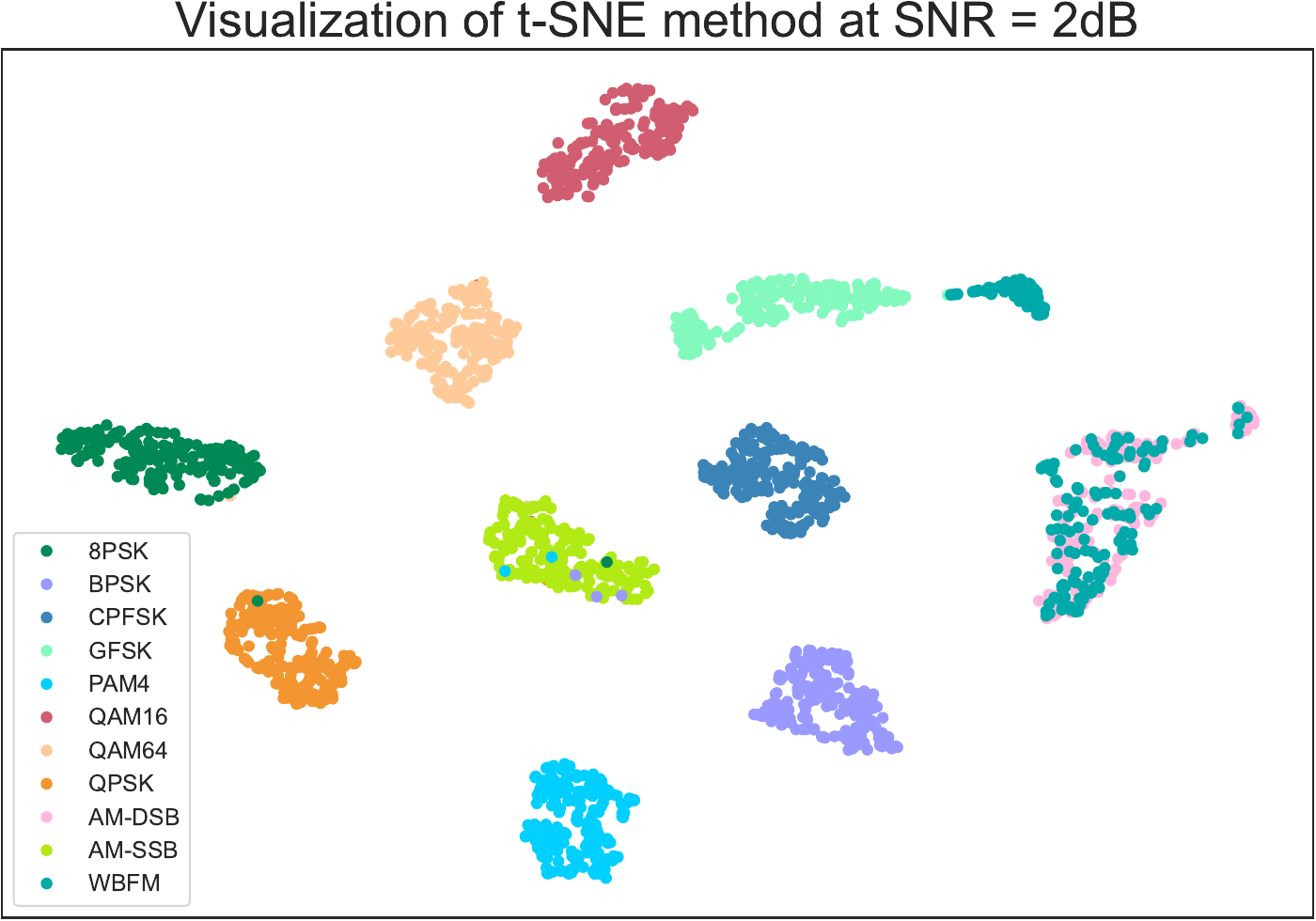}
            \includegraphics[width=0.32\textwidth]{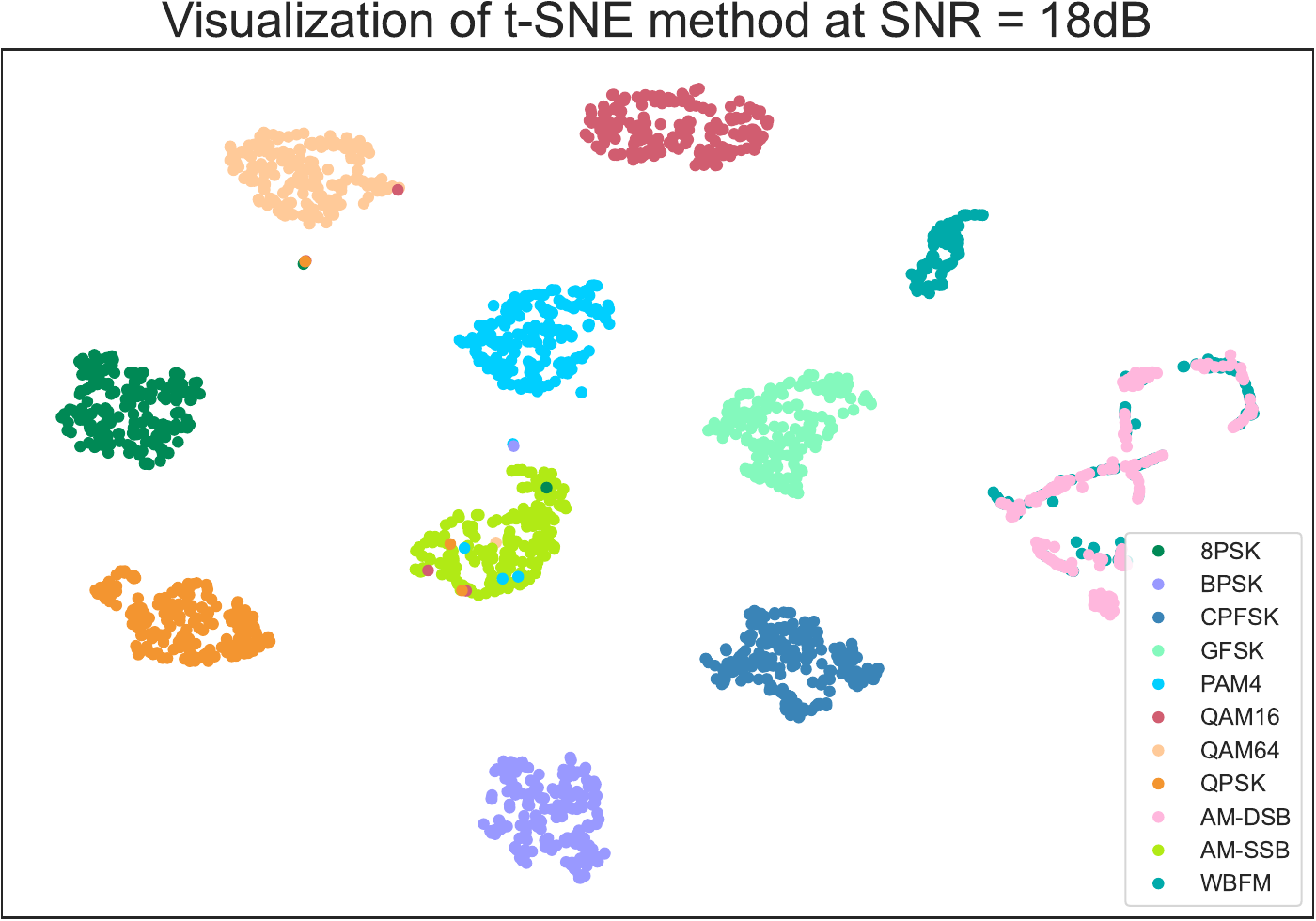}
            }
        \\
        \subfloat[RML22]{
            \includegraphics[width=0.32\textwidth]{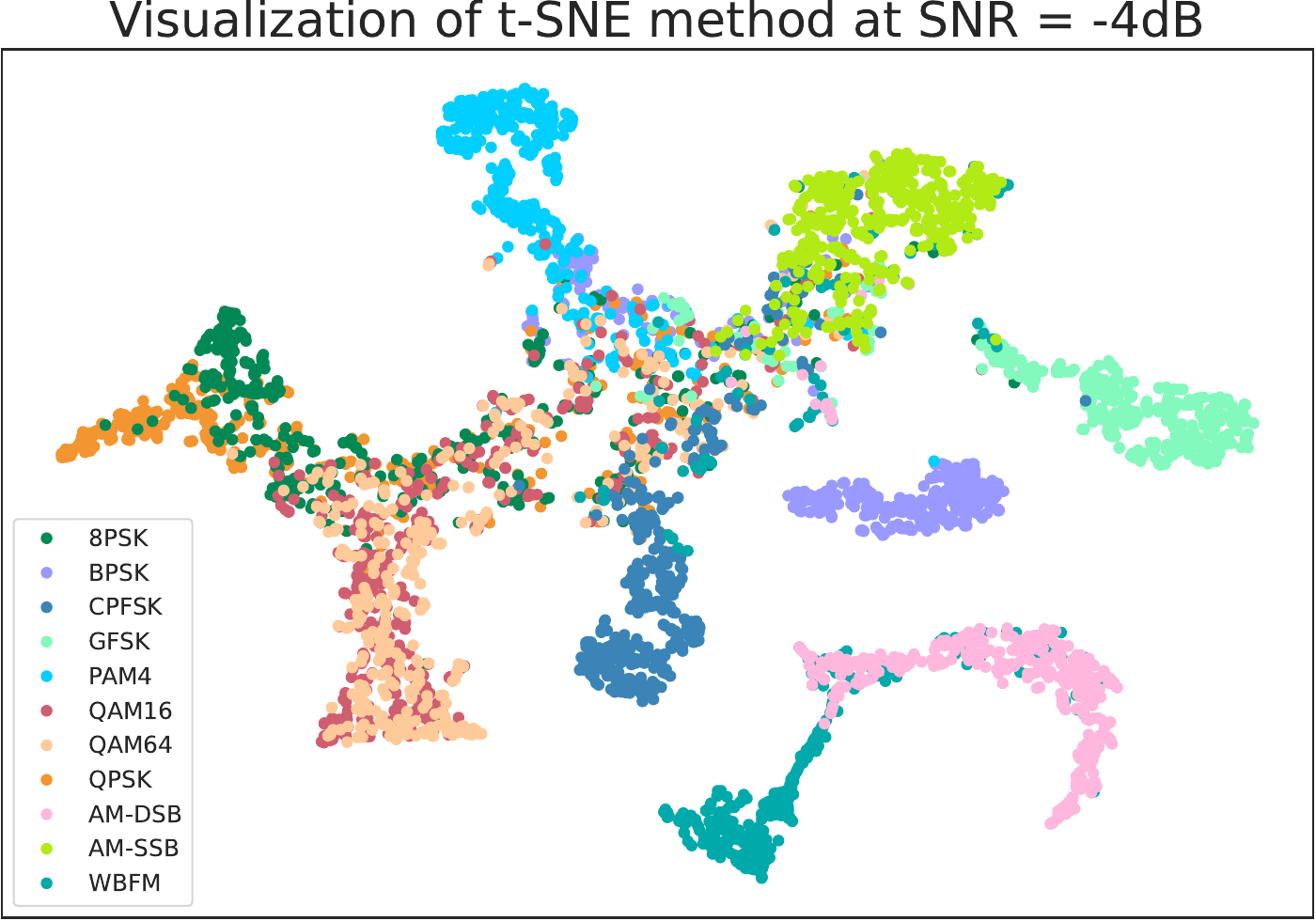}
		\includegraphics[width=0.32\textwidth]{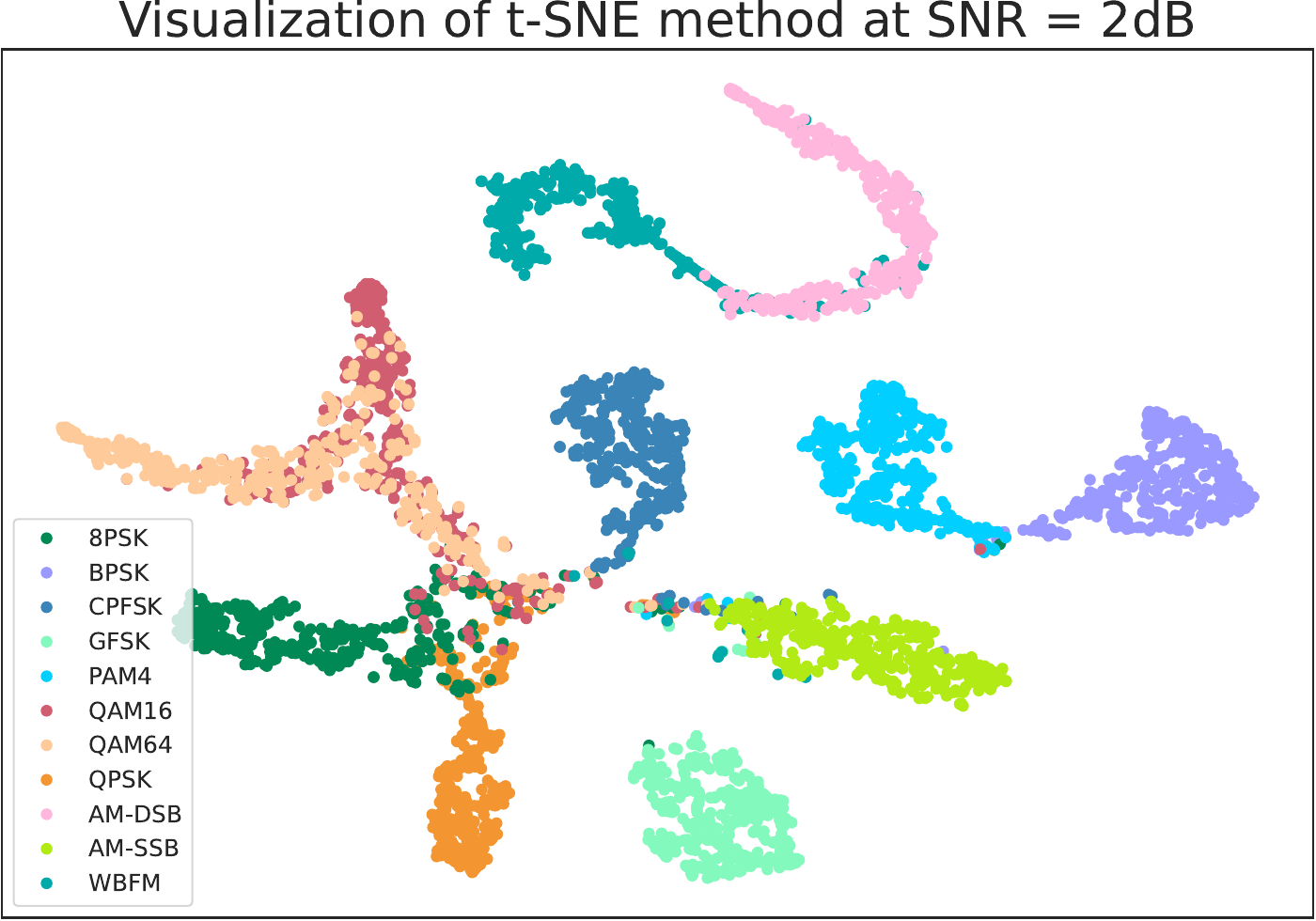}
            \includegraphics[width=0.32\textwidth]{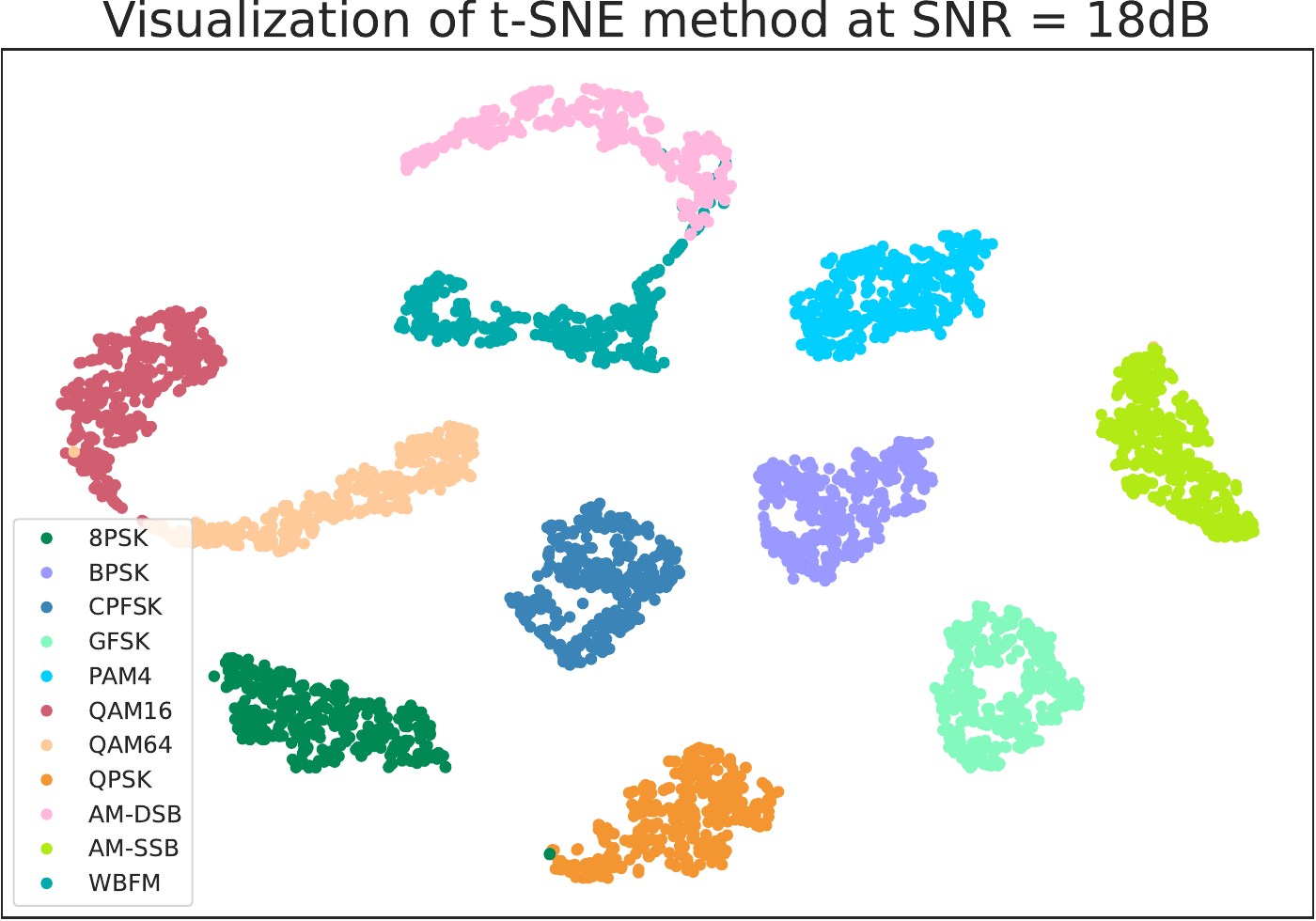}
            }  
        \caption{Visualization of fully connected layer output with t-SNE.}
	\label{fig:tsne} 
\end{figure*}

Finally, to assess the challenges of urban intelligent transportation under channel conditions characterized, we conduct t-SNE \cite{tsne} visualizations on the two datasets. Fig. \ref{fig:tsne} compares the visualization results on RML2016.10a and RML22 datasets at -4dB, 2dB, and 18dB SNRs. 
In the RML2016.10a dataset, the shapes of the clusters roughly show circular or compact structures, and as the SNR increases, the distribution of the signal categories gradually becomes divisible, and the boundaries between the clusters are more and more obvious, which suggests that the signal effective features have been learned.
In the RML22 dataset, the distribution of signals with different modulation types is more dispersed, with significant overlap between clusters and certain clusters exhibiting a large extended range. As the SNR increases, although the distribution of clusters becomes clearer, there is still a greater degree of extension with fuzzy boundaries, suggesting that this dataset is more challenging to learn in terms of feature representation. This further indicates that research on AMR for urban transportation systems is necessary in order to increase the accuracy and reliability of communication.

\section{conclusion}
In this paper, we propose an STF-GCN that enhances feature recognition by simultaneously learning representations in the spatial, temporal, and frequency domains through an adaptive correlation-based adjacency matrix. The proposed PoolGAT layer reduces the computational cost by using an attentional mechanism to coarsen and retain key features. We demonstrate the generalizability of our approach through comparative experiments with two different channels, and the strength of the simulation results highlights the effectiveness of our approach. In Furthermore, we explore the additional challenges faced by urban channel in our visualization analysis by comparing with LOS channel. In future work, we will explore the application of AMR in Multiple-Input Multiple-Output (MIMO) systems and effective solutions to reduce multipath, delay, and Doppler effects significantly to accommodate realistic scenarios required for complex urban communications.
\bibliographystyle{IEEEtran}
\bibliography{ref}

\end{document}